\documentclass[12pt]{article}
\usepackage{mathpazo}
\usepackage{amsmath}
\usepackage{graphicx}
\usepackage{enumerate}
\usepackage{natbib}
\usepackage{url} 

\usepackage[noend]{algcompatible}
\RequirePackage[OT1]{fontenc}
 \usepackage{color, colortbl}

 \usepackage[ruled,vlined]{algorithm2e}
 \usepackage{algpseudocode}
\usepackage{rotating}
\usepackage{fancyvrb}
\usepackage{animate}
\usepackage{xcolor}
\usepackage{listings}
\usepackage{pdflscape}
\usepackage{animate}
\usepackage{media9}
\usepackage{graphicx}
\usepackage{amssymb}
\usepackage{amsthm}
\usepackage{mathabx}
\usepackage{color}
\usepackage{amsopn}
\usepackage{amsmath}
\usepackage{curves}
\usepackage{mathabx}
\usepackage{pstricks}
\usepackage{color}
\usepackage{pstcol}
\usepackage{pst-plot}
\usepackage{pst-tree}
\usepackage{pst-eps}
\usepackage{multido}
\usepackage{pst-node}
\usepackage{pst-3d}
\usepackage{colordvi}
\usepackage{multirow}
\usepackage{pst-grad}
\usepackage{colortbl}
 \usepackage{lscape}
 \usepackage{xr-hyper}
 \usepackage{hyperref}

\definecolor{GGG}{RGB}{220,230,220}

\theoremstyle{plain}
\newtheorem{theorem}{Theorem}[section]

\newtheorem{definition}[theorem]{Definition}

\newcommand{\blind}{1}

\begin{document}

\def\spacingset#1{\renewcommand{\baselinestretch}%
{#1}\small\normalsize} \spacingset{1}


\if1\blind
{
	\title{\bf Nonparametric estimation of highest density regions for COVID-19}
  \author{P. Saavedra-Nieves\hspace{.2cm}\\Universidade de Santiago de Compostela}
  \date{}
  \maketitle  
} \fi

\if0\blind
{
  \bigskip
  \bigskip
  \bigskip
  \begin{center}
\end{center}
  \medskip
} \fi

\bigskip

\begin{abstract}
\noindent Highest density regions refer to level sets containing points of relatively high density. Their estimation from a random sample, generated from the underlying density, allows to determine the clusters of the corresponding distribution. This task can be accomplished considering different nonparametric perspectives. From a practical point of view, reconstructing highest density regions can be interpreted as a way of determining hot-spots, a crucial task for understanding COVID-19 space-time evolution. In this work, we compare the behavior of classical plug-in methods and a recently proposed hybrid algorithm for highest density regions estimation through an extensive simulation study. Both methodologies are applied to analyze a real data set about COVID-19 cases in the United States.
 
\end{abstract}

\noindent%
{\it Keywords:} Bootstrap, cluster, COVID-19, highest density region, kernel density estimation, $r-$convexity
 
 \tableofcontents
\spacingset{1.5} 

\section{Introduction}\label{intro}
As noted by the World Health Organization (WHO)\footnote{\url{https://www.who.int/}}, coronavirus disease usually known as COVID-19 is an infectious pathology caused by a newly discovered virus. Most people infected experience mild to moderate respiratory illness and recover without requiring special treatment, elderly people, and those with severe medical problems such as cardiovascular disease, diabetes, chronic respiratory disease, and cancer are more likely to develop serious illness.

 The origin of the virus is located in Wuhan City, Hubei Province, China at the end of 2019. China officially reported a total of 84025 cases and 4637 deaths in May 2020. European and American governments were unable to stop its fast expansion due to its high degree of contagiousness. In fact, the WHO on March 11th, declared the coronavirus a global pandemic. Figure 1 in Section 1 of Supplementary Material (SM) shows the daily evolution of COVID-19 confirmed cases distribution around the world from January 22nd and May 5th through an interactive map. The area of each red circle is proportional to the number of new confirmed cases. It was elaborated from available data on a GitHub repository\footnote{\url{https://github.com/CSSEGISandData/COVID-19}} of Johns Hopkins University \footnote{\url{https://systems.jhu.edu/research/public-health/ncov/}}. It should be noted that confirmed cases include presumptive positive cases and probable cases, in accordance with Centers for Disease Control and Prevention\footnote{\url{https://www.cdc.gov/}} guidelines as of April 2020.

The capacity of health systems has been overwhelmed in many countries, and COVID-19 became a public health problem. One of the most difficult issues to deal with was the lack of personal protective equipments for medical staff leading to a considerable number of infections among them. Shortage of disease tests or masks to be used by population in order to avoid the increases of infections were other important questions that authorities had to solve. More efficient solutions to manage the expansion of the virus could be provided by public entities of a country if areas where coronavirus incidence is extremely high were detected. In particular, real-time evolution analysis of these subpopulations or clusters would provide useful information to rule decisions such as prioritize the supply of sanitary material in a specific time moment or facilitate masks to the populations that lives in zones where an incidence peak of the virus is observed. Therefore, estimating regions where the coronavirus incidence is remarkably high is one of the main aims of this work. Mathematically,
this can be seen as a density level set estimation problem. Formally, given a random sample of points $\mathcal{X}_n=\{X_1,\cdots,X_n\}$ from a density $f$, level set estimation theory deals with the problem of reconstructing, for a given level $t>0$, the unknown set  \begin{equation}\label{conjuntonivel1}G(t)=\{x\in\mathbb{R}^d:f(x)\geq t\}.\end{equation}
 \cite{hartigan75} puts forward the idea of connecting the concept of clusters with the connected components of the level set defined in (\ref{conjuntonivel1}). The generalization of this definition in the directional setting and for general manifolds can be found in \cite{saavedra2020nonparametric} and \cite{rinaldo}, respectively. Note that the number of these clusters is usually smaller than the number of modes (local maxima of $f$). These two concepts are closely related although due to its geometrical nature, the concept of cluster is generally easier to handle. One non-minor problem of the definition in (\ref{conjuntonivel1}) is that it relies on the user-specified level $t$ and which poses some drawbacks for practical purposes. Additionally, the areas of the distribution support
 where $f$ is close to zero deserve less interest since the probability of finding observations from the underlying distribution is usually very low. Then, an alternative definition for density level sets is considered next. Given $\tau\in (0,1)$, it is defined the highest density region (HDR) of probability $1-\tau$ as the level set
 \begin{equation}\label{conjuntonivel2}
 L(\tau)=\{x\in\mathbb{R}^d:f(x)\geq f_\tau\}
 \end{equation}where\vspace{-.4cm}
 \begin{equation}\label{umbral}
 f_\tau=\sup
 \left\{y\in(0,\infty):\int_{G(y)}f(t) dt\geq1-\tau\right\}.
 \end{equation}
Equation (\ref{umbral}) shows that $f_\tau$ is equal to the largest threshold such that the set $L(\tau)$ has a probability greater than or equal to $1-\tau$ with
 respect to the distribution induced by $f$. Note that $L(\tau)$ is equal to the minimum volume level set with probability content at least $1-\tau$, see \cite{pol3} and \cite{gar}. In practice, if $\hat{G}(t)$ denotes an estimator of the level set $G(t)$ defined in (\ref{conjuntonivel1}) obtained from $\mathcal{X}_n$, $f_\tau$ can be estimated as
 $$\hat{f}_\tau=\max\{t>0:\mathbb{P}_n(\hat{G}(t))\geq 1-\tau\}$$where
 $\mathbb{P}_n$ denotes the empirical probability measure induced by $\mathcal{X}_n$.

\begin{figure}
  \hspace{.2cm}\includegraphics[height=125pt,width=158pt]{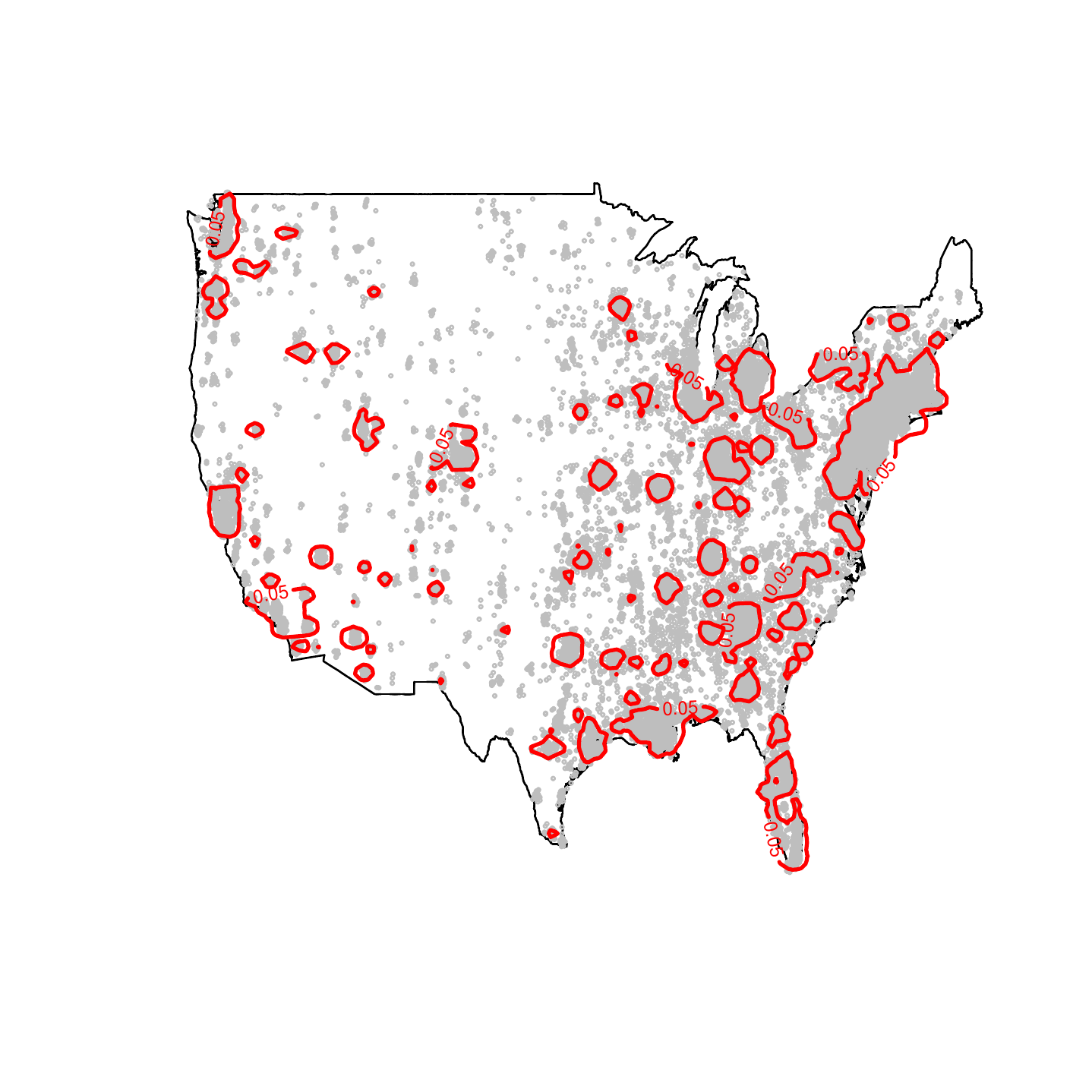}\hspace{-.8cm}\includegraphics[height=125pt,width=158pt]{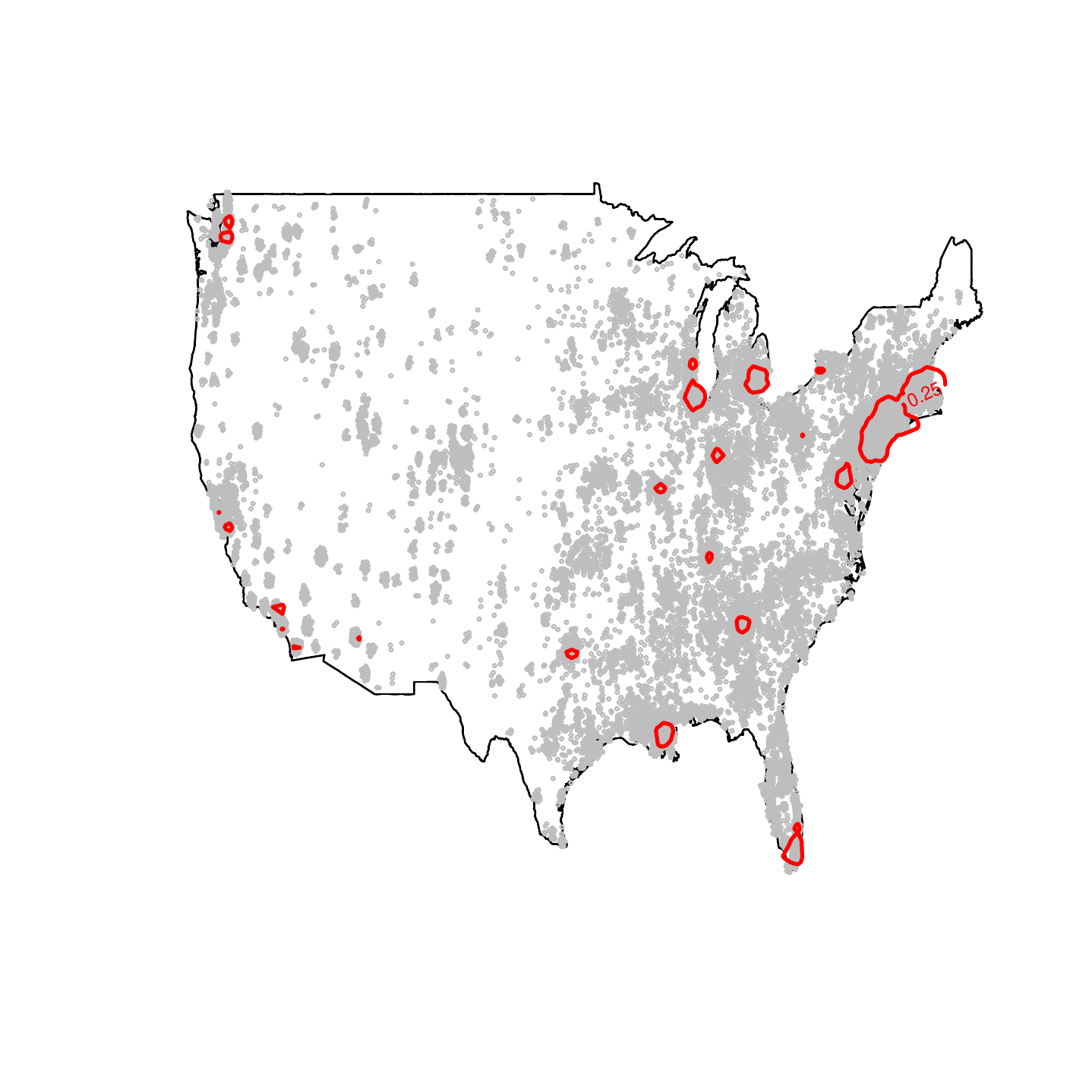}\hspace{-.8cm}\includegraphics[height=125pt,width=158pt]{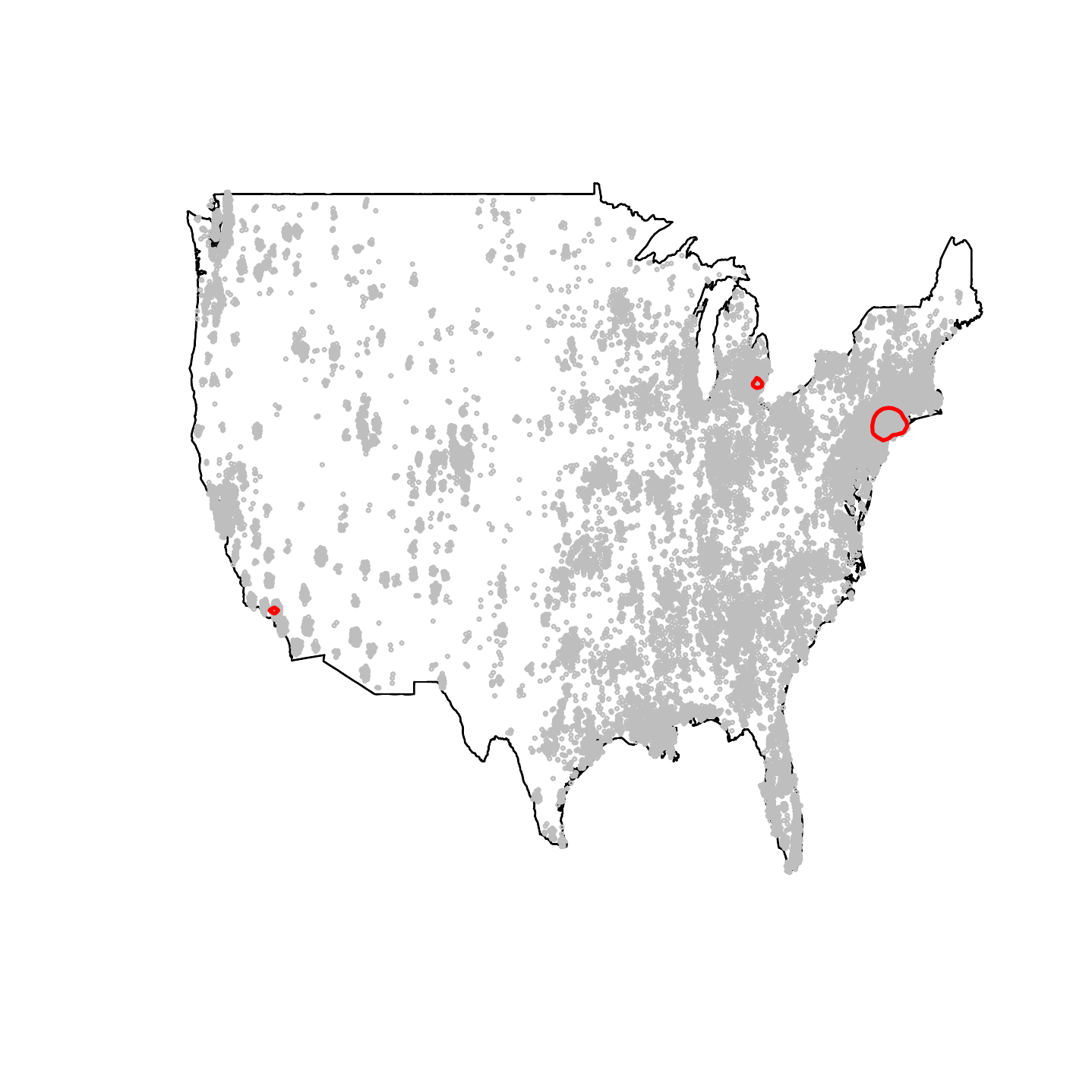}\vspace{-.85cm}\\
  	\caption{Contours of HDRs for $\tau=0.05$ (left), $\tau=0.25$ (center) and $\tau=0.5$ (right).}\label{contor}
  \end{figure}

Figure \ref{contor} shows the geographical coordinates (slightly jittered) of confirmed cases of coronavirus reported in the United States. This data set is also available in the GitHub repository of Johns Hopkins University. As an illustration, contours of the HDRs have been represented for different values of $\tau$. Observe that the effective support of $f$ is represented for values of $\tau$
  close to zero. However, for higher values of $\tau$, HDR coincides with the domain concentrated around the greatest modes or $\tau-$clusters of the distribution. Therefore, the largest incidence peaks of COVID-19 can be determined from $\tau$ values close to one. Of course, HDRs could be also estimated globally at world level. However, the lack of uniformity in data registers between countries and even between different administrative divisions in a same country hampers the possibility of carrying out a global analysis.

    Nonparametric estimation of HDRs has been considered in the statistical literature from three different perspectives, namely plug-in algorithms, excess mass approaches and hybrid methods. See \cite{rosaa} for a review on this topic. The most simple and common reconstruction is provided by plug-in algorithms. Specifically, they propose to estimate the HDR $L(\tau)$ as $\hat{L}(\tau) = \{f_n\geq \hat{f}_\tau\}$ where $f_n$ is usually the kernel density estimator, which is defined at point $x$ as
    $$f_n(x)=\frac{1}{nH^d}\sum_{i=1}^n K\left(\frac{x-X_i}{H}\right),
    $$where $H $ is a bandwidth and $K$, a kernel function. According to \cite{hynd}, $\hat{f}_\tau$ could be determined as the $\tau-$quantile of $f_n(\mathcal{X}_n)$. In this work, we will focus in the hybrid methodology which is a \emph{hybrid} of a plug-in approach, because it smooths $\mathcal{X}_n$, and excess mass methods, because it assumes some shape condition on the HDR to be reconstructed. A classical hybrid method, called granulometric smoothing method, is proposed in \cite{r2}. Designed for reconstructing HDRs, it assumes that $L(\tau)$ and the closure of its complement are both $r-$convex. The geometric condition of $r-$convexity is a much more flexible geometrical property than  convexity as it will be shown later. In fact, this shape property allows to consider more than one connected component, an interesting feature for practical purposes. However, the real value of $r$ is usually unknown as the theoretical HDR to be reconstructed. \cite{rosaa} establish an alternative hybrid method for estimating HDRs also under $r-$convexity assumptions but, unlike granulometric algorithm, suggesting a consistent estimator of the parameter $r$ which plays an important role on the geometry of the HDR. Although satisfactory theoretical results have been obtained for this recent proposal, its practical performance has not been checked yet. An extensive simulation study is carried out in this work to analyze its behavior. In addition, this method is used in practice for estimating HDRs for COVID-19 in the United States on a weekly bases, providing a space-time evolution description of the pandemic spread.

  This paper is organized as follows. Some mathematical background is introduced in Section \ref{mathtools}. First, geometric assumptions on the HDR and the optimal value of $r$ to be estimated from the sample $\mathcal{X}_n$ are discussed. Then, the main theoretical aspects of hybrid method for reconstructing HDRs proposed in \cite{rosaa} are reviewed. In Section \ref{4}, we describe in detail the practical implementation of the algorithm. In Section \ref{simus}, the performance of this method will be checked through simulations. Additionally, our proposal will be used in practice in order to reconstruct HDRs for COVID-19 in Section \ref{realdataanalysis}. Then, the time and spatial evolution of its clusters can be studied. Finally, conclusions are exposed in Section \ref{conclusions}. Sections 1 and 2 of SM contain interactive representations for the real data results.

  \section{Mathemathical tools}\label{mathtools}

    Usually required geometric assumptions on the HDRs and its nonparametric estimation will be reviewed next. In particular, the role of the usually unknown parameter $r$ will be discussed. Furthermore, the hybrid method proposed in \cite{rosaa} and its main theoretical aspects are briefly reviewed.

  HDRs are assumed to be $r-$convex for some $r>0$ in \cite{rosaa}. Definition \ref{rconvexi} establishes formally the definition of this geometric property.
  \begin{definition}\label{rconvexi}
  	A closed set $A\subset\mathbb{R}^d$ is said to be $r-$convex, for some $r>0$, if $A=C_{r}(A)$, where
  	$$C_{r}(A)=\bigcap_{\{B_r(x):B_r(x)\cap
  		A=\emptyset\}}\left(B_r(x)\right)^c$$
  	denotes the $r-$convex hull of $A$ and $B_r(x)$, the open ball with
  	center $x$ and radius $r$, whereas $T^c$ denotes the complementary of $T$.
  \end{definition}

\begin{figure}[h!]
  	\hspace{-.85cm}\includegraphics[height=150pt,width=175pt]{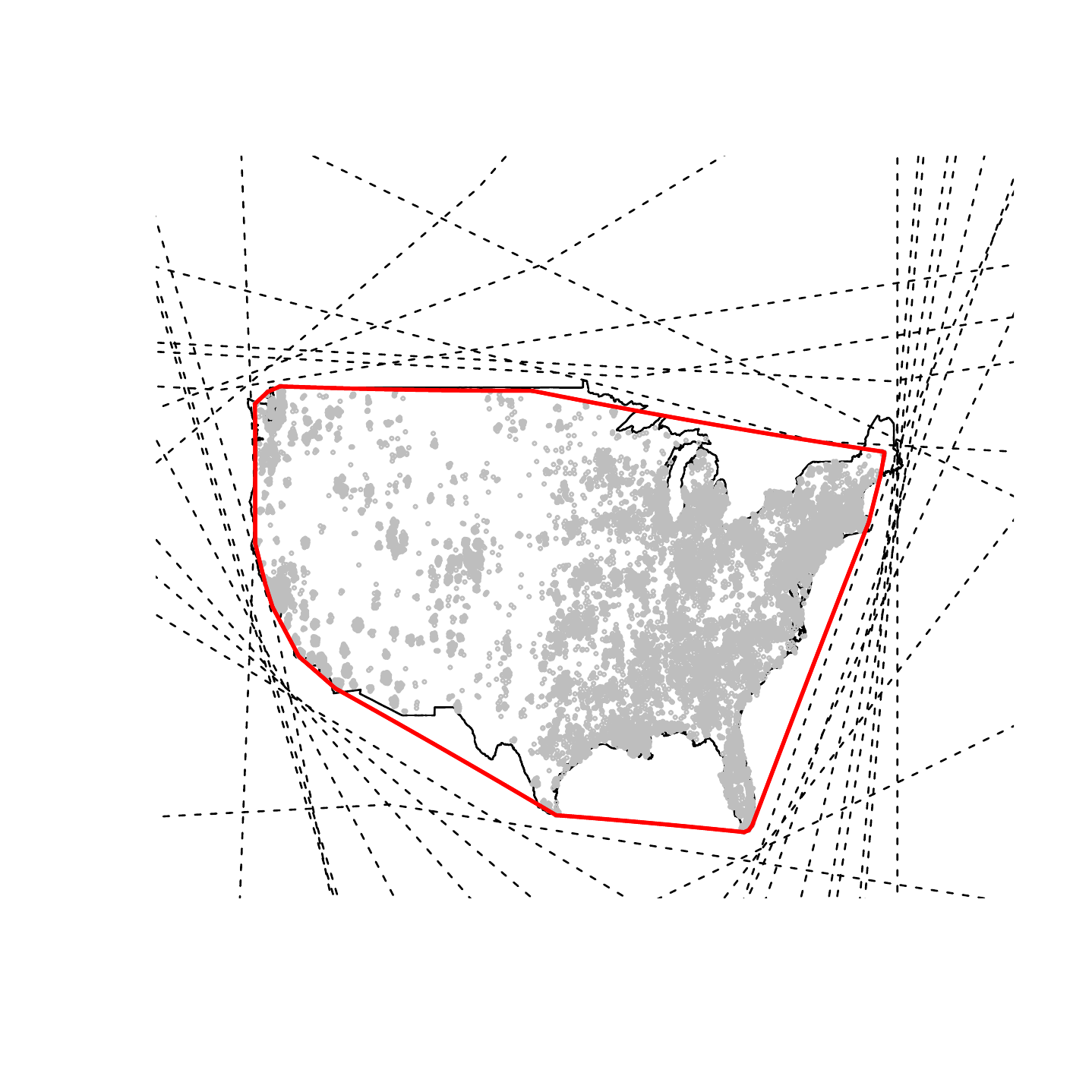}\hspace{-.8cm}\includegraphics[height=150pt,width=175pt]{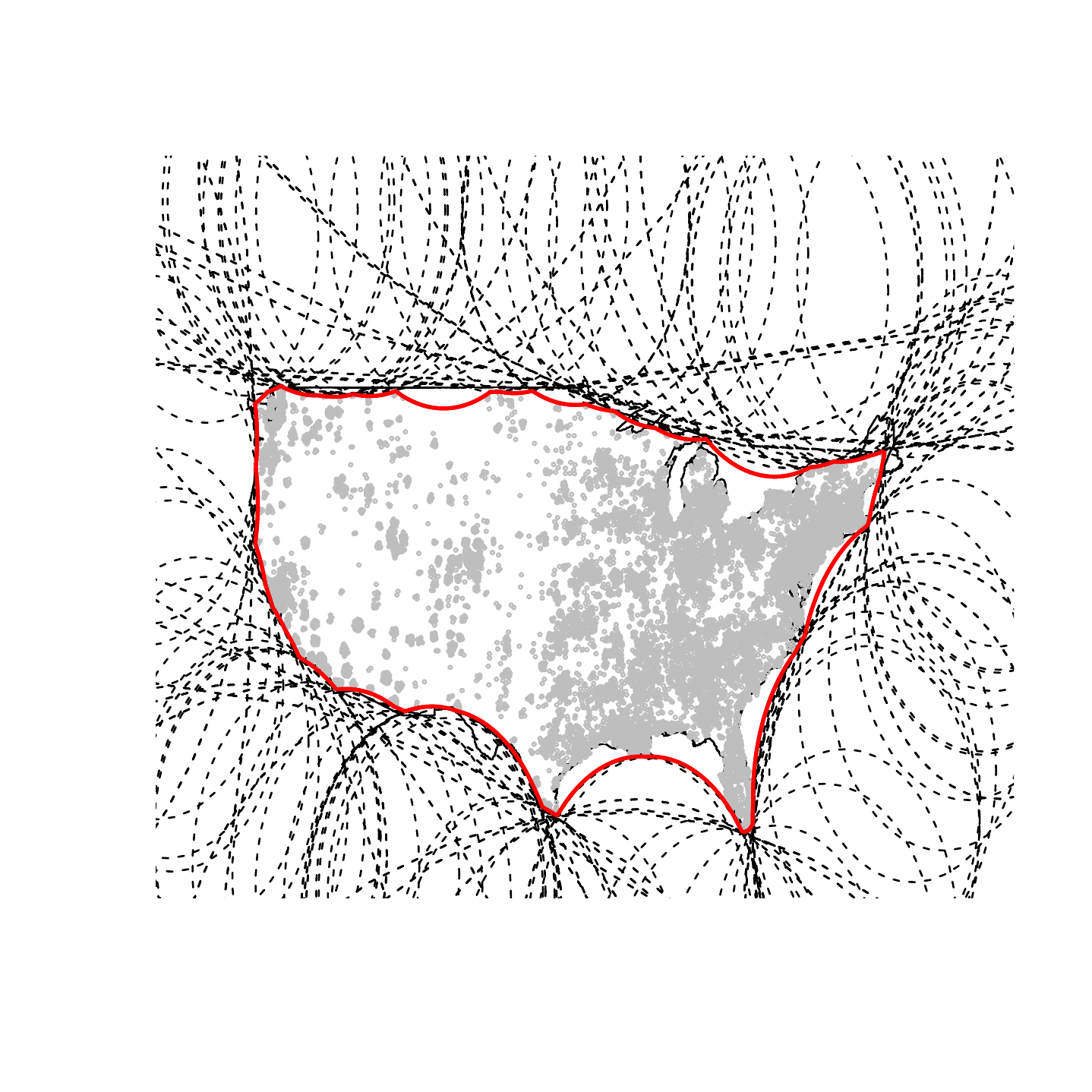}\hspace{-.9cm}\includegraphics[height=150pt,width=175pt]{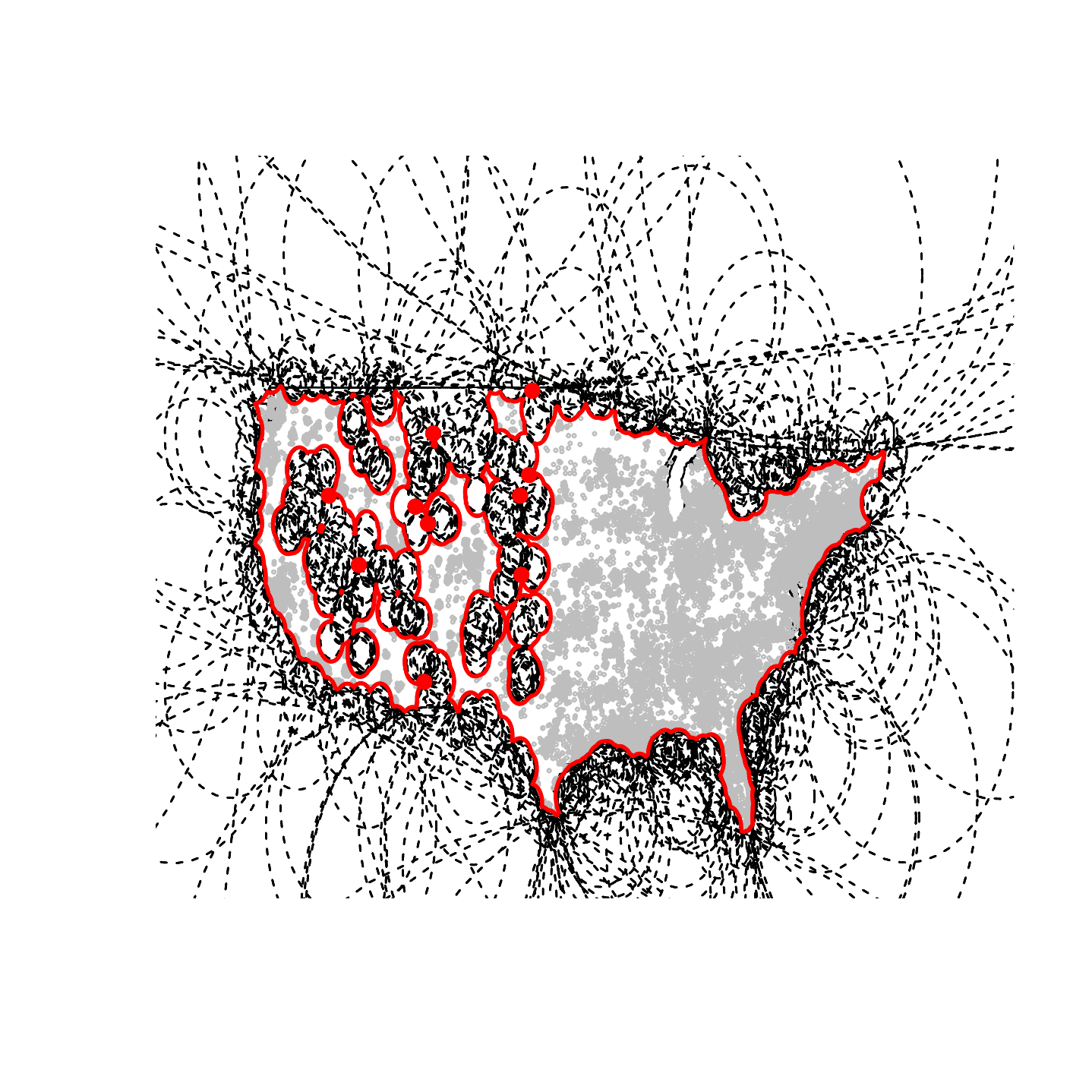}\vspace{-1cm}\\  
  	\caption{$C_{r}(\mathcal{X}_{n})$ (red color) taking $r=1000$ (left), $r=10$ (center) and $r=1$ (right). }\label{fig111}
  \end{figure}

  In practice, $C_{r}(\mathcal{X}_n)$ can be computed as the intersection of the complements of all open balls of radius larger than or equal to $r$ that do not intersect $\mathcal{X}_n$. In particular, if $r$ is equal to infinity, $C_r(A)$ is equal to the convex hull of $A$. In Figure \ref{fig111}, the computation of $C_{r}(\mathcal{X}_{n})$ is shown considering the  geographical coordinates of confirmed cases of coronavirus reported in the United States from March 26th to April 1st as $\mathcal{X}_{n}$ for $r=1000$, $r=10$ and  $r=1$. It is equal to the intersection of the complements of all black open balls represented. Note that 
  $C_{1000}(\mathcal{X}_{n})$ is practically equal to the convex hull. However, if we select $r=1$, the number of connected components increases considerably. In fact, there are multiple isolated points.

  The parameter $r$ can be seen as a shape index that controls the level of fragmentation of the hull. Therefore, it plays the role of a smoothing parameter with a considerable influence on the estimations. Hence, if a HDR is assumed to be $r-$convex and the subset $\mathcal{X}_n^f(f_{\tau})=\{X_i\in \mathcal{X}_n:f(X_i)\geq f_{\tau}\}$ inside the HDR was known, the most natural estimator would be the $r-$convex hull of $\mathcal{X}_n^f(f_{\tau})$. It is the smallest $r-$convex set containing $\mathcal{X}_n$. However, $r$ and $\mathcal{X}_n^f(f_{\tau})$ are usually unknown in practice as the theoretical HDR $L(\tau)$. Then, estimating both of them from $\mathcal{X}_n$ are a first key step for reconstructing HDRs. As regards $r$, it easy to check that if a set is $r-$convex then it will be also $r^*-$convex for all $0<r^*\leq r $. Then, it is necessary to determine precisely  which is the optimal value of $r$ to be estimated. \cite{rosaa}, propose to estimate the highest value of $r$ which verifies that $L(\tau)$ is $r-$convex. Concretely, if $L(\tau)$ is a compact, nonempty, nonconvex and $r-$convex HDR for some $r>0\label{r0t}$, it is defined
 	\begin{equation}\label{estimadorr0levelsetu}
  	r_0(f_{\tau})=\sup\{\gamma>0:C_\gamma(L(\tau))=L(\tau)\}.
  	\end{equation}
 Of course, the definition in (\ref{estimadorr0levelsetu}) depends on $\tau$. Moreover, the case when $L(\tau)$ is convex is not considered because, in such particular case, $ r_0(f_{\tau})$ is non finite. Then, the most natural reconstruction of the HDR $L(\tau)$ would be the convex hull instead of the $r-$convex hull. \cite{rosaa} also prove that, under a mild regularity condition, $L(\tau)$ is $r_0(t)-$convex and, therefore, it is the optimal value to be estimated. This property, which is also needed to guarantee the consistency results for the HDR estimator, is slightly stronger than $r-$convexity:\vspace{-.1cm}\\

  ($R_{\lambda}^r$) A closed ball of radius $\lambda>0$ rolls freely in $G(t)$ (inside) and a closed ball of\\ \textcolor[rgb]{1.00,1.00,1.00}{$ $ $ $ $ $ $ $ $ $ $ $ ($R_{\lambda}^r$)}radius $r>0$ rolls freely in $\overline{G(t)^c}$ (outside).\vspace{.3mm}\\
  
  Following \cite{r2} and \cite{r33}, a closed ball of radius $\gamma>0$ rolls freely in a closed set $A$ if for each boundary point $b\in\partial A$ there exists $x\in\mathbb{R}^d$ such that $b\in B_\gamma[x]\subset A$ where $B_\gamma[x]$ denotes the closed ball of radius $\gamma$ centering at $x$.  
  
  Sets fulfilling property ($R_{\lambda}^r$) show interesting conditions which make them easier to handle. For instance, their frontiers are smooth and, under the condition ($R_{\lambda}^r$), the $r$-rolling condition implies the $r-$convexity. Following \cite{cue22}, this is not true in general.
  
  Under several regularity assumptions on the density $f$ and on its usual kernel density estimator $f_n$, \cite{rosaa} suggest to estimate consistently the unknown index $r_0(f_\tau)$ as
  	\begin{equation}\label{estimadorr0hat}
  	\hat{r}_0(\hat{f}_\tau)=\sup\{\gamma>0:C_\gamma(\mathcal{X}_n^+(\hat{f}_\tau))\cap \mathcal{X}_n^-(\hat{f}_\tau)=\emptyset\},\vspace{-.2cm}
  	\end{equation}where
  	$$\mathcal{X}_n^+(\hat{f}_\tau)=\{X\in\mathcal{X}_n: f_n(X)\geq \hat{f}_\tau+D_n\},\mbox{ }\mathcal{X}_n^-(\hat{f}_\tau)=\{X\in\mathcal{X}_n: f_n(X)< \hat{f}_\tau-D_n\}\mbox{ and }
  	\label{mas2}$$and $D_n$ denotes a non negative sequence that verifies $D_n\stackrel{\text{a.s}}{\to}0$ and that bounds the quantity $|f-f_n|$. In fact, $D_n$ could be chosen as $D_n=M_n(\log(n)/n)^{p/(d+2p)}$ for $M_n\to\infty$.

  The original random sample $\mathcal{X}_n$ is divided into three subsamples, $\mathcal{X}_n^+(\hat{f}_\tau)$, $\mathcal{X}_n^-(\hat{f}_\tau)$ and $\mathcal{X}_n\setminus \left(\mathcal{X}_n^+(\hat{f}_\tau)\cup\mathcal{X}_n^-(\hat{f}_\tau)\right)$. Following \cite{rosaa}, it can be checked that, given $f_\tau>0$, $ \mathcal{X}_n^+(f_\tau)$ and $\mathcal{X}_n^-(f_\tau)$ provide feasible information on the HDR and its complementary, respectively.  Then, it would be natural to consider $C_{\hat{r}_0(\hat{f}_\tau)}(\mathcal{X}_n^+(\hat{f}_\tau))$ as an estimator for the HDR $L(\tau)$. However, the consistency cannot be guaranteed in this case, at least under general conditions. Then, \cite{rosaa} propose $\hat{L}(\tau)=C_{r_n(\hat{f}_\tau)}(\mathcal{X}_n^+(\hat{f}_\tau))$ as the resulting estimator of the HDR where $r_n(\hat{f}_\tau)=\nu \hat{r}_0(\hat{f}_\tau)$ for a fixed value $\nu\in (0,1)$. Remember that if $L(\tau)$ is $r-$convex is also $r^*-$convex for $0<r^*\leq r$ and, therefore, it is verified that $L(\tau)=C_r(L(\tau))=C_{r^*}(L(\tau))$. Then, it is expected that, for $n$ large enough, $L(\tau)$ was $r_n(t)-$convex. Theorems 3.3 in \cite{rosaa} shows that this hybrid method achieves minimax rates for Hausdorff metric and distance in measure, up to log factors, in most cases where the minimax rates are known, see \cite{mam95}. These distances between sets are commonly used in order to assess the performance of set estimators calculating the distance between them and the corresponding theoretical HDR. Let $A$
  and $C$ be two closed, bounded, nonempty subsets of $\mathbb{R}^{d}$. The
  Hausdorff distance between $A$ and $C$ is defined by\vspace{-0.13cm}
  $$
  d_{H}(A,C)=\max\left\{\sup_{a\in A}d(a,C),\sup_{c\in C}d(c,A)\right\},\vspace{-0.13cm}
  $$
  where $d(a,C)=\inf\{\|a-c\|:c\in C\}$ and $\|\mbox{ }\|$ denotes the Euclidean norm.
  Another possibility, if $A$ and $C$ are two bounded and Borel sets then the distance in measure between $A$ and $C$ is defined by $d_{\mu}(A,C)=\mu(A\triangle C)$, where $\mu\label{lebesgue2}$ denotes the Lebesgue measure and $\triangle$, the symmetric difference, that is, $A\triangle C=(A \setminus C)\cup(C \setminus A). $

  Despite these satisfactory theoretical achievements, a practical implementation of this method had to be proposed avoiding its dependency on the unknown sequence $D_n$. Details on the algorithm established in \cite{rosaa} are discussed in what follows.

 \section{Hybrid algorithm for nonparametric HDRs estimation}\label{4}
The reconstruction of the HDR $L(\tau)$ from the random sample $\mathcal{X}_n$ relies strongly on the specifications of the sets $\mathcal{X}_n^+(\hat{f}_\tau)$ and $\mathcal{X}_n^-(\hat{f}_\tau)$ whose definitions involve the unknown sequence $D_n$. According to \cite{rosaa}, it could be any stochastic upper bound of the absolute difference between $f$ and its kernel estimator $f_n$. Although this bounding problem is already considered in \cite{hall} and \cite{gine2}, results obtained from simulations were not entirely satisfactory. Therefore, an automatic bootstrap algorithm was established in \cite{rosaa} for computing $D_n$ focusing on the problem of HDRs reconstruction. 

This hybrid procedure, fully described in Algorithm \ref{al1}, is closely related to the definition of $f_\tau$ established in (\ref{umbral}). Given $\tau\in (0,1)$, we will select the biggest value $0<\overline{\tau}\leq \tau $ such that the associated HDR estimator contains a proportion of sample points bigger than or equal to $1-\tau$. Although the performance of this approach has not been checked conveniently in practice, it avoids the direct estimation of the sequence $D_n$.

Following \cite{rosaa}, the reconstruction of HDR corresponding to each $\overline{\tau}$ is established from the subsets of $\mathcal{X}_n$ denoted by $\mathcal{X}_n^+(\hat{f}_{\overline{\tau}}^+)$ and $\mathcal{X}_n^-(\hat{f}_{\overline{\tau}}^-)$ in Algorithm \ref{al1}. Note that the corresponding thresholds $\hat{f}_{\overline{\tau}}^+$ and $\hat{f}_{\overline{\tau}}^-$ are calculated from the bootstrap procedure described below. 

Specifically, a total of $B$ bootstrap resamples $\mathcal{X}_{1,n}^*,...,\mathcal{X}_{B,n}^*$ of size $n$ must be generated from $f_n$. For each of these ones, the proportion of points inside and outside the known region $\{f_n\geq (f_n)_{\overline{\tau}}\}$ in the bootstrap world must be determined. In this work, the threshold $(f_n)_{\overline{\tau}}$ is calculated following the method proposed in \cite{hynd}. Given $p$, $\overline{\tau}_{+}$ and $\overline{\tau}_{-}$ are calculated from the $B$ values of the probability contents of the HDR and its complementary, respectively. Then, $\hat{f}_{\overline{\tau}}^+$ and $\hat{f}_{\overline{\tau}}^-$ are determined as it is described in Algorithm \ref{al1} and their selection is performed attending to the bootstrap probability contents of the HDR and its complementary. If the proportion of points in $\mathcal{X}_n$ contained in $\hat{L}(\overline{\tau})$ is greater than or equal to $1-\tau$, the searching procedure is finished. Otherwise, a smaller $\overline{\tau}$ should be considered. Note that, $D_n$ could be computed as the maximum value between $\hat{f}_{\overline{\tau}}^+-\hat{f}_{\tau}$ and  $\hat{f}_{\tau}-\hat{f}_{\overline{\tau}}^-$. Then, the direct estimation of the sequence $D_n$ is not necessary here.

\begin{figure}[h!]   \vspace{-8cm}
	$\hspace{1.75cm}$\begin{picture}(0,0)
	\put(-110,-470){\includegraphics[scale=.6]{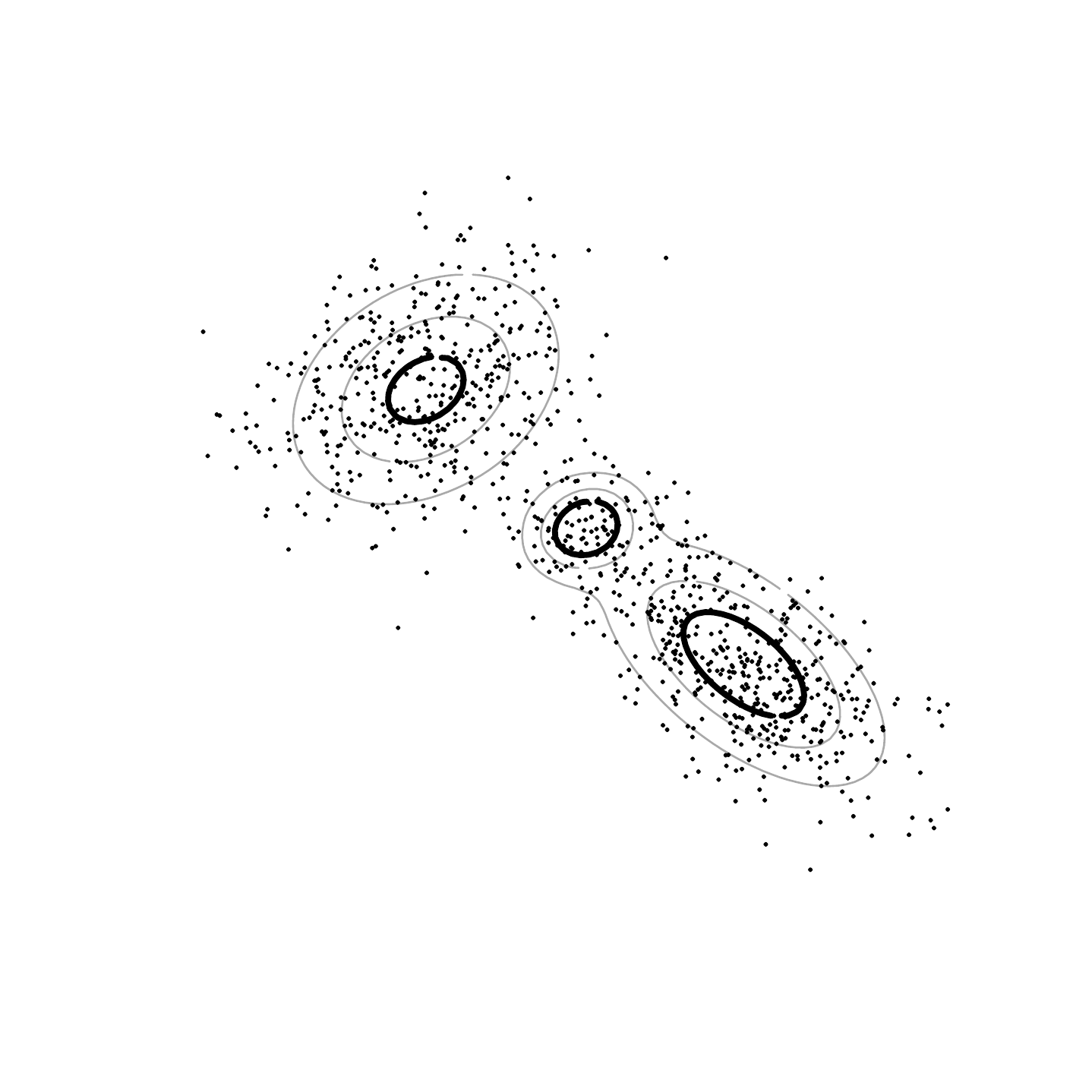}}
	\put(40,-470){\includegraphics[scale=.6]{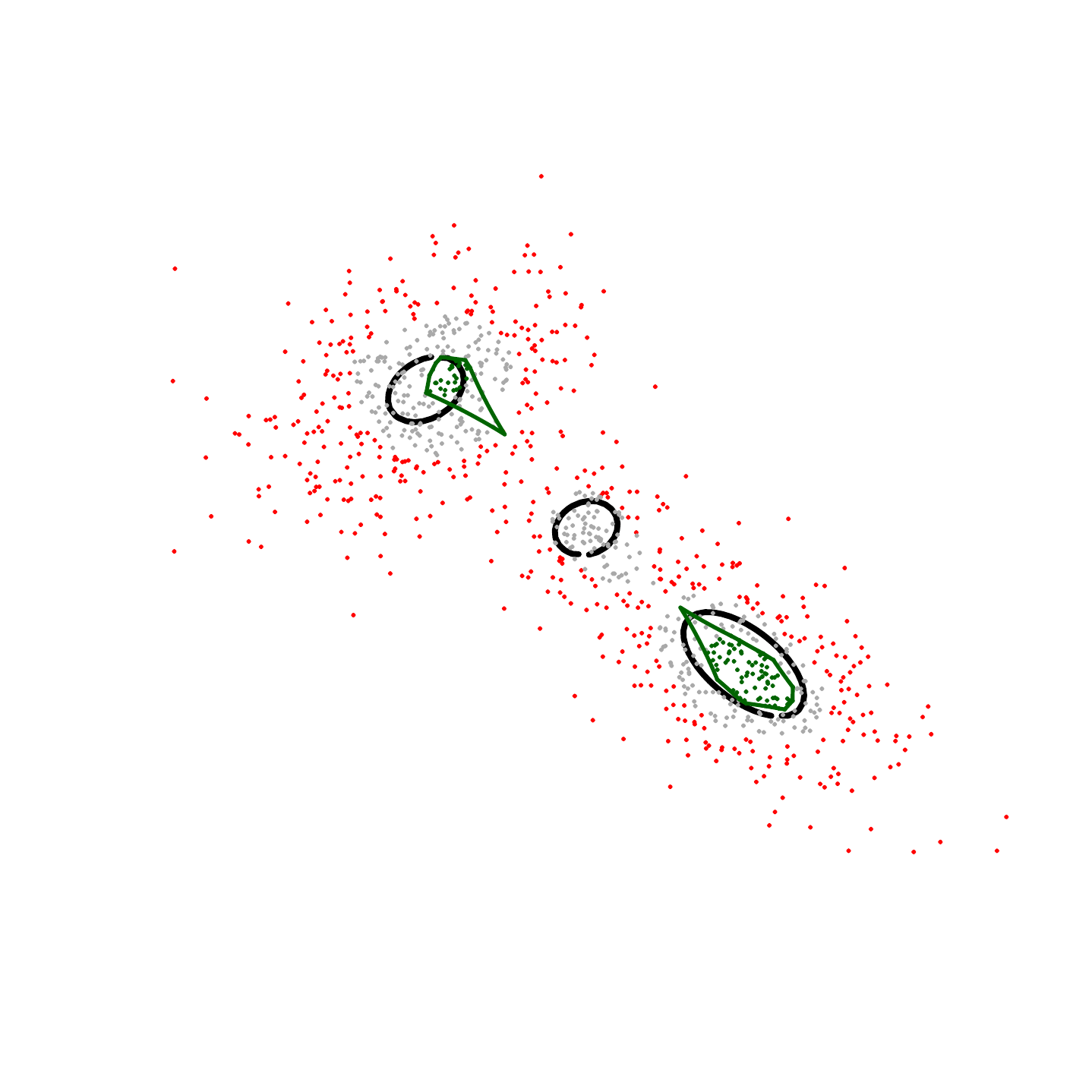}}
	\put(190,-470){\includegraphics[scale=.6]{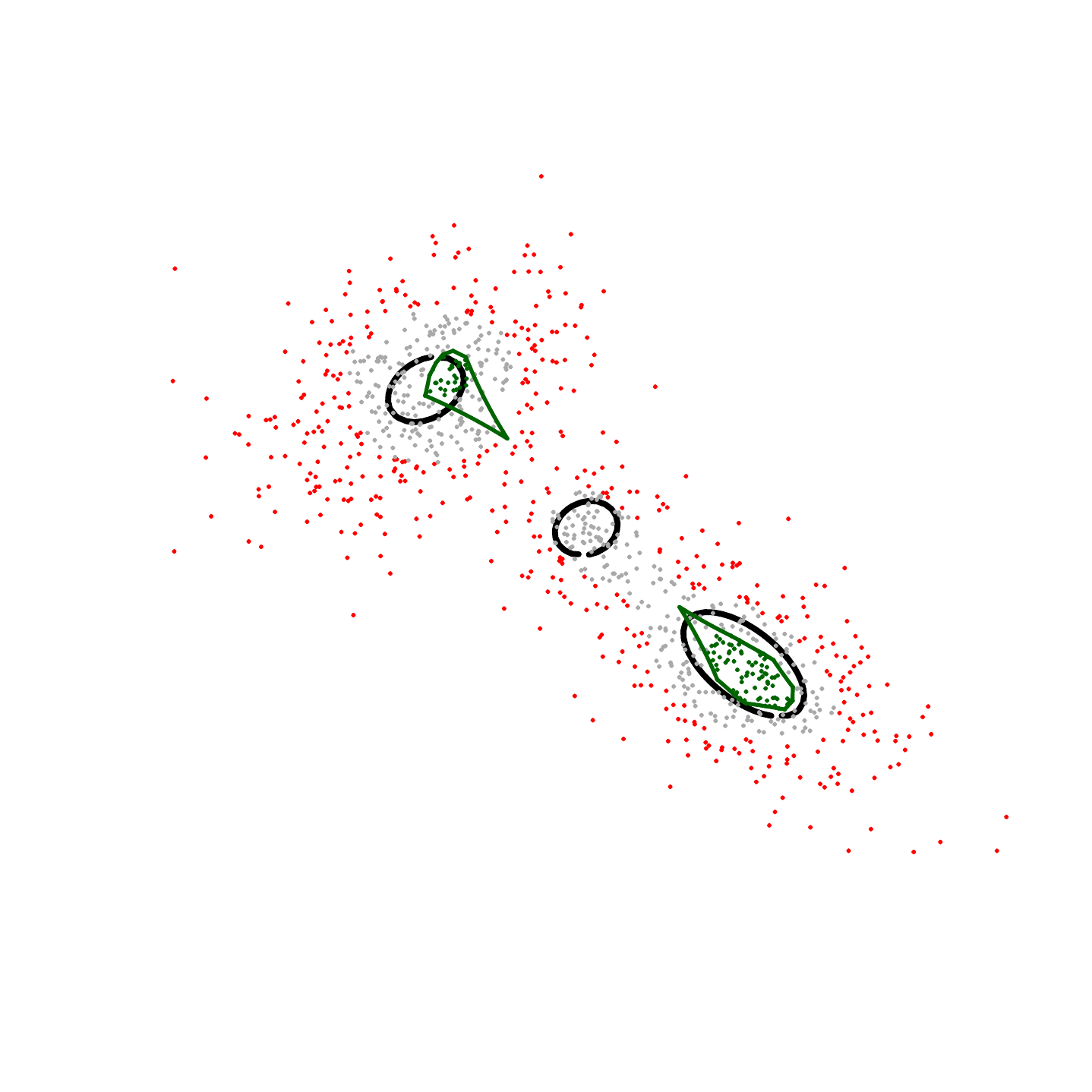}}
	
	\put(-30,-250){ \textbf{Phase 0}} 
	\put(120,-250){ \textbf{Phase 1}} 	 
	\put(280,-250){ \textbf{Phase 2}}

	\put(-110,-630){\includegraphics[scale=.6]{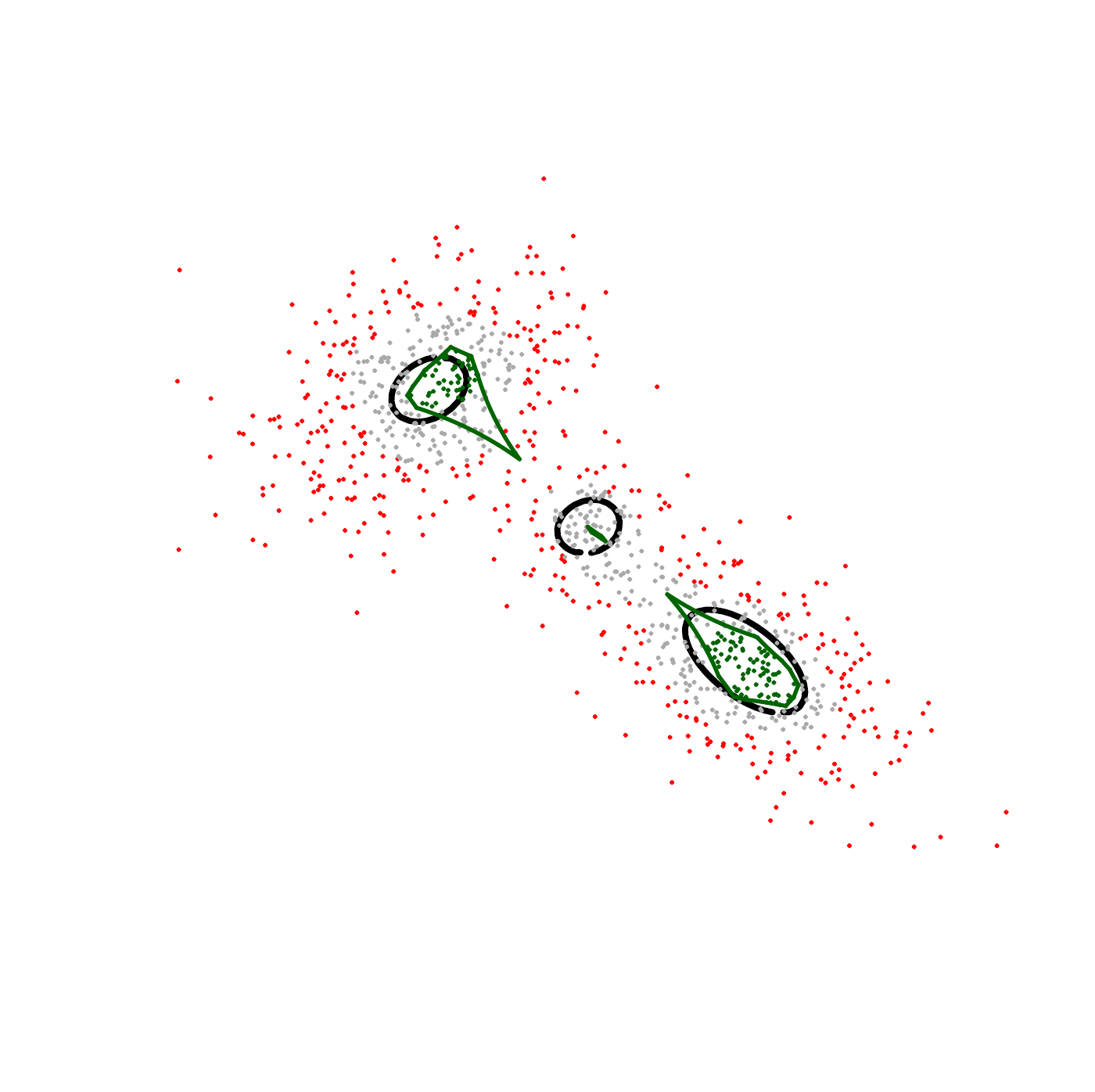}}
	\put(40,-630){\includegraphics[scale=.6]{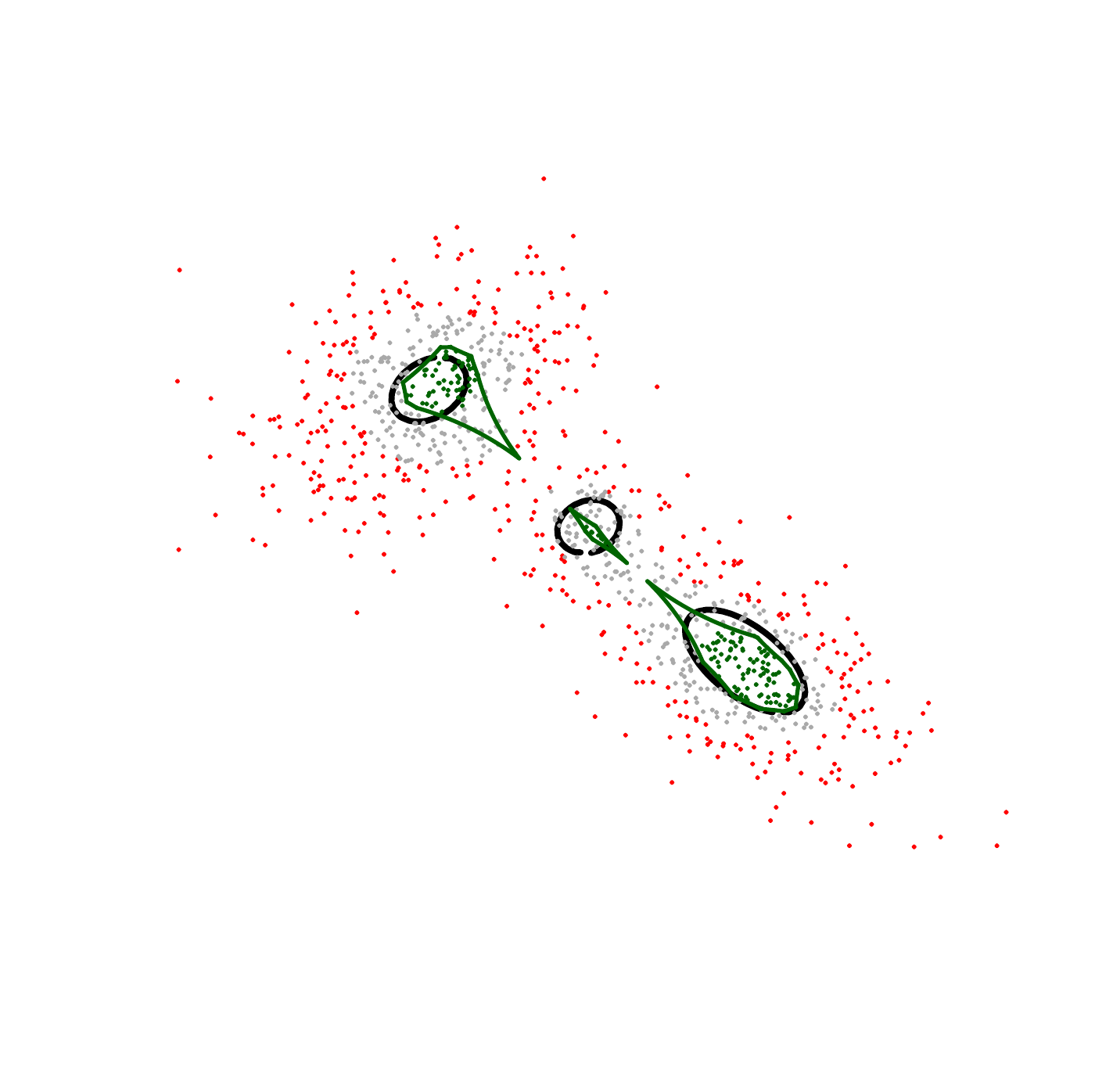}}
	\put(190,-630){\includegraphics[scale=.6]{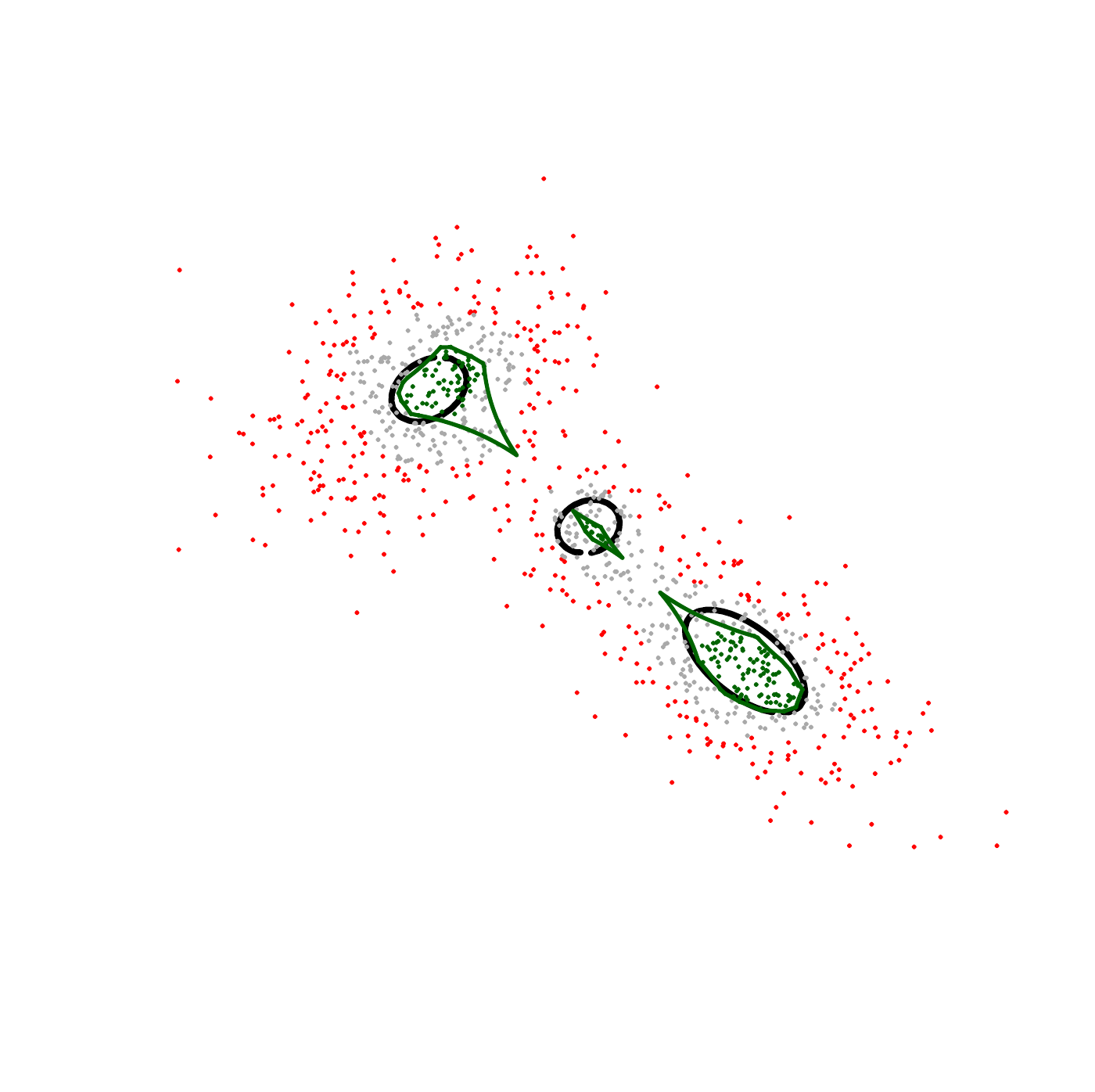}}
	\put(-30,-425){ \textbf{Phase 3}} 
	\put(120,-425){ \textbf{Phase 4}} 	 
	\put(280,-425){ \textbf{Phase 5}}

	\end{picture}\vspace{20cm}
	\caption{Phases of Algorithm \ref{al1} when a HDR $L(\tau)$ for $\tau=0.8$ is estimated from a random sample $\mathcal{X}_{1000}$ generated from a trimodal density. From phase 1 to phase 5,  $\mathcal{X}_n^+(\hat{f}_{\overline{\tau}}^+)$ and $C_{\hat{r}_0(\hat{f}_{\overline{\tau}})}(\mathcal{X}_n^+(\hat{f}_{\overline{\tau}}^+))$ are represented using green color; $\mathcal{X}_n^-(\hat{f}_{\overline{\tau}}^-)$, using red color and $\mathcal{X}_n\setminus (\mathcal{X}_n^+(\hat{f}_\tau^+)\cup \mathcal{X}_n^-(\hat{f}_\tau^-))$, using gray color.}\label{algorithmphases}
\end{figure}
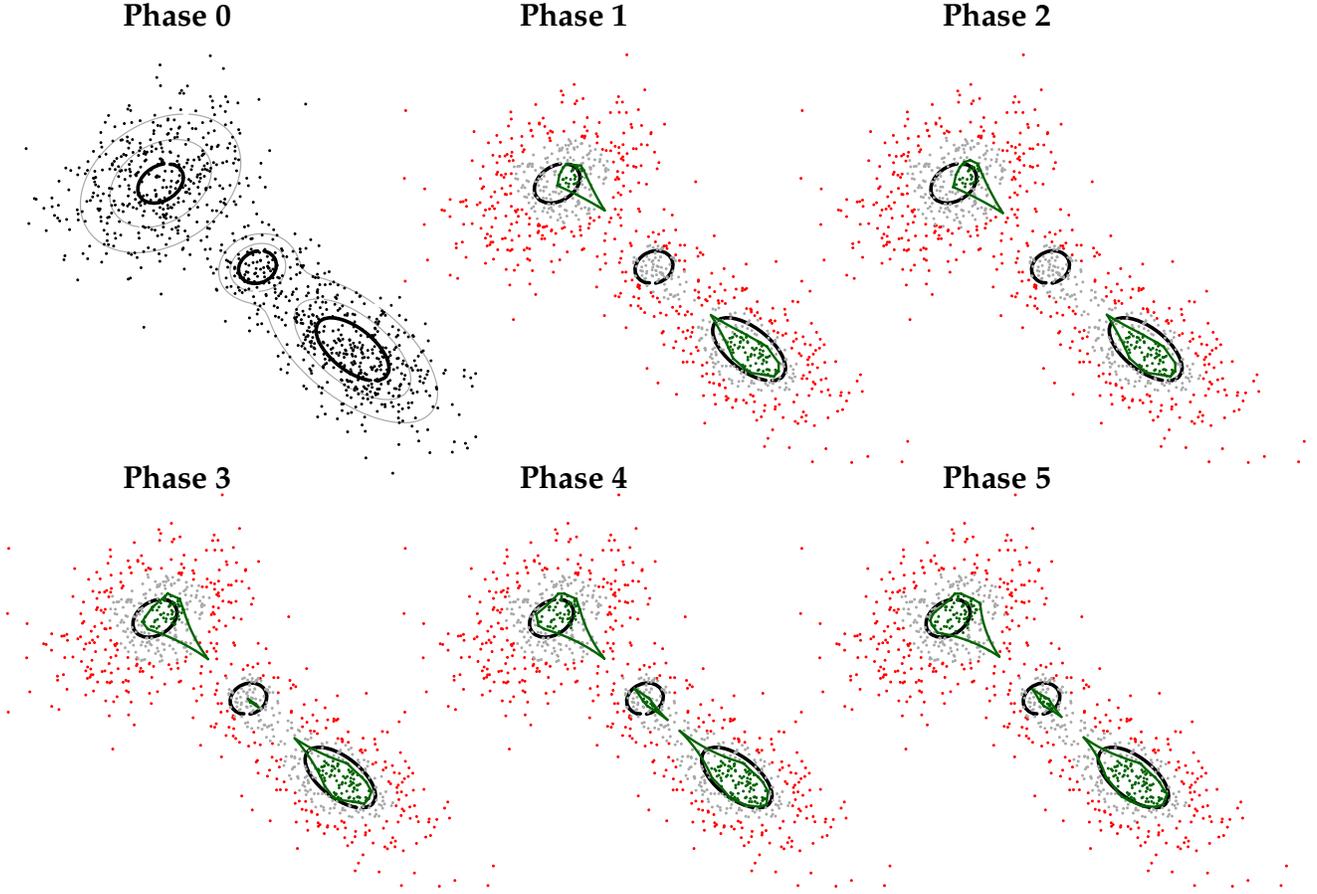

Figure \ref{algorithmphases} shows the phases of Algorithm \ref{al1} when a HDR for $\tau=0.8$ is estimated from a random sample $\mathcal{X}_{1000}$ generated from a trimodal density that will be also consider as a model for simulations in Section \ref{simus}. In this illustration, $B=100$, $p=0.05$ and \textit{step}$=0.025$\\

\begin{algorithm}[H]\label{al1}
	\SetAlgoLined\vspace{1mm}
	\textbf{Inputs}: $\tau$, $\mathcal{X}_n$, $B$, $p$ and \textit{step}.\\
	1. Calculate the kernel density estimador $f_n$ from $\mathcal{X}_n$.\\
	2. Initialize $P=0$ and $\overline{\tau}=\tau$.\\
	\While{$P<(1-\tau)$,}{
		$\square$  	Determinate the threshold $(f_n)_{\overline{\tau}}$. \\
		\ForEach{$i \in \{1,...,B\}$}{
			$ \circ$ Generate a bootstrap sample of size $n$ from $f_n$, $\mathcal{X}_{i,n}^*$.\\
			$ \circ$ Determinate $f_n(\mathcal{X}_{i,n}^*)$.\\
			$ \circ$ Calculate $\overline{\tau}_{i,+}^{*}=\frac{\#\{f_n^{*}(\mathcal{X}_{i,n}^{*})\geq (f_n)_{\overline{\tau}} \}}{n}$ and $\overline{\tau}_{i,-}^{*}=\frac{\#\{f_n^{*}(\mathcal{X}_{i,n}^{*})< (f_n)_{\overline{\tau}}\}}{n}$.
		}
		$ \square$ Determine $\overline{\tau}_{-}$ and $\overline{\tau}_{+}$ as the $p-$quantiles of the vectors $\overline{\tau}_{i,-}^*$ and $\overline{\tau}_{i,+}^*$, respectively.\\
		$ \square$ Calculate the thresholds $\hat{f}_{\overline{\tau}}^{-}$ and $\hat{f}_{\overline{\tau}}^+$ as the $\overline{\tau}_{-}$ and $(1-\overline{\tau}_{+})$-quantiles of $f_n(\mathcal{X}_n)$, respectively.\\
		$ \square$ Obtain the subsets of $\mathcal{X}_n$:
		\small{
			$$\mathcal{X}_n^+(\hat{f}_{\overline{\tau}}^+ )=\{X\in\mathcal{X}_n:f_n(X)\geq \hat{f}_{\overline{\tau}}^+\},\mbox{ }\mathcal{X}_n^-(\hat{f}_{\overline{\tau}}^-)=\{X\in\mathcal{X}_n:f_n(X)< \hat{f}_{\overline{\tau}}^-\}$$
			$$\mbox{and } \mathcal{X}_n\setminus (\mathcal{X}_n^+(\hat{f}_\tau^+)\cup \mathcal{X}_n^-(\hat{f}_\tau^-)). $$
		}	
	
		$ \square$ Compute $\hat{r}_0(\hat{f}_{\overline{\tau}})$ from the sets established in the previous step.\\
		$ \square$ Estimate $L(\overline{\tau})$ as $\hat{L}(\overline{\tau})=C_{\hat{r}_0(\hat{f}_{\overline{\tau}})}(\mathcal{X}_n^+(\hat{f}_{\overline{\tau}}^+))$.\\
		$ \square$ Update:\\
		\hspace{.2cm} \textcolor{gray}{$\circ$} $P$ as the proportion of points of $\mathcal{X}_n$ in $\hat{L}(\overline{\tau})$.\\
		\hspace{.2cm} \textcolor{gray}{$\circ$} $\overline{\tau}=\overline{\tau}-\mbox{\textit{step}}$.
	}
	\textbf{Output}: $\hat{L}(\overline{\tau})$.
	\caption{Full hybrid algorithm in order to estimate the HDR $L(\tau)$} 
\end{algorithm}\newpage

$ $\\Note that phase 0 only shows the theoretical contours of the trimodal density for $\tau=0.8$ (black color) and two additional smaller values (gray color) in order to describe the density conveniently. Furthermore, it contains the random sample $\mathcal{X}_{1000}$ (black color) used for reconstructing $L(\tau)$. Apart from the theoretical HDR, next phases contains the sets $\mathcal{X}_n^+(\hat{f}_{\overline{\tau}}^+)$ and $C_{\hat{r}_0(\hat{f}_{\overline{\tau}})}(\mathcal{X}_n^+(\hat{f}_{\overline{\tau}}^+))$ (green color), $\mathcal{X}_n^-(\hat{f}_{\overline{\tau}}^-)$ (red color) and $\mathcal{X}_n\setminus (\mathcal{X}_n^+(\hat{f}_\tau^+)\cup \mathcal{X}_n^-(\hat{f}_\tau^-))$ (gray color). It is convenient to remember  that the information on $L(\tau)$ and its complementary is provided by the sets $\mathcal{X}_n^+(\hat{f}_{\overline{\tau}}^+)$ and  $\mathcal{X}_n^-(\hat{f}_{\overline{\tau}}^-)$, respectively. However, sample points in $\mathcal{X}_n\setminus (\mathcal{X}_n^+(\hat{f}_\tau^+)\cup \mathcal{X}_n^-(\hat{f}_\tau^-))$ can be considered as a non-informative doubtful set. Although the connected components of the theoretical HDR are overestimated in some areas, the algorithm succeeds in detecting the three greatest modes or $\tau-$clusters of the distribution.

The code of Algorithm \ref{al1} can be obtained from the
author upon request. It should be noted that R packages \texttt{ks} and \texttt{alphahull} were used in order to compute the kernel density estimator and the set $C_{\hat{r}_0(\hat{f}_{\overline{\tau}})}(\mathcal{X}_n^+(\hat{f}_{\overline{\tau}}^+))$. For more details, see \cite{duong2007ks} and \cite{pateiro2009alphahull}, respectively.

\section{Simulation study}\label{simus}

\begin{figure}[h!]   \vspace{-8cm}
	$\hspace{1.75cm}$\begin{picture}(0,0)
	\put(-50,-370){\includegraphics[scale=.3]{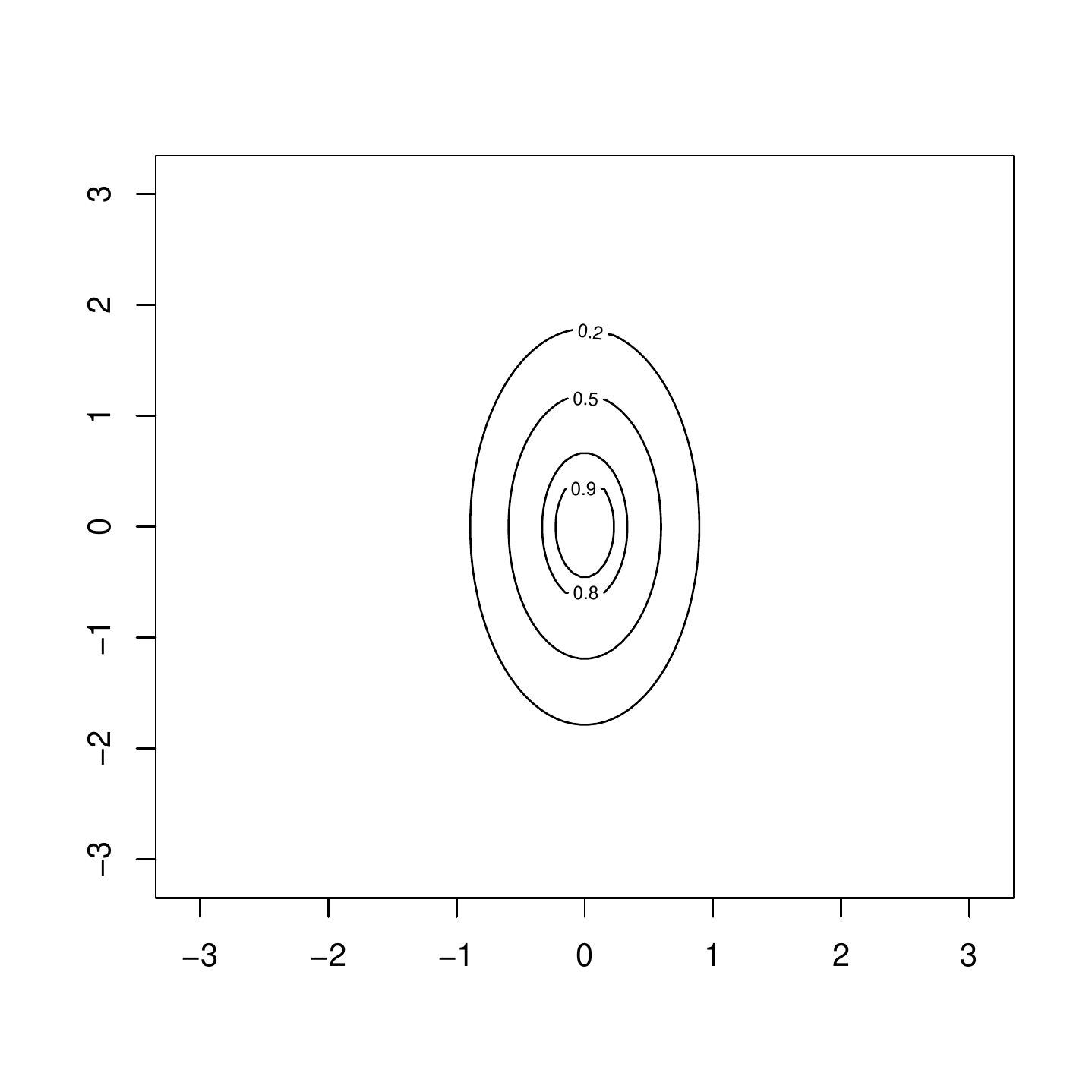}}
	\put(100,-370){\includegraphics[scale=.3]{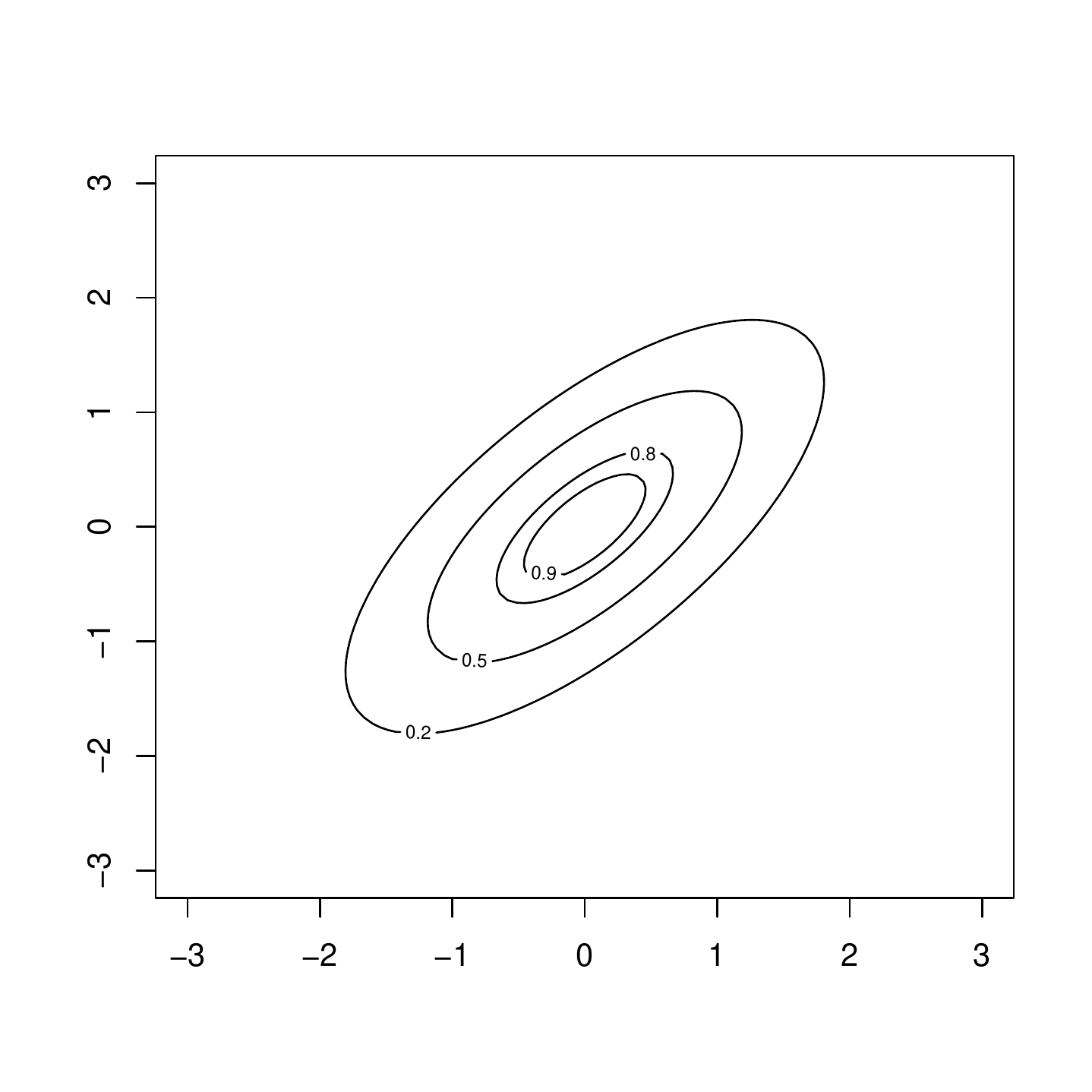}}
	\put(250,-370){\includegraphics[scale=.3]{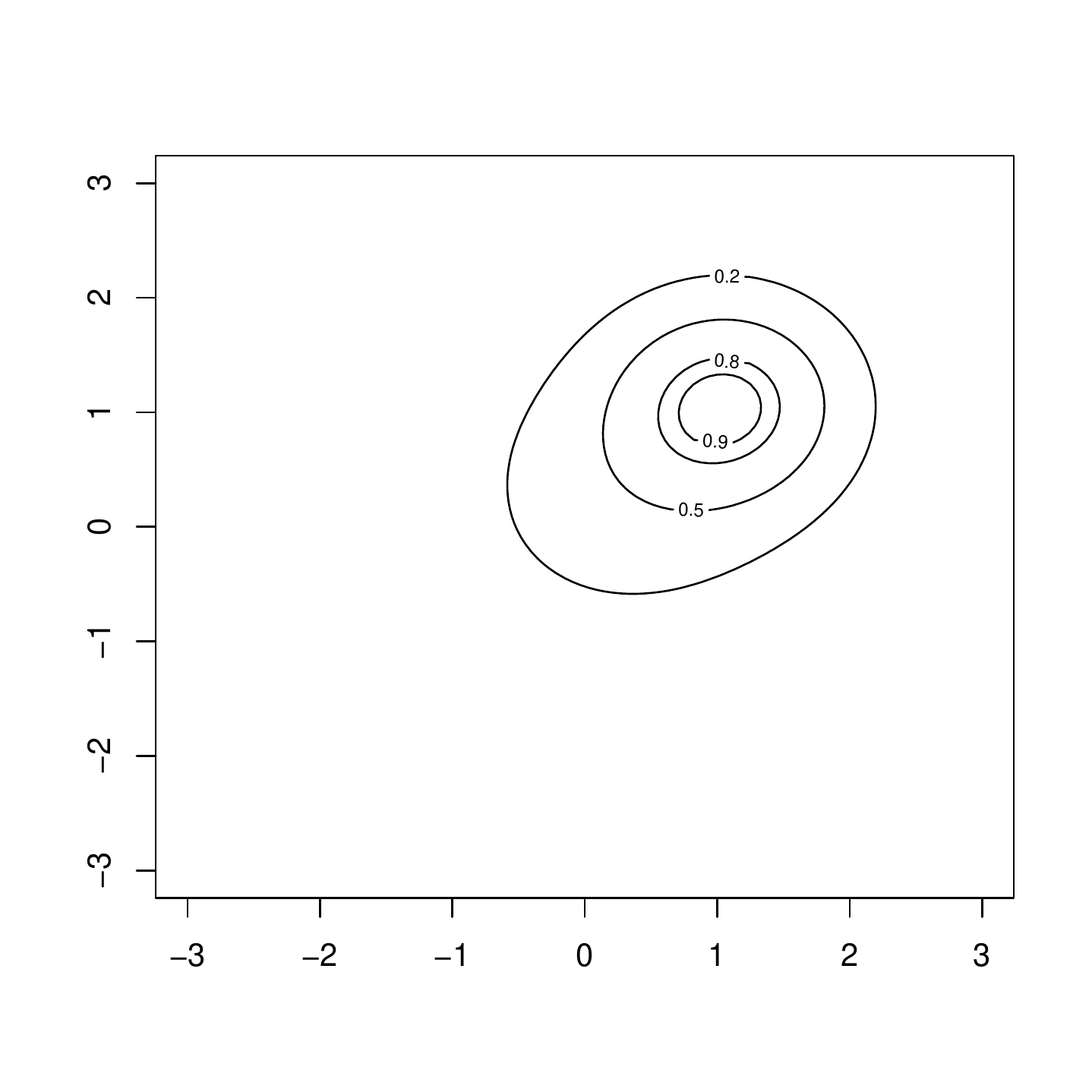}}
	
	\put(-55,-250){\small{\textbf{Uncorrelated Normal (1)}}} 
	\put(103,-250){\small{\textbf{Correlated Normal (2)}}} 	 
	\put(285,-250){\small{\textbf{Skewed (3)}}}

	\put(-50,-490){\includegraphics[scale=.3]{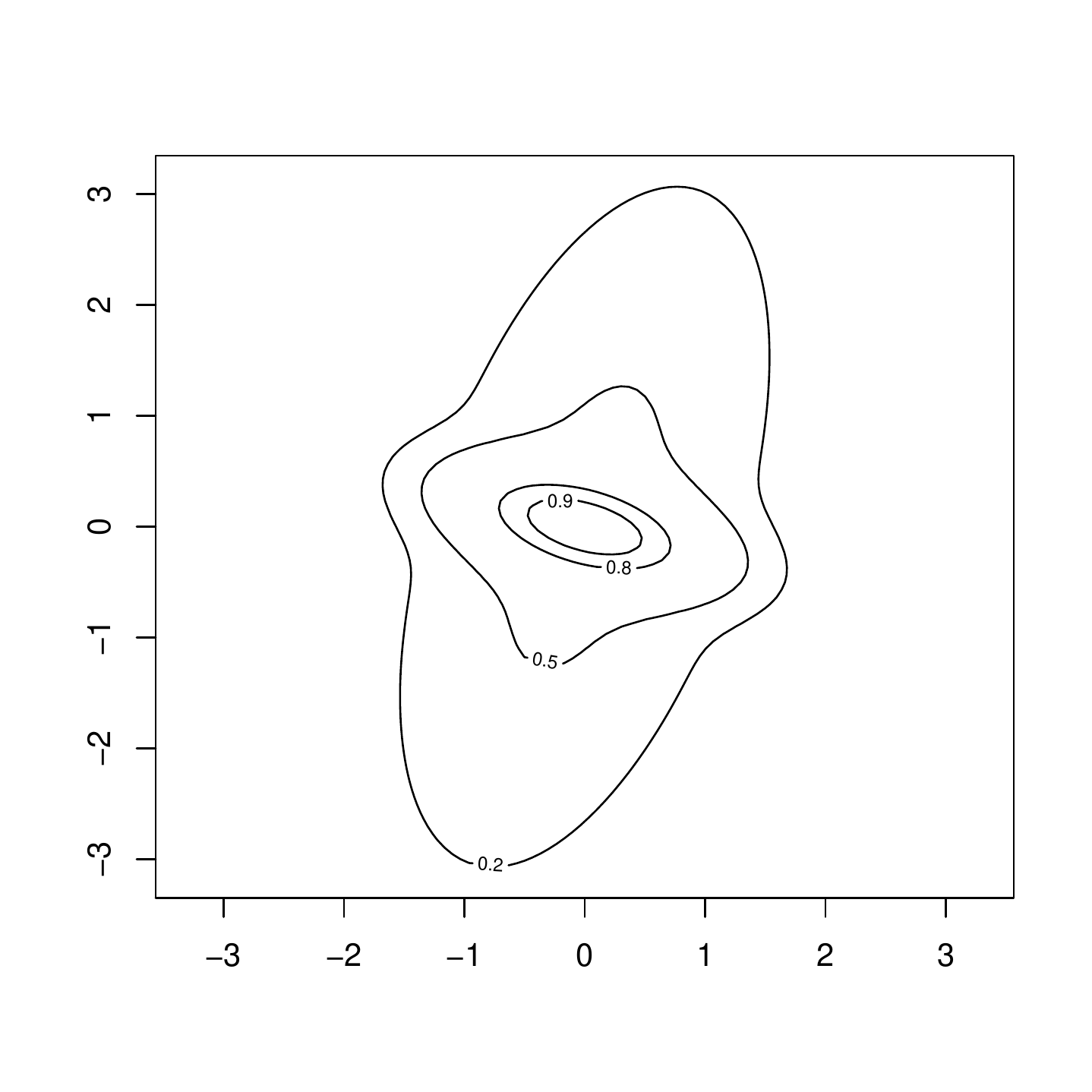}}
	\put(100,-490){\includegraphics[scale=.3]{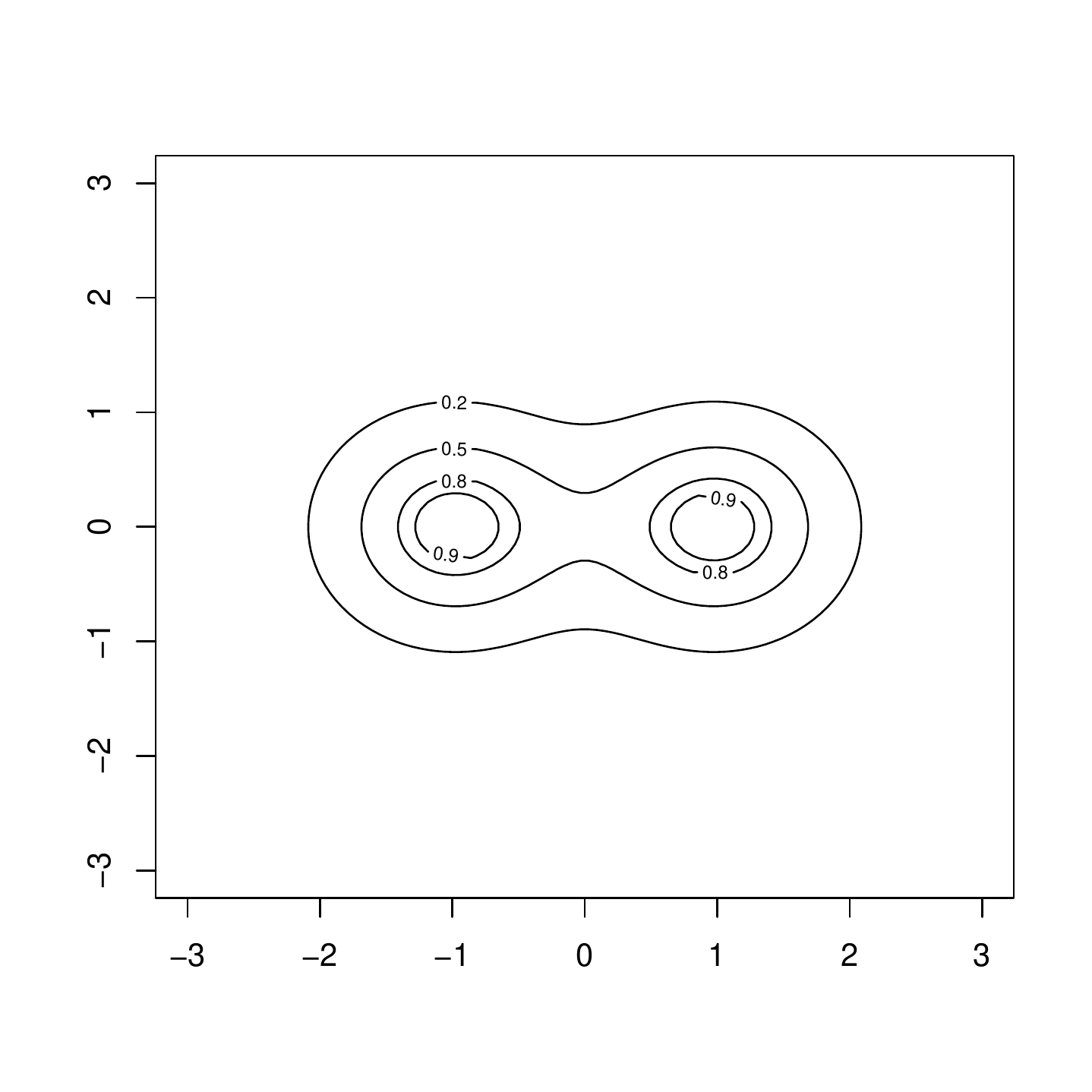}}
	\put(-16,-375){\small{\textbf{Kurtotic (4)}} }
	\put(129,-375){\small{\textbf{Bimodal I (5)}} }	 
	
	\put(250,-490){\includegraphics[scale=.3]{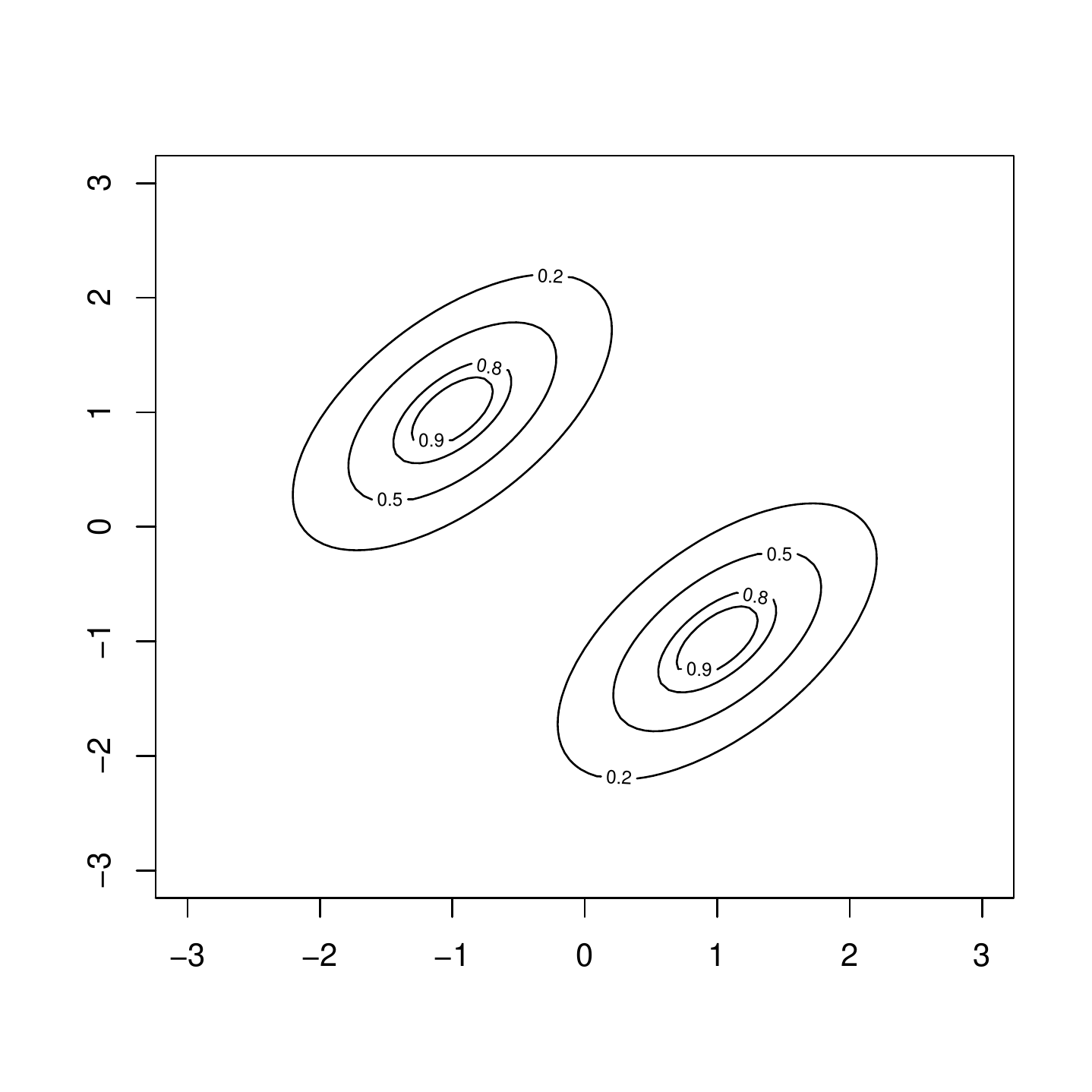}}
	\put(275,-375){\small{\textbf{Bimodal III (6)}}}	 
	\put(-50,-615){\includegraphics[scale=.3]{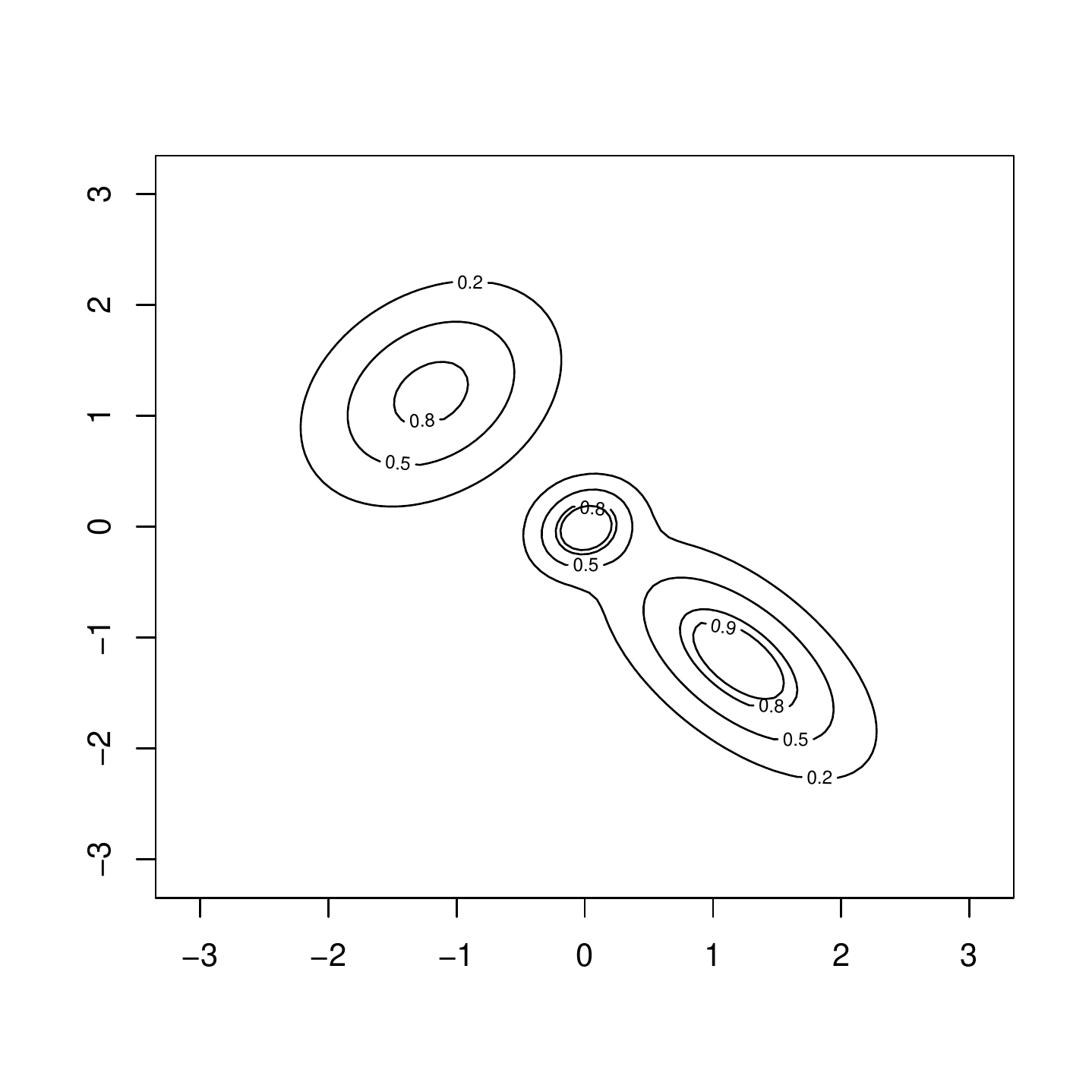}}
	
	\put(-22,-500){\small{\textbf{Trimodal I (7)}} }

	\put(100,-615){\includegraphics[scale=.3]{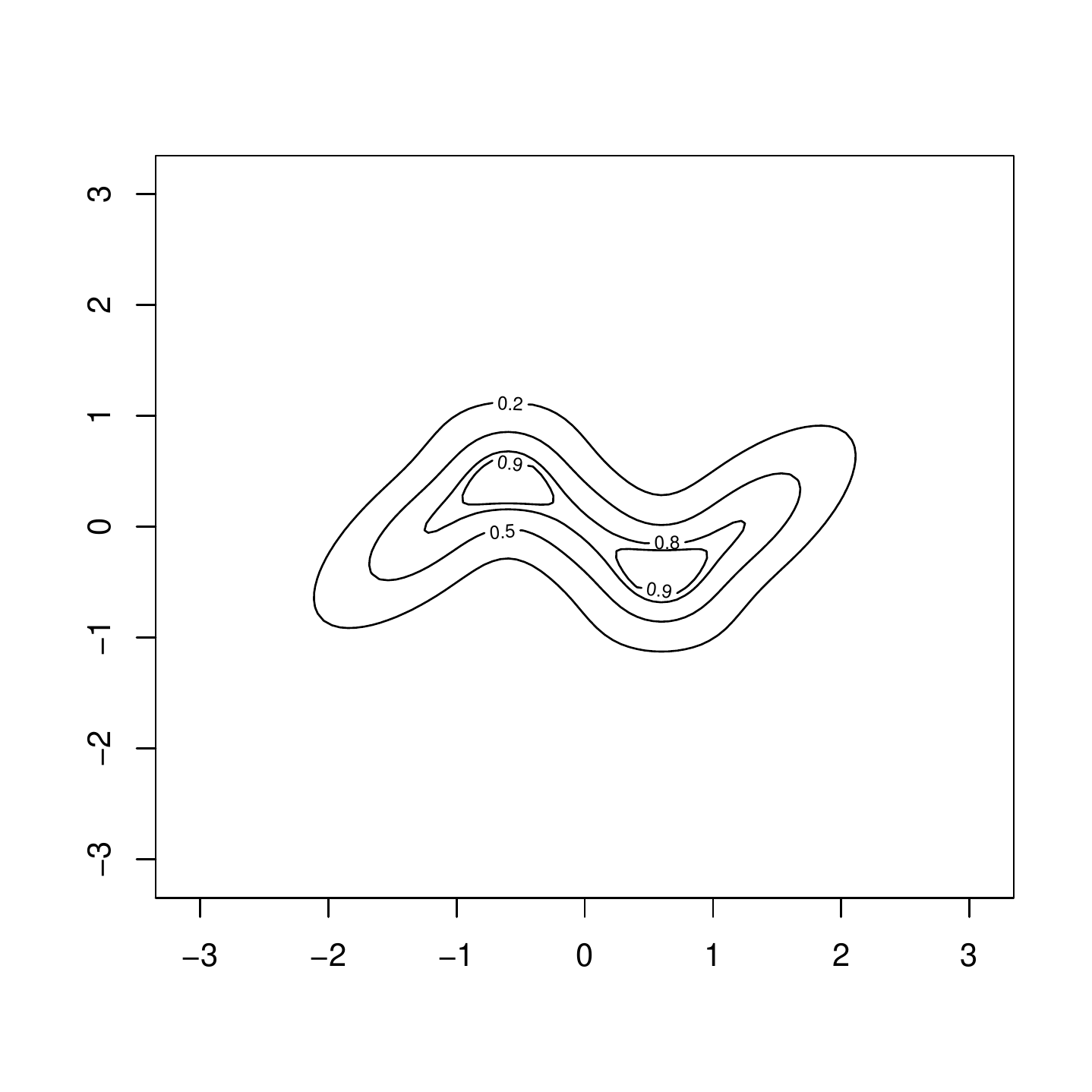}}
	\put(250,-615){\includegraphics[scale=.3]{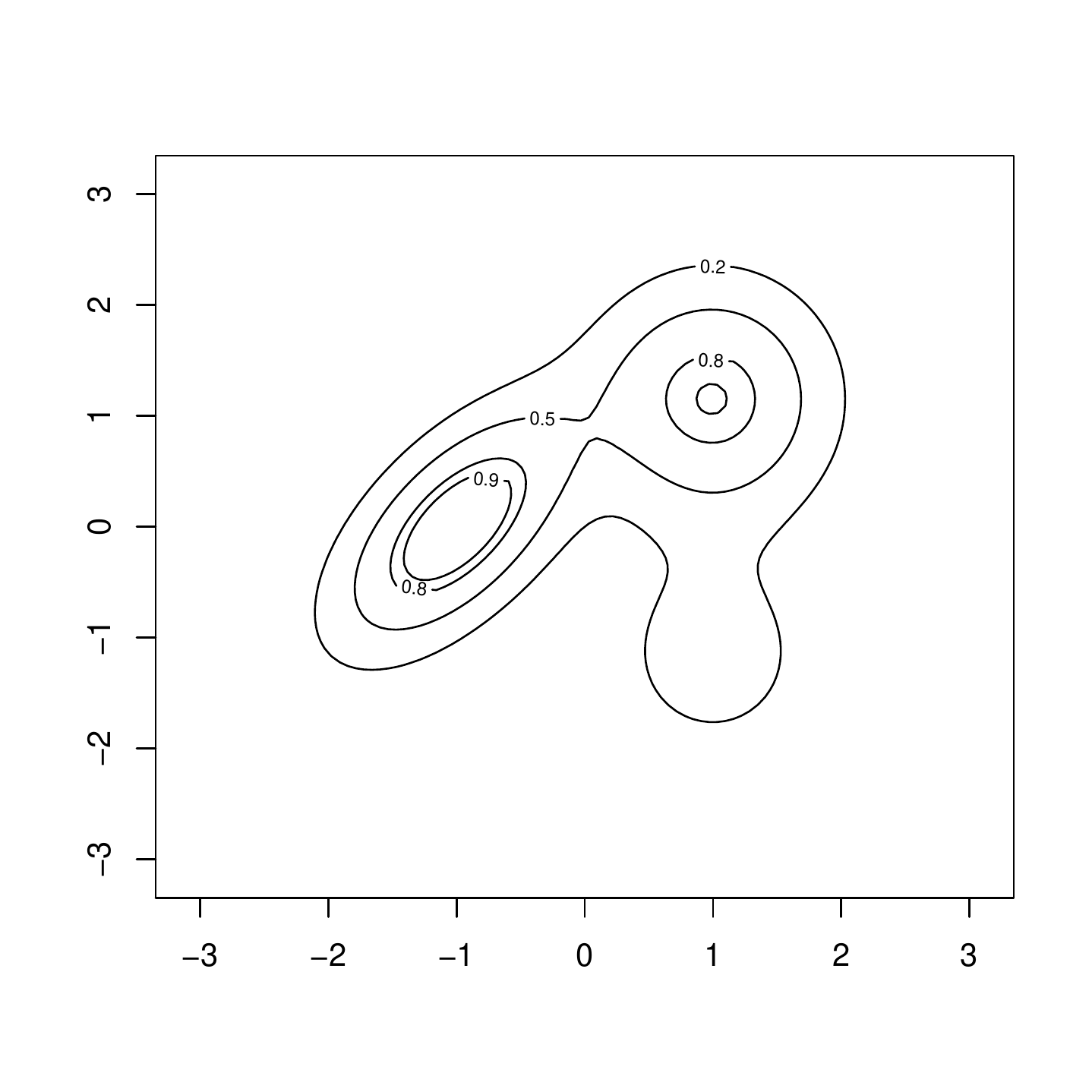}}
	\put(121,-500){ \small{\textbf{Trimodal II (8)}} }
	\put(268,-500){ \small{\textbf{Trimodal III (9)}} }	 
	\end{picture}\vspace{21cm}
	\caption{Contours of nine bivariate normal mixture densities corresponding to $\tau=0.2$, $\tau=0.5$,  $\tau=0.8$ and $\tau=0.9$.}\label{simulationmodels}
\end{figure}
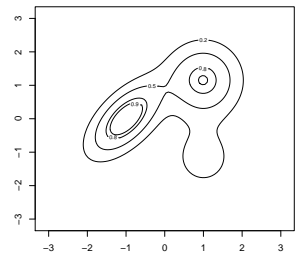

The behavior of the HDRs estimator established in Algorithm \ref{al1} will be analyzed through a simulation study. Nine normal mixture models (see \cite{wand}) are considered in this study. Figure \ref{simulationmodels} contains the contours for $\tau=0.2$, $\tau=0.5$, $\tau=0.8$ and $\tau=0.9$ of the corresponding HDRs. So, the number of connected components for each value of $\tau$ can be checked easily.

Algorithm \ref{al1} described in Section \ref{4} is compared with two plug-in methods emerging from the consideration of different bandwidth selectors. Concretely, a plug-in selector $H_1$ generalizing the proposal in \cite{wand1994multivariate} and the cross-validation selector presented in \cite{duong2005cross} are used. Moreover, the Hausdorff distance and the distance in measure are the two error criteria considered in order to analyze the performance of these HDRs estimation methods.

A total of $250$ random samples of sizes $n = 500$, $n = 1000$ and $n=2000$ were generated for each
model. Note that the largest samples sizes are considered in order to analyze the behavior of the algorithm attending to the enormous amounts of data available for COVID-19. From each sample, HDRs were reconstructed using the three cited methods for $\tau=0.2$, $\tau=0.5$, $\tau=0.8$ and $\tau=0.9$. Remark that the last two values of $\tau$ correspond to the greatest modes which will be particularly interesting for the real-data analysis of the COVID-19 distribution. Two different values for the number of bootstrap resamples $B$ ($100$ and $250$), six different values of the probability $p$ ($p_1=0.05$, $p_2=0.1$, $p_3=0.15$, $p_4=0.2$, $p_5=0.25$ and $p_6=0.3$) and just a single value of parameter \textit{step} ($0.025$) were considered. Some simulations were also performed when the parameter \textit{step} takes the value $0.01$ but non significant differences were detected. The pilot bandwidth used in order to determine the kernel density estimator in Algorithm \ref{al1} is $H_1$.

\begin{table}[h!]\centering
	\caption{For models from 1 to 9 and for the hybrid method taking $p=p_3$, quotient ($q_1$) of the Hausdorff error means for $n=500$ and $n=1000$ when $\tau=0.5$ and $B=100$; quotient ($q_2$) of Hausdorff error means for $n=500$ and $n=1000$ when $\tau=0.8$ and $B=250$ quotient ($q_3$) of Hausdorff error means for $B=100$ and $B=250$ when $n=500$ and $\tau=0.5$.}\label{q1}
	\vspace{.6cm}\begin{tabular}{cccccccccc}
		\hline
		Models& \textbf{1}&\textbf{2}&\textbf{3}&\textbf{4}&\textbf{5}&\textbf{6}&\textbf{7}&\textbf{8}&\textbf{9}\\
		\hline
		$q_1$&1.33& 1.25& 1.20& 1.14& 1.03& 1.53&1.00& 1.17& 1.07\\
		$q_2$& 1.24 &1.24& 1.21& 1.28& 1.08 &1.39 &1.06& 1.06& 1.13\\
		$q_3$&1.00 &1.04 &1.00 &1.00& 0.95& 0.98& 0.98& 0.98& 0.98\\
		\hline		
	\end{tabular}
\end{table}   

For each method and each sample, we measure the estimation error approximating the Hausdorff
distance between the boundaries of the estimated and the theoretical sets. A grid of points on the frontiers of these two sets can be easily obtained for the different estimation methods considered in this work. Note that the theoretical HDR and its estimator could be very close in Hausdorff
distance and still have quite different shapes. Such anomalies typically arise when boundaries of these sets are far apart. Hence, a natural way to reinforce the notion of visual proximity between the two sets is to impose the proximity of the respective boundaries. Additionally, the distance in measure is also approximated from the generation of a uniform sample of
size $2e+05$ on a square $[-3,3]\times [-3,3]$ containing the contours of the considered models.

Both error criteria provide similar results. However, distance in measure usually presents smaller errors when the Algorithm \ref{al1} shows a poor behavior with a milder penalization of its performance. For this reason, only results for Hausdorff distance are shown.

 \begin{table}[h!]\centering
 	\caption{Means and standard deviations of $250$ errors in Hausdorff distance for $\tau=0.8$, $n=500$ and $B=250$.}\label{tau08n500B250dh}\vspace{.6cm}	\small{
 		$\hspace{-1.2cm}$\begin{tabular}{ccccccccccccccccccc}
 			\hline
 			Models	&\multicolumn{2}{c}{\textbf{1}}&\multicolumn{2}{c}{\textbf{2}}&\multicolumn{2}{c}{\textbf{3}}&\multicolumn{2}{c}{\textbf{4}}&\multicolumn{2}{c}{\textbf{5}}&\multicolumn{2}{c}{\textbf{6}}&\multicolumn{2}{c}{\textbf{7}}&\multicolumn{2}{c}{\textbf{8}}&\multicolumn{2}{c}{\textbf{9}}\\
 			&M& SD&M& SD&M& SD&M& SD&M& SD&M& SD&M& SD&M& SD&M& SD \\  \hline
 			$p_1$&0.21&0.07&0.26&0.09&0.17&0.05&0.23&0.06&0.55&0.07&0.48&0.13&0.73&0.44&0.53&0.15&0.63&0.15\\
 			$p_2$&0.21&0.07&0.25&0.09&0.17&0.05&0.23&0.06&0.55&0.07&0.47&0.13&0.74&0.47&0.53&0.14&0.62&0.14\\
 			
 			$p_3$&0.21&0.07&0.26&0.09&0.17&0.05&0.23&0.06&0.54&0.07&0.46&0.12&0.75&0.48&0.53&0.15&0.62&0.16\\
 			
 			$p_4$&0.21&0.07&0.26&0.09&0.17&0.05&0.22&0.06&0.54&0.07&0.46&0.12&0.75&0.48&0.53&0.15&0.62&0.14\\
 			
 			$p_5$&0.21&0.07&0.25&0.09&0.17&0.05&0.23&0.06&0.55&0.07&0.46&0.12&0.76&0.50&0.53&0.15&0.62&0.13\\
 			
 			$p_6$&0.21&0.07&0.26&0.09&0.17&0.05&0.23&0.06&0.55&0.07&0.46&0.12&0.74&0.46&0.53&0.15&0.62&0.15\\
 			
 			$H_1$&0.21&0.07&0.25&0.09&0.17&0.05&0.23&0.06&0.43&0.15&0.29&0.11&1.01&0.59&0.44&0.14&0.35&0.25\\
 			
 			$H_2$&0.19&0.07&0.23&0.09&0.16&0.05&0.22&0.06&0.42&0.16&0.27&0.10&1.02&0.58&0.44&0.15&0.37&0.29\\
 			\hline
 			$p_3/H_1$&1.00&1.00&1.04&1.00&1.00&1.00&1.00&1.00&1.26&0.47&1.59&1.09&0.74&0.81&1.20&1.07&1.77&0.64\\
 			
 			$p_3/H_2$&1.11&1.00&1.13&1.00&1.06&1.00&1.05&1.00&1.29&0.44&1.70&1.20&0.74&0.83&1.20&1.00&1.68&0.55\\				
 			\hline		
 	\end{tabular}}
 \end{table}

 Although simulation results for $\tau=0.2$ and $\tau=0.5$ are not shown in this section, the information that they provide is briefly exposed next. In particular, if $\tau=0.2$ and $n=1000$, plug-in and hybrid method provide similar Hausdorff results for unimodal models from 1 to 4. In particular, hybrid method is the most competitive for densities 3 and 4 characterized its skewness and its kurtosis, respectively. From models 6 to 9, hybrid results are remarkably poor. Concretely, the hybrid error means is sometimes more than twice the plug-in error means. The same comparison is also considered when $\tau=0.5$ and $n=1000$. Again, results for the four first models are similar. For models from 6 to 9, the Hausdorff mean errors decrease considerably reinforcing the idea that when bigger values $\tau$ are selected, hybrid method is more precise. Quotients $q_1$ and $q_2$ exposed in Table \ref{q1} show that the hybrid algorithm improves its behavior considerably when the sample size increases. Remember that $q_1$ denotes the quotient the Hausdorff error means for $n=500$ and $n=1000$ when $\tau=0.5$ and $B=100$ and $q_2$, when $\tau=0.8$ and $B=250$. In particular, errors for the bimodal density have been reduced a 53\% and a 39\% when $\tau=0.5$ and $\tau=0.8$, respectively. The error reduction is less remarkable for models 8 and 9 but, it should be noted, that they are really complex models to be estimated, at least, for the considered values of $n$. The influence of the parameter $B$ denoting the number of bootstrap resamples considered in Algorithm \ref{al1} can be checked through quotient $q_3$ of Hausdorff error means for $B=100$ and $B=250$ when $n=500$ and $\tau=0.5$ contained in Table \ref{q1}. Then, the effect of $B$ is non significant. It only remains to ckeck the influence of  probability $p$. The analysis of simulations for $\tau=0.8$ and $\tau=0.9$, that will be exposed 
 immediately, will show that it also plays a non relevant role.

 Table \ref{tau08n500B250dh} shows the average (M) and the standard deviations (SD) of the $250$ Hausdorff estimation errors
 obtained for the three mentioned HDRs estimation methods when $\tau=0.8$ from samples of sizes $n = 500$ and $B=250$, respectively. The first six rows contain the results of Algorithm \ref{al1} for the considered values of $p$. Results clearly show that this parameter has not influence on the estimations given the higher flexibility of the hybrid method. Rows seven and eight contain the mean errors for plug-in methods with bandwidths $H_1$ and $H_2$ respectively, with similar results. Note that they present similar results. The two last rows show, for comparisons, the error means quotients of Algorithm \ref{al1} when $p=p_3$ and the plug-in methods. Note that the hybrid algorithm is competitive except to models 6 and 9. However, it offers the most competitive option if HDRs are reconstructed for model 7.

 \begin{table}[h!]
 	\caption{Means and standard deviations of $250$ errors in Hausdorff distance for $\tau=0.8$, $n=1000$ and $B=250$.}\vspace{0.6cm}\label{tau08n1000B250}\centering	\small{
 		$\hspace{-1.2cm}$\begin{tabular}{ccccccccccccccccccc}
 			\hline
 			Models&\multicolumn{2}{c}{\textbf{1}}&\multicolumn{2}{c}{\textbf{2}}&\multicolumn{2}{c}{\textbf{3}}&\multicolumn{2}{c}{\textbf{4}}&\multicolumn{2}{c}{\textbf{5}}&\multicolumn{2}{c}{\textbf{6}}&\multicolumn{2}{c}{\textbf{7}}&\multicolumn{2}{c}{\textbf{8}}&\multicolumn{2}{c}{\textbf{9}}\\
 			&M& SD&M& SD&M& SD&M& SD&M& SD&M& SD&M& SD&M& SD&M& SD \\  \hline
 			$p_1$&0.17&0.05&0.21&0.07&0.13&0.04&0.17&0.05&0.51&0.05&0.33&0.07&0.70&0.34&0.51&0.15&0.56&0.10\\
 			$p_2$&0.17&0.05&0.21&0.07&0.13&0.04&0.17&0.05&0.50&0.05&0.33&0.06&0.71&0.35&0.50&0.14&0.56&0.10\\
 			
 			$p_3$&0.17&0.05&0.21&0.07&0.14&0.04&0.18&0.05&0.50&0.05&0.33&0.06&0.71&0.35&0.50&0.16&0.55&0.07\\
 			
 			$p_4$&0.17&0.05&0.21&0.07&0.14&0.04&0.18&0.05&0.51&0.05&0.32&0.06&0.69&0.32&0.50&0.15&0.55&0.10\\
 			
 			$p_5$&0.17&0.05&0.21&0.07&0.13&0.04&0.17&0.05&0.51&0.05&0.32&0.06&0.70&0.32&0.50&0.16&0.55&0.11\\
 			
 			$p_6$&0.17&0.05&0.21&0.07&0.13&0.04&0.17&0.05&0.50&0.05&0.32&0.06&0.71&0.36&0.49&0.14&0.55&0.10\\
 			
 			$H_1$&0.17&0.06&0.21&0.07&0.14&0.04&0.19&0.05&0.32&0.11&0.22&0.08&0.76&0.50&0.38&0.13&0.27&0.18\\
 			
 			$H_2$&0.17&0.05&0.20&0.07&0.13&0.04&0.18&0.05&0.32&0.12&0.20&0.07&0.78&0.50&0.38&0.13&0.26&0.18\\
 			\hline
 			$p_3/H_1$&1.00&0.83&1.00&1.00&1.00&1.00&0.95&1.00&1.56&0.45&1.50&0.75&0.93&0.70&1.32&1.23&2.04&0.39\\
 			
 			$p_3/H_2$&1.00&1.00&1.05&1.00&1.08&1.00&1.00&1.00&1.56&0.42&1.65&0.86&0.91&0.70&1.32&1.23&2.12&0.39\\
 			\hline		
 	\end{tabular}} 
 \end{table}    
 \begin{figure}[h!]
 	$\hspace{-.9cm}$	\includegraphics[scale=.3]{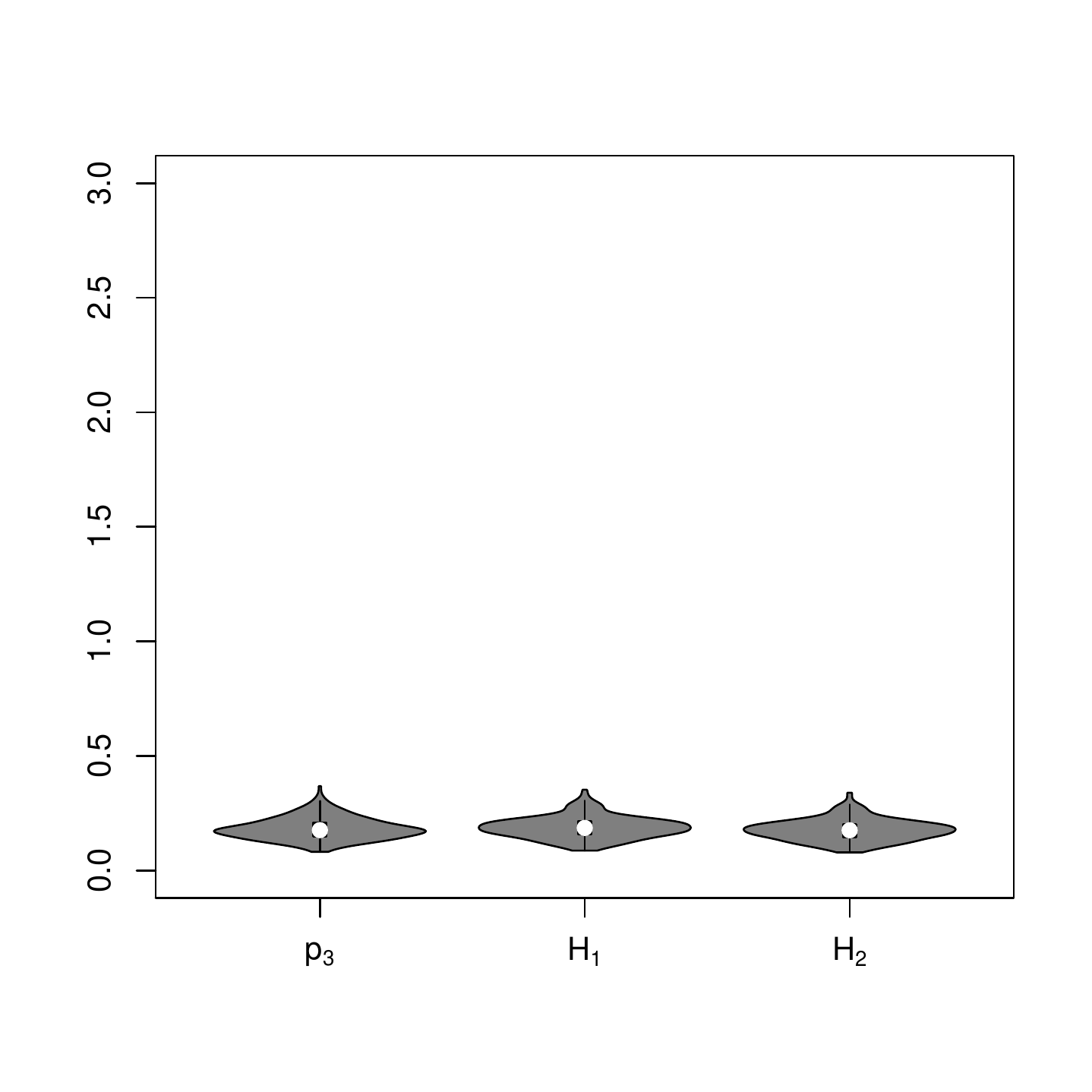}\includegraphics[scale=.3]{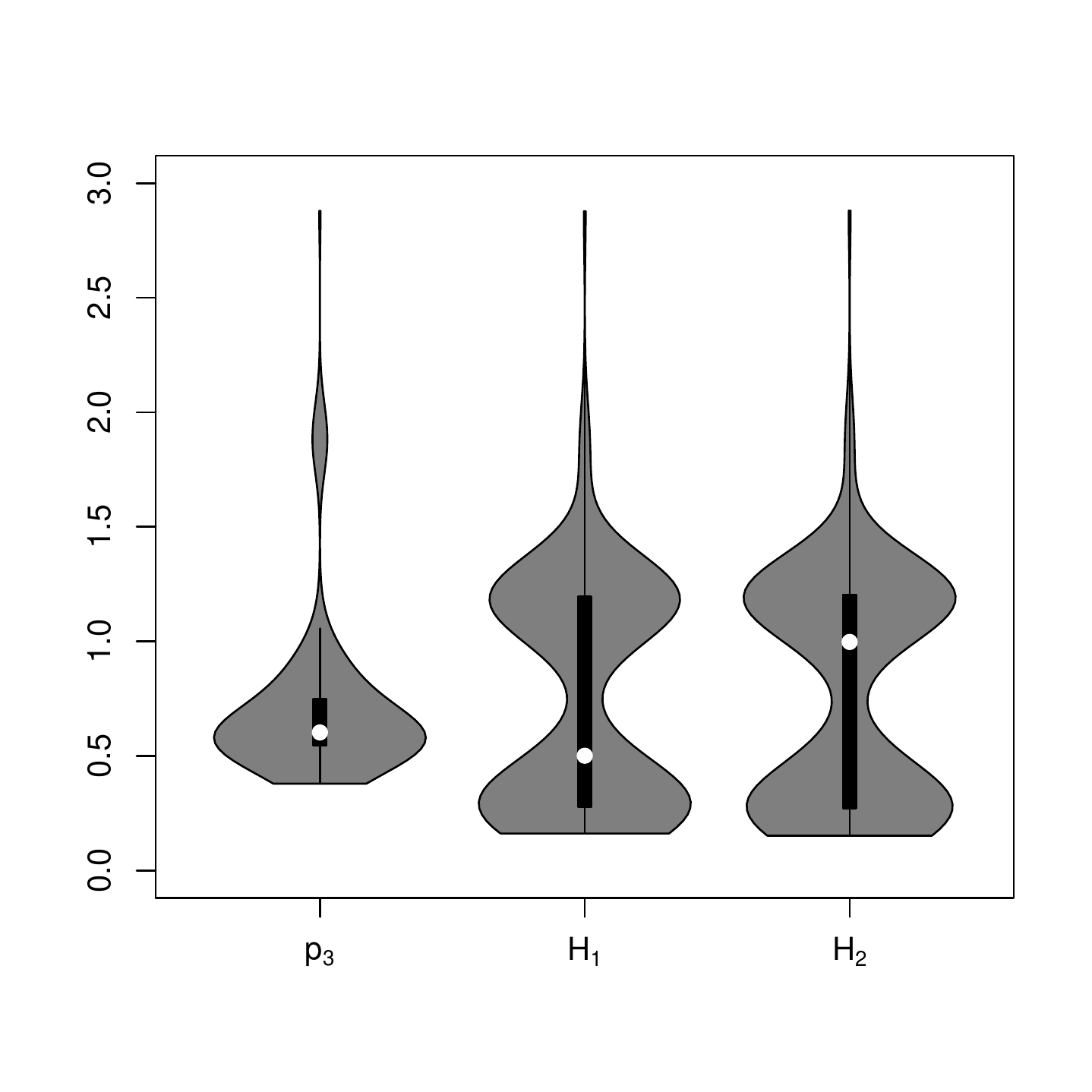}\includegraphics[scale=.3]{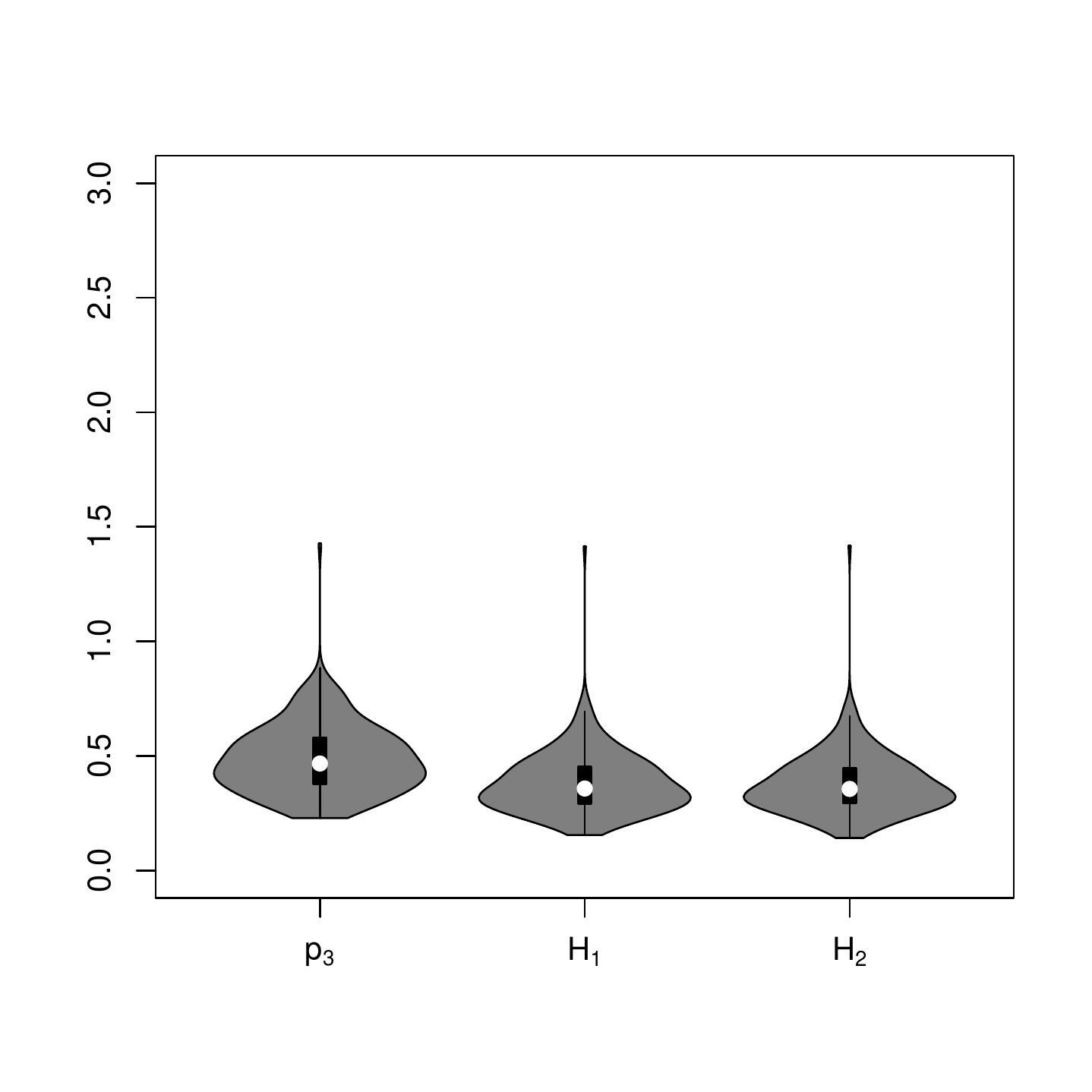}\includegraphics[scale=.3]{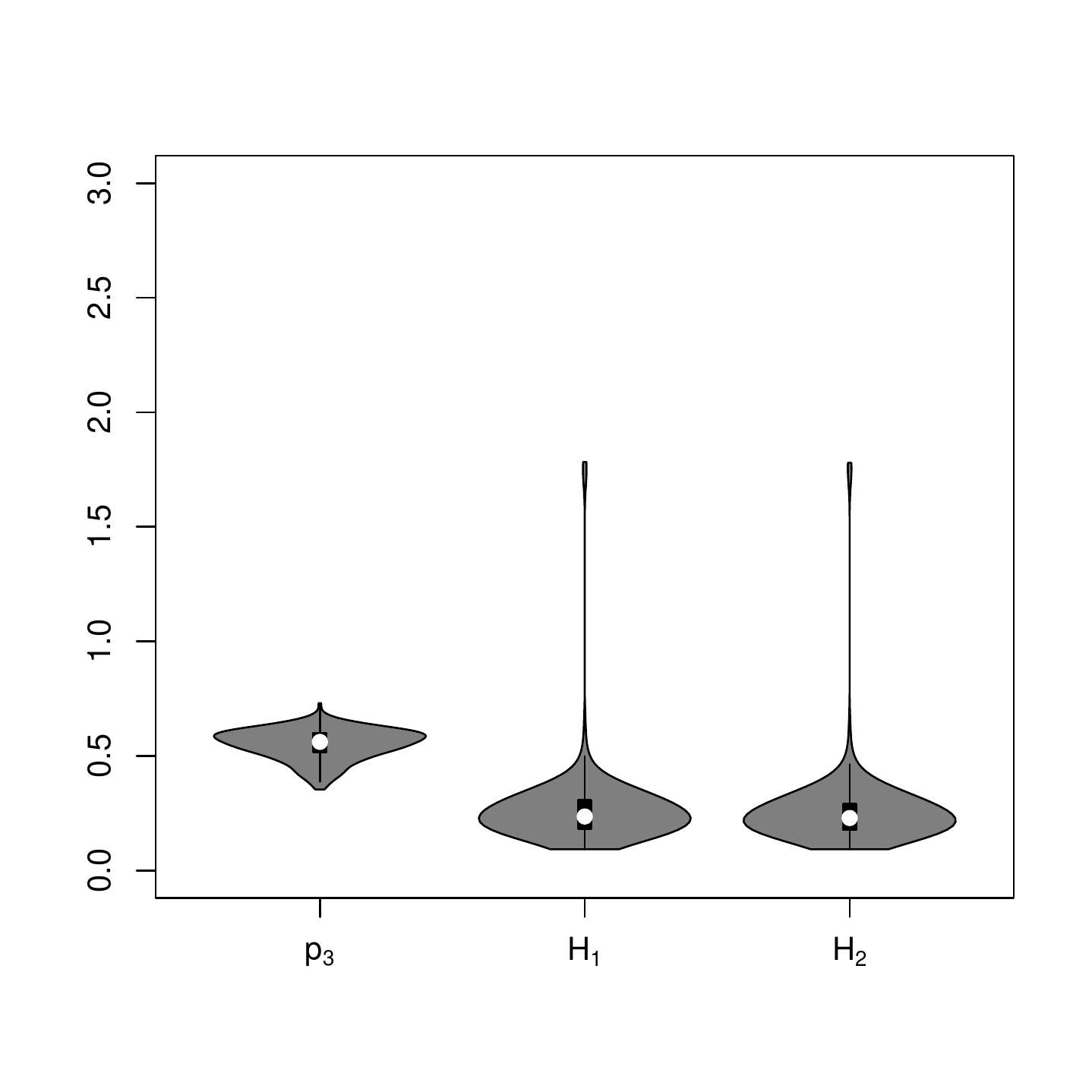}\vspace{-1.5cm}\\
 	\caption{Violin plots of Hausdorff errors of Algorithm \ref{al1} (selecting $p_3$) and plug-in methods (with bandwidths $H_1$ and $H_2$) for models 4, 7, 8 and 9 when $\tau=0.8$, $n=1000$ and $B=250$.}\label{v08}
 \end{figure}

$$ $$

 \begin{table}[h!]\centering
 	\caption{Means and standard deviations of $250$ errors in Hausdorff distance for $\tau=0.9$, $n=1000$ and $B=100$.}\vspace{0.6cm}\label{tau09n1000B100}	\small{
 		$\hspace{-1.2cm}$\begin{tabular}{ccccccccccccccccccc}
 			\hline
 			Models&\multicolumn{2}{c}{\textbf{1}}&\multicolumn{2}{c}{\textbf{2}}&\multicolumn{2}{c}{\textbf{3}}&\multicolumn{2}{c}{\textbf{4}}&\multicolumn{2}{c}{\textbf{5}}&\multicolumn{2}{c}{\textbf{6}}&\multicolumn{2}{c}{\textbf{7}}&\multicolumn{2}{c}{\textbf{8}}&\multicolumn{2}{c}{\textbf{9}}\\
 			&M& SD&M& SD&M& SD&M& SD&M& SD&M& SD&M& SD&M& SD&M& SD \\  \hline
 			$p_1$&0.19&0.07&0.22&0.09&0.14&0.04&0.16&0.06&0.67&0.21&0.44&0.38&1.58&1.00&0.46&0.19&0.90&0.49\\
 			$p_2$&0.18&0.07&0.22&0.09&0.14&0.04&0.15&0.05&0.67&0.21&0.42&0.32&1.58&1.00&0.46&0.20&0.88&0.47\\
 			$p_3$&0.19&0.07&0.22&0.09&0.14&0.05&0.16&0.05&0.67&0.21&0.43&0.37&1.60&1.00&0.46&0.20&0.88&0.48\\
 			$p_4$&0.19&0.07&0.22&0.09&0.14&0.05&0.16&0.05&0.68&0.23&0.42&0.35&1.58&1.00&0.45&0.19&0.89&0.49\\
 			$p_5$&0.18&0.07&0.22&0.09&0.14&0.05&0.16&0.05&0.67&0.21&0.41&0.31&1.53&0.99&0.46&0.20&0.90&0.50\\
 			$p_6$&0.18&0.07&0.22&0.09&0.14&0.04&0.16&0.06&0.66&0.19&0.42&0.32&1.60&1.00&0.46&0.20&0.89&0.49\\
 			$H_1$&0.18&0.07&0.22&0.09&0.14&0.05&0.16&0.06&0.40&0.33&0.36&0.50&1.87&0.85&0.41&0.23&0.69&0.71\\
 			$H_2$&0.17&0.06&0.21&0.08&0.14&0.04&0.15&0.06&0.39&0.34&0.34&0.50&1.88&0.84&0.40&0.24&0.74&0.74\\
 			\hline
 			$p_3/H_1$&1.06&1.00&1.00&1.00&1.00&1.00&1.00&0.83&1.68&0.64&1.19&0.74&0.86&1.18&1.12&0.87&1.28&0.68\\
 			$p_3/H_2$&1.12&1.17&1.05&1.12&1.00&1.25&1.07&0.83&1.72&0.62&1.26&0.74&0.85&1.19&1.15&0.83&1.19&0.65\\
 			\hline		
 	\end{tabular}}
 \end{table}

 \begin{table}[h!]\centering
 	\caption{Means and standard deviations of $250$ errors in Hausdorff distance for $\tau=0.9$, $n=2000$ and $B=100$.}\vspace{0.6cm}\label{tau09n2000B100}	\small{
 		$\hspace{-1.2cm}$\begin{tabular}{ccccccccccccccccccc}
 			\hline
 			Models	&\multicolumn{2}{c}{\textbf{1}}&\multicolumn{2}{c}{\textbf{2}}&\multicolumn{2}{c}{\textbf{3}}&\multicolumn{2}{c}{\textbf{4}}&\multicolumn{2}{c}{\textbf{5}}&\multicolumn{2}{c}{\textbf{6}}&\multicolumn{2}{c}{\textbf{7}}&\multicolumn{2}{c}{\textbf{8}}&\multicolumn{2}{c}{\textbf{9}}\\
 			&M& SD&M& SD&M& SD&M& SD&M& SD&M& SD&M& SD&M& SD&M& SD \\  \hline
 			
 			$p_1$&0.15&0.05&0.18&0.07&0.12&0.04&0.13&0.04&0.60&0.05&0.30&0.07&1.24&0.76&0.39&0.13&0.79&0.54\\
 			$p_2$&0.15&0.05&0.18&0.07&0.11&0.04&0.12&0.04&0.61&0.12&0.30&0.07&1.29&0.80&0.38&0.11&0.74&0.52\\
 			$p_3$&0.15&0.06&0.18&0.07&0.11&0.04&0.11&0.04&0.59&0.06&0.29&0.07&1.28&0.78&0.39&0.13&0.77&0.54\\
 			$p_4$&0.15&0.05&0.18&0.07&0.11&0.04&0.11&0.04&0.60&0.12&0.29&0.07&1.28&0.78&0.39&0.13&0.76&0.54\\
 			$p_5$&0.15&0.05&0.18&0.07&0.11&0.04&0.11&0.04&0.59&0.06&0.29&0.07&1.27&0.79&0.39&0.13&0.74&0.53\\
 			$p_6$&0.15&0.05&0.18&0.07&0.12&0.04&0.12&0.04&0.59&0.06&0.29&0.07&1.30&0.82&0.39&0.12&0.76&0.54\\
 			$H_1$&0.15&0.06&0.18&0.07&0.12&0.04&0.12&0.05&0.27&0.18&0.21&0.09&1.43&0.84&0.34&0.14&0.59&0.69\\
 			$H_2$&0.14&0.06&0.18&0.07&0.12&0.04&0.12&0.04&0.26&0.18&0.20&0.08&1.40&0.83&0.33&0.14&0.61&0.71\\
 			\hline
 			$p_3/H_1$&1.00&1.00&1.00&1.00&0.92&1.00&0.92&0.80&2.19&0.33&1.38&0.78&0.90&0.93&1.15&0.93&1.31&0.78\\
 			$p_3/H_2$&1.07&1.00&1.00&1.00&0.92&1.00&0.92&1.00&2.27&0.33&1.45&0.88&0.91&0.94&1.18&0.93&1.26&0.76\\
 			\hline		
 	\end{tabular}}
 \end{table}

 \begin{figure}[h!]
 	$\hspace{-.9cm}$	\includegraphics[scale=.3]{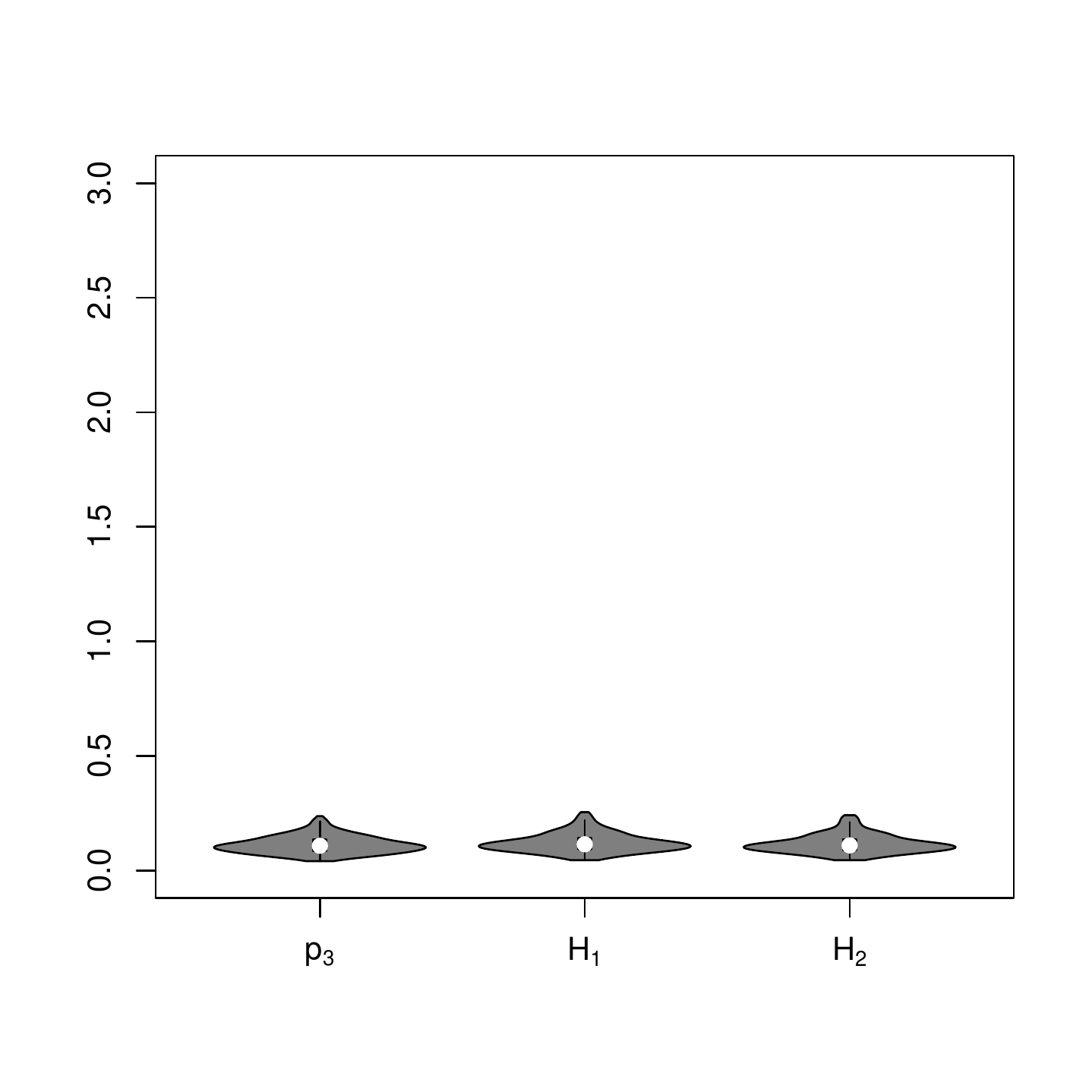}\includegraphics[scale=.3]{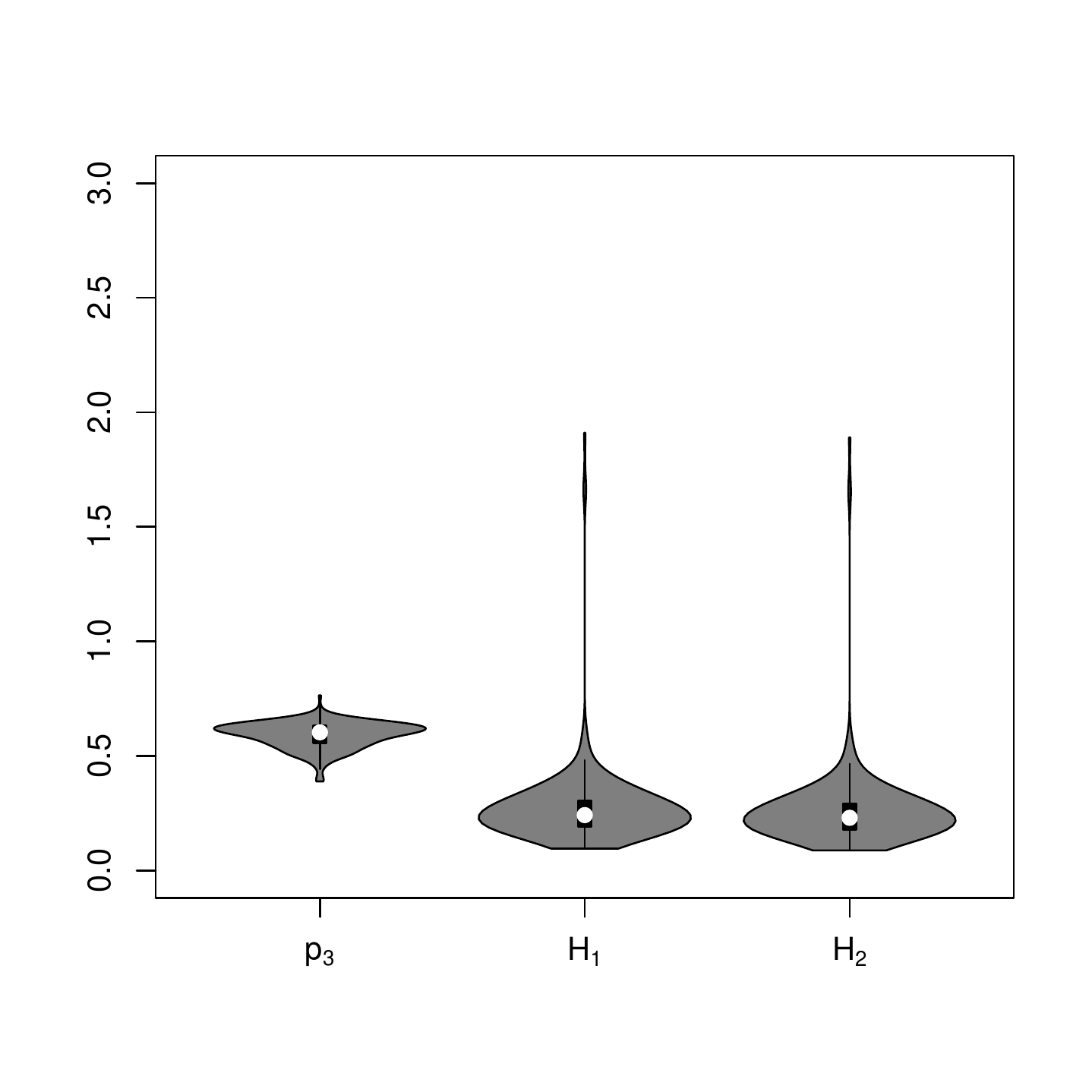}\includegraphics[scale=.3]{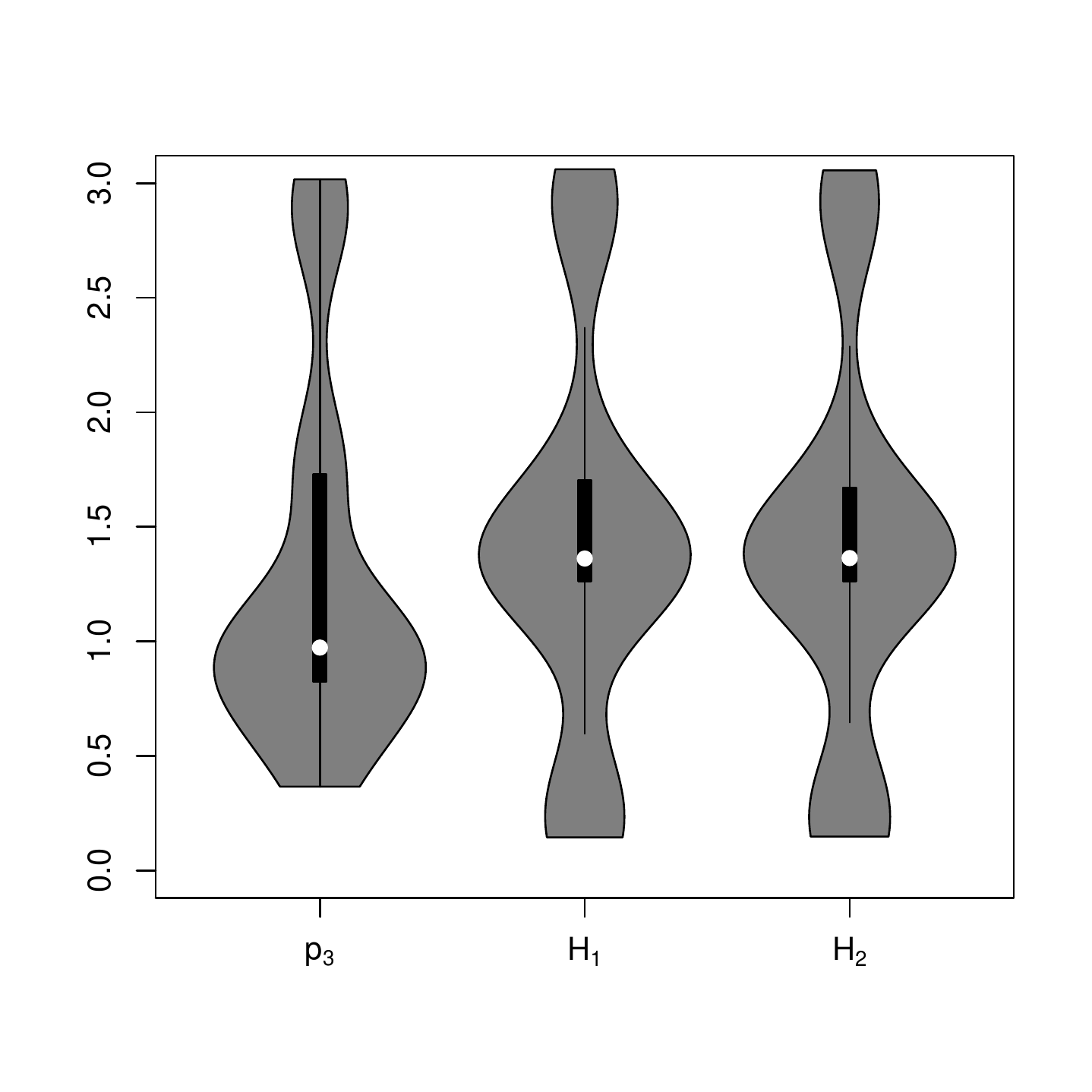}\includegraphics[scale=.3]{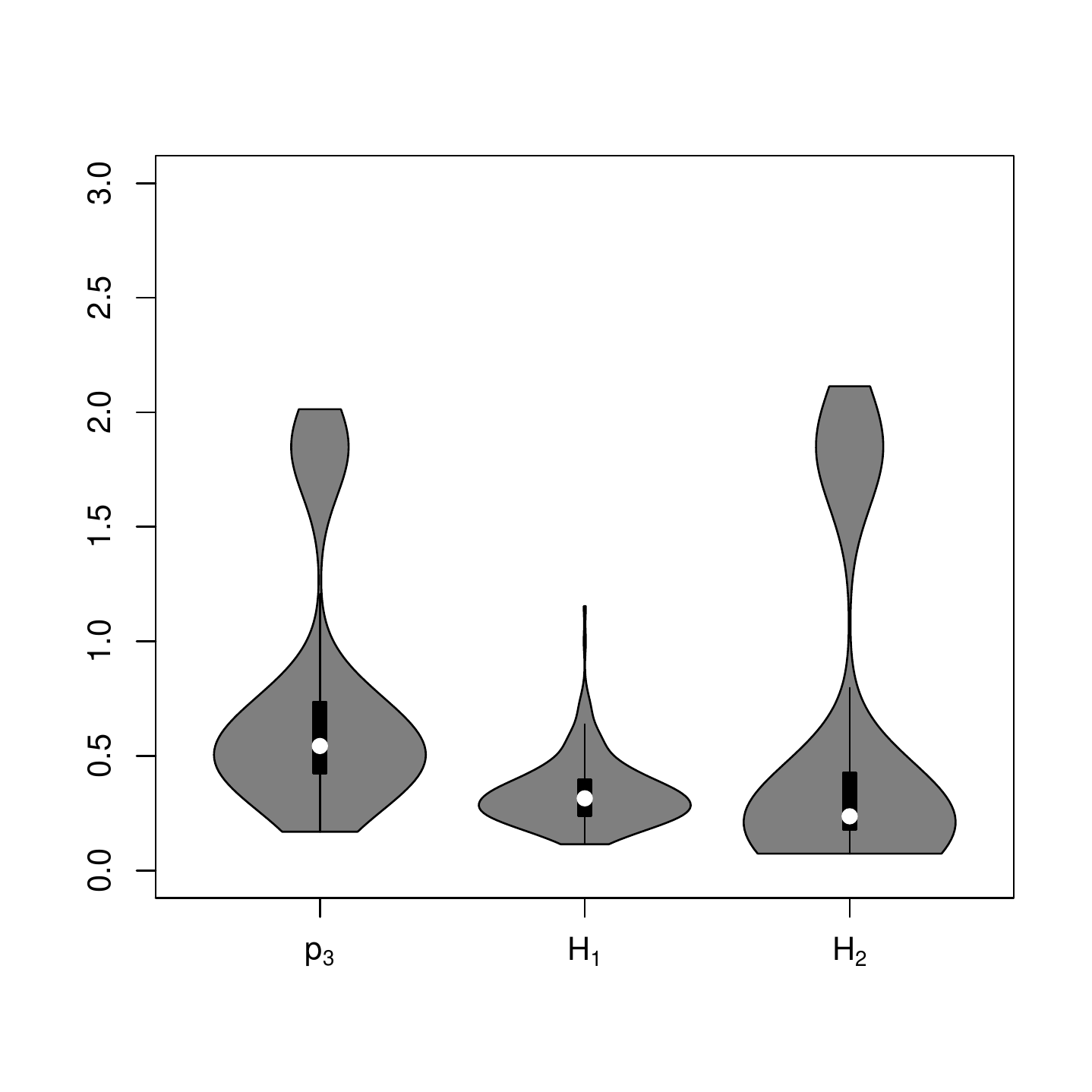}\vspace{-1.5cm}\\
 	\caption{Violin plots for Hausdorff errors of Algorithm \ref{al1} (selecting $p_3$) and plug-in methods (with bandwidths $H_1$ and $H_2$) for models 3, 5, 7 and 9 when $\tau=0.9$, $n=2000$ and $B=100$.}\label{v09}
 \end{figure}


 Table \ref{tau08n1000B250} shows the results obtained when $\tau=0.8$, $n = 1000$ and $B=250$. Although plug-in methods are clearly more competitive for models 5, 6 and 9, Algorithm \ref{al1} present a good performance for all the other densities. Despite plug-in methods improves their results considerably, the hybrid algorithm is again the best option for estimating the HDRs of model 7. This feature can be also seen in Figure \ref{v08}. Again, it is observed that the selection of $p$ is not relevant the simulated scenarios. The distance in measure improves slightly the hybrid errors. For instance, when distance in measure is used, $p_3/H_1$ is equal to $1.67$ for density 9.

 Simulations for $\tau=0.9$, $n =1000$ and $B=100$ are summarized in Table \ref{tau09n1000B100}. In this case, the hybrid method offers competitive results for all models except for the bimodal model 5. Additionally, the comparison of the nineth rows in Tables \ref{tau08n1000B250} and \ref{tau09n1000B100} show that the hybrid algorithm reduce considerably their results from models from 6 to 9. For instance, when $\tau=0.8$, its error is twice the plug-in error for density 9; although the HDR has also two connected components, when $\tau=0.9$, the hybrid error only increases a 28\%.







Table \ref{tau09n2000B100} shows the results for $\tau=0.9$, $n = 2000$ and $B=100$. Note that means are considerably smaller than errors shown in Table \ref{tau09n1000B100} for $n=1000$. However, plug-in methods become more competitive comparatively from models 5 to 9. This can be also seen in Figure \ref{v09}. Again, the consideration of the distance in measure makes smaller the hybrid errors. Concretely, when distance in measure is considered, $p_3/H_1$ is equal to $1.71$ and $1.26$ for densities 5 and 6, respectively.


Some general conclusions can be summarized from simulations. First, both plug-in methods considered present competitive and very similar results for any of the considered densities and values of $\tau$. As regards the method exposed in Algorithm \ref{al1}, it is important to note that it does not depend strongly on the selection of input parameters $B$, $p$ and \textit{step}. Moreover, it is checked that its performance clearly improves when bigger values of $\tau$ and $n$ are considered. These issues justify the application of this hybrid method for COVID-19 analysis because, in such setting, the identification of distribution modes (when $\tau$ is large) is specially relevant and the sample sizes are usually huge. Although results for smaller values of $\tau$ were not satisfactory, if $\tau$ is close to zero, this problem could be treated alternatively using support estimation methods such as the one proposed in \cite{rodriguez2019extent}. Finally, it is important to remark that the simplicity of plug-in methods and their good global results are not good enough reasons to reject Algorithm \ref{al1}. Both methods provide HDRs reconstructions of completely different nature. If information on the shape of the HDR is known, plug-in methods does not incorporate it in the estimator. Additionally, one of the most remarkable characteristic of this hybrid algorithm is that the resulting estimators present interesting mathematical properties such as easy-to-handle frontiers.

 \section{Space-time evolution of COVID-19 clusters for confirmed cases in the United States}\label{realdataanalysis}

Real-time detection of extremely high incidence of COVID-19 areas is a natural step within a proper space-time characterization of the disease evolution. This type of analysis enables the identification of regions where medical services and materials need to be reinforced. Moreover, it is also possible to detect areas where the coronavirus has suffered a setback or, on the contrary, it continues expanding. In particular, if concrete measures have been taken in order to stop its expansion in a specific area, the
 analysis of its spatial distribution could determinate their degree of efficacy.

 From the real data set already introduced in Section \ref{intro}, the time and spatial evolution of HDRs and their clusters for the coronavirus in the United States will be analyzed on a weekly basis between February and May 2020 using a plug-in method with bandwidth $H_1$ and also Algorithm \ref{al1}. For the last one, the number of bootstrap resamples generated is $B=250$, the value of $p$ is equal to $0.25$ and the value of \textit{step}, $0.025$. Specifically, a total of ten samples corresponding to the ten weeks studied were considered with sample sizes equal to $89$, $1007$, $8007$, $58673$, $149814$, $221738$, $214264$, $204644$, $209343$ and $167116$, respectively. Note that, except for the first two weeks when the number of data smaller, samples sizes are larger than $2000$, one of the reference values in the simulation study. According to the results shown in Section \ref{simus}, the algorithm analyzed in this work improves its performance when the sample size increases  and high values of $\tau$ are considered. Therefore, we expect that it was at least as competitive as plug-in methods in these scenarios.

\begin{figure}[h!]
\includegraphics[height=175pt,width=220pt]{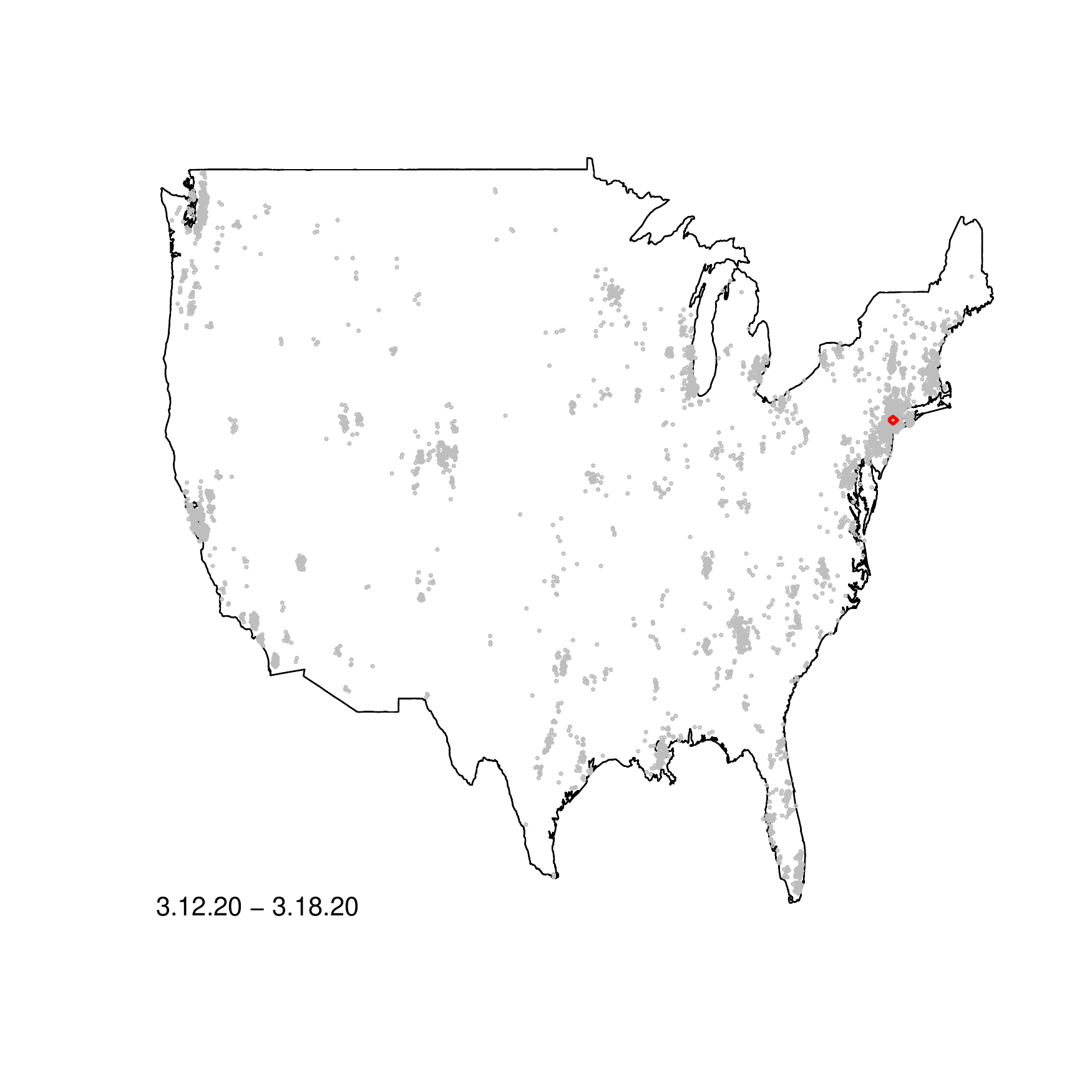}\includegraphics[height=175pt,width=220pt]{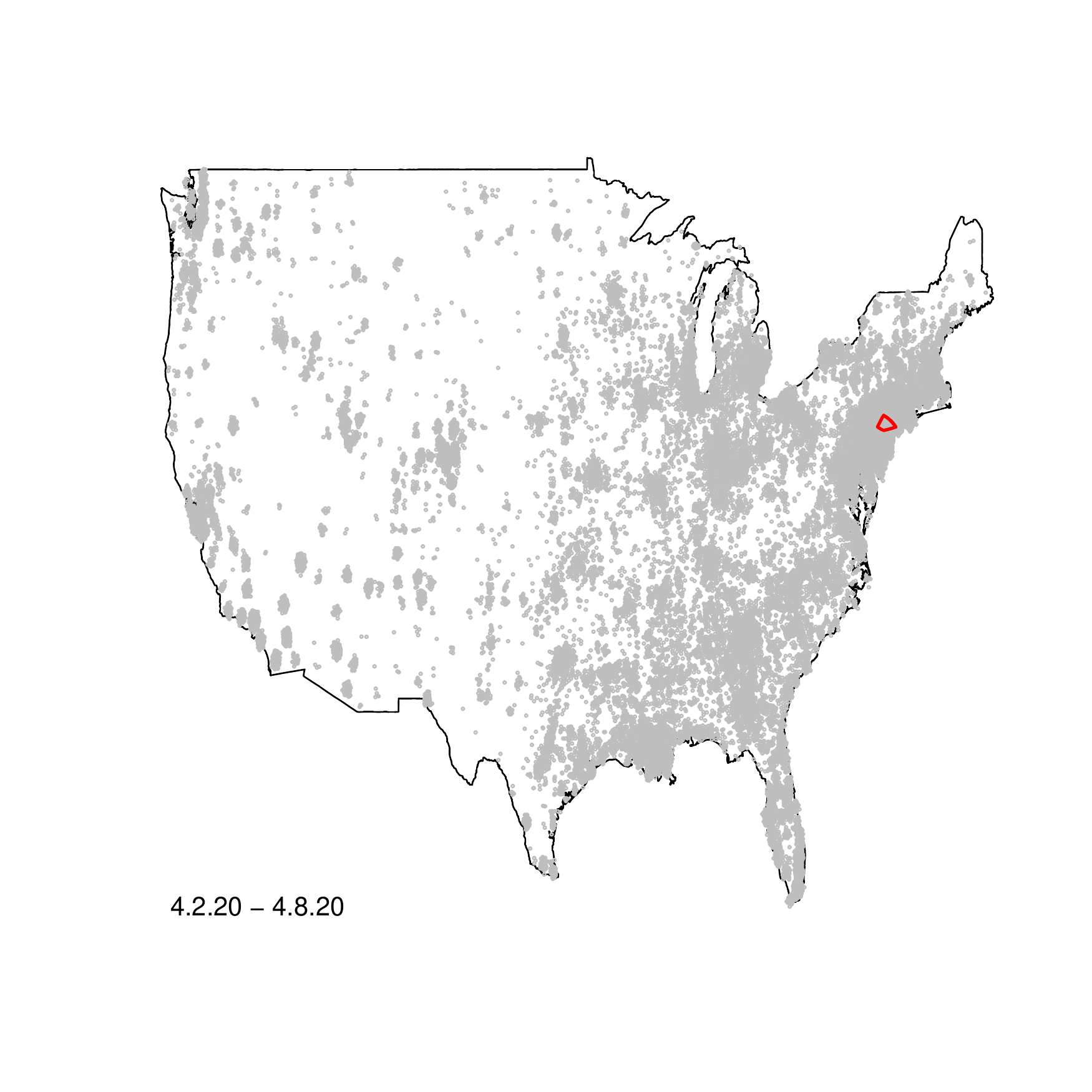}\\
\includegraphics[height=175pt,width=220pt]{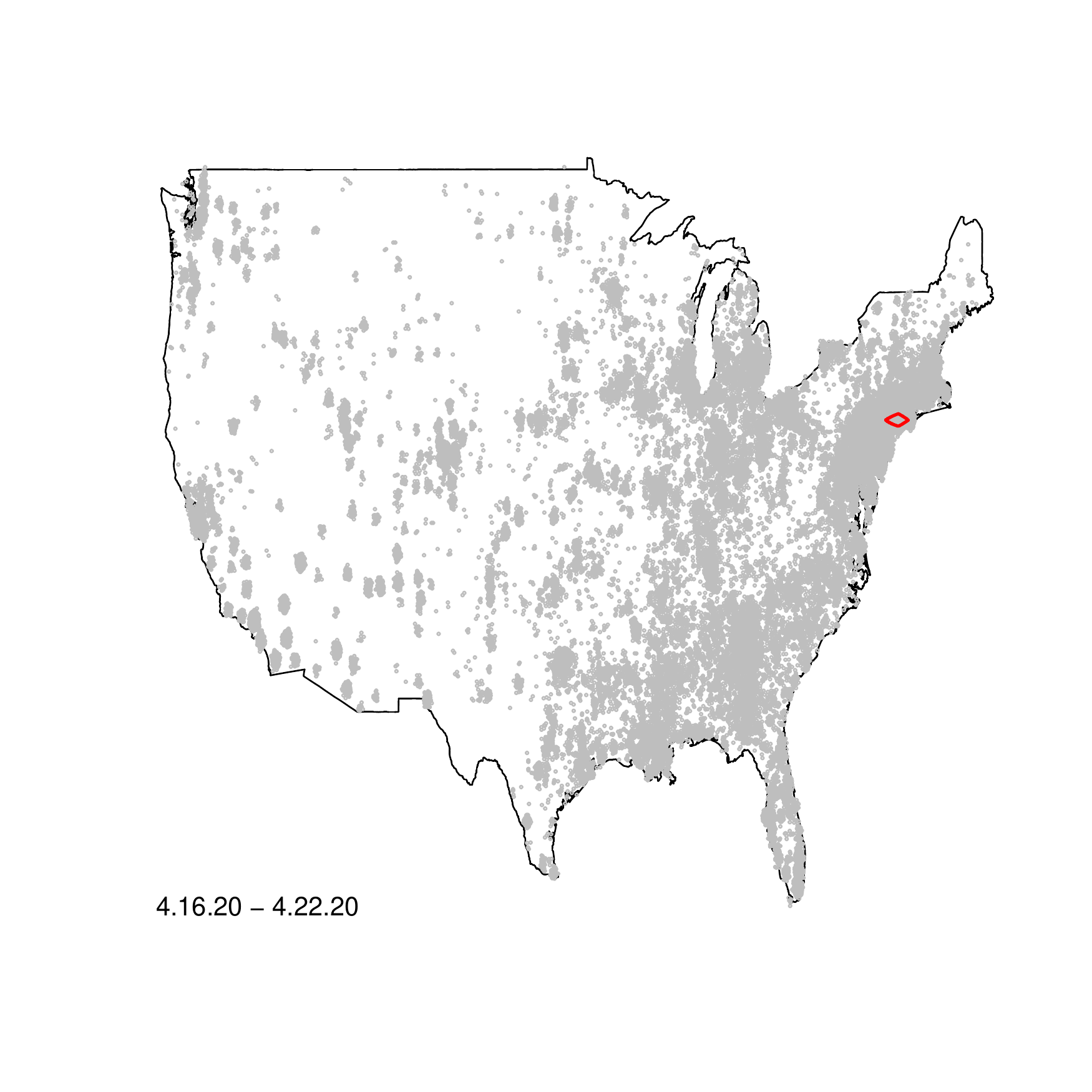}\includegraphics[height=175pt,width=220pt]{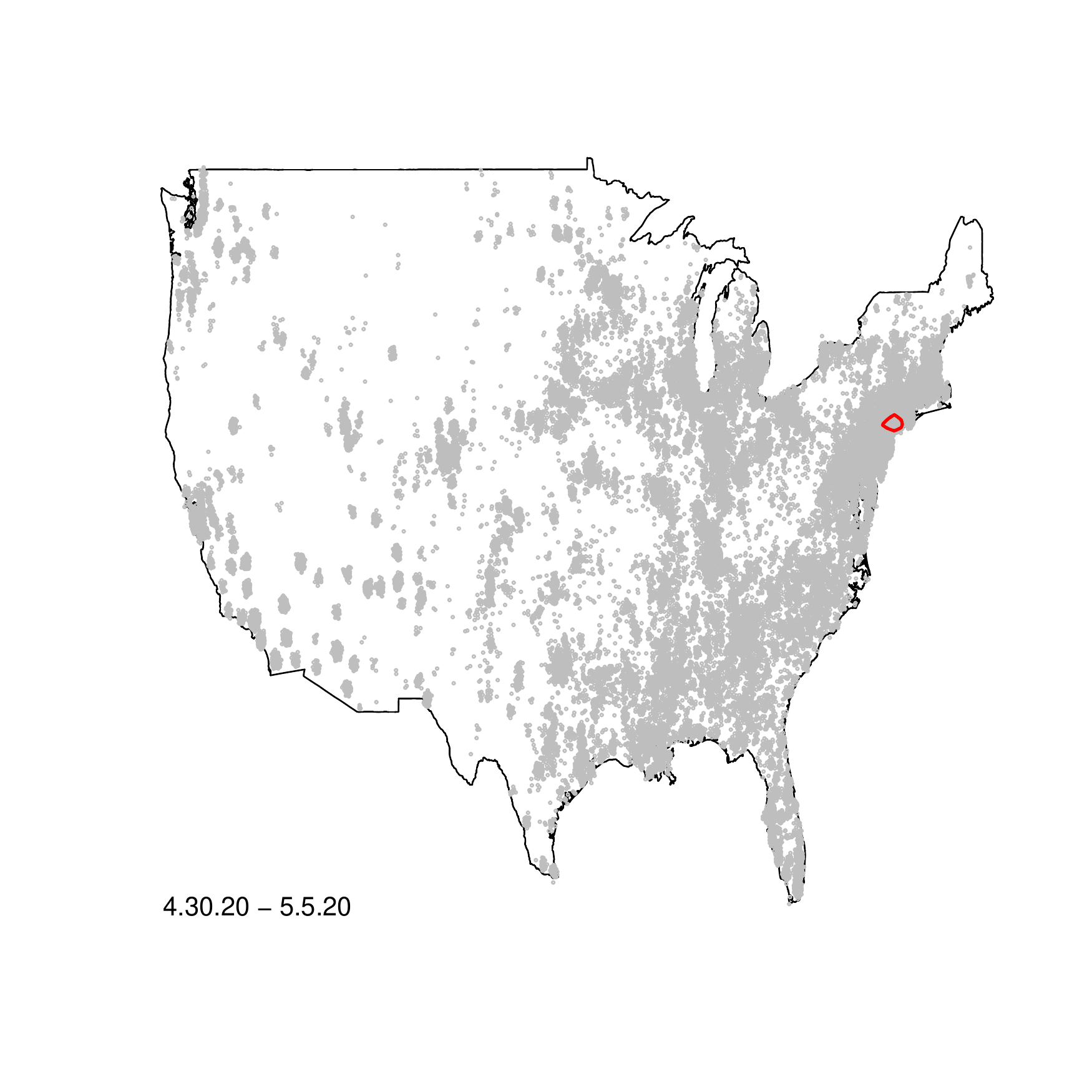}\vspace{-.3cm}\\
	\caption{Time and spatial evolution of HDRs when $\tau=0.9$ for confirmed COVID-19 cases in the United States by week between February 2020 and May 2020.}\label{coveeuu11}
\end{figure}

 Figure 1 in Section 2 of SM shows an interactive representation of weekly HDRs reconstructions $C_{\hat{r}_0(\hat{f}_\tau)}(\mathcal{X}_n^+
(\hat{f}_\tau))$ for $\tau=0.9$ obtained using the
 Algorithm \ref{al1}. Specifically, the animation covers the period from February 27th to May 5th. Remember that each one of the ten estimators contain at least the 10\% of confirmed cases. Therefore, the largest modes or clusters of COVID-19 distribution are identified. In the first week, a single cluster is detected in Seattle. However, only a week later, an additional risky area is located on New York. Note that New York becomes the only cluster detected from third to tenth week. Therefore, it is the city with the biggest numbers of cases which make sense because 
 it is also the most populous city in the United States. As a summary, Figure \ref{coveeuu11} shows the HDRs estimator for the third, 
 sixth, eighth and tenth weeks. 

\begin{figure}[h!]
	\includegraphics[height=175pt,width=220pt]{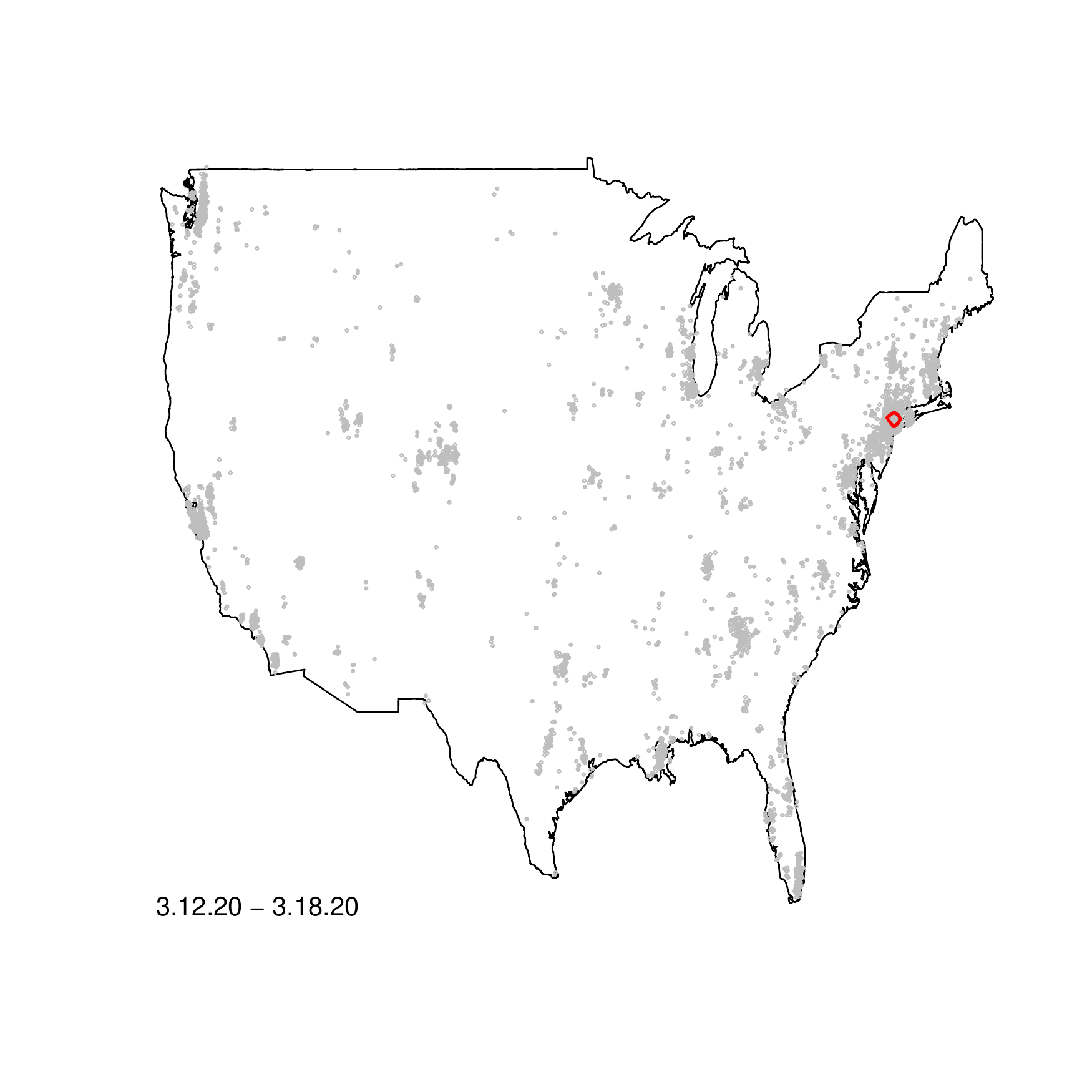}\includegraphics[height=175pt,width=220pt]{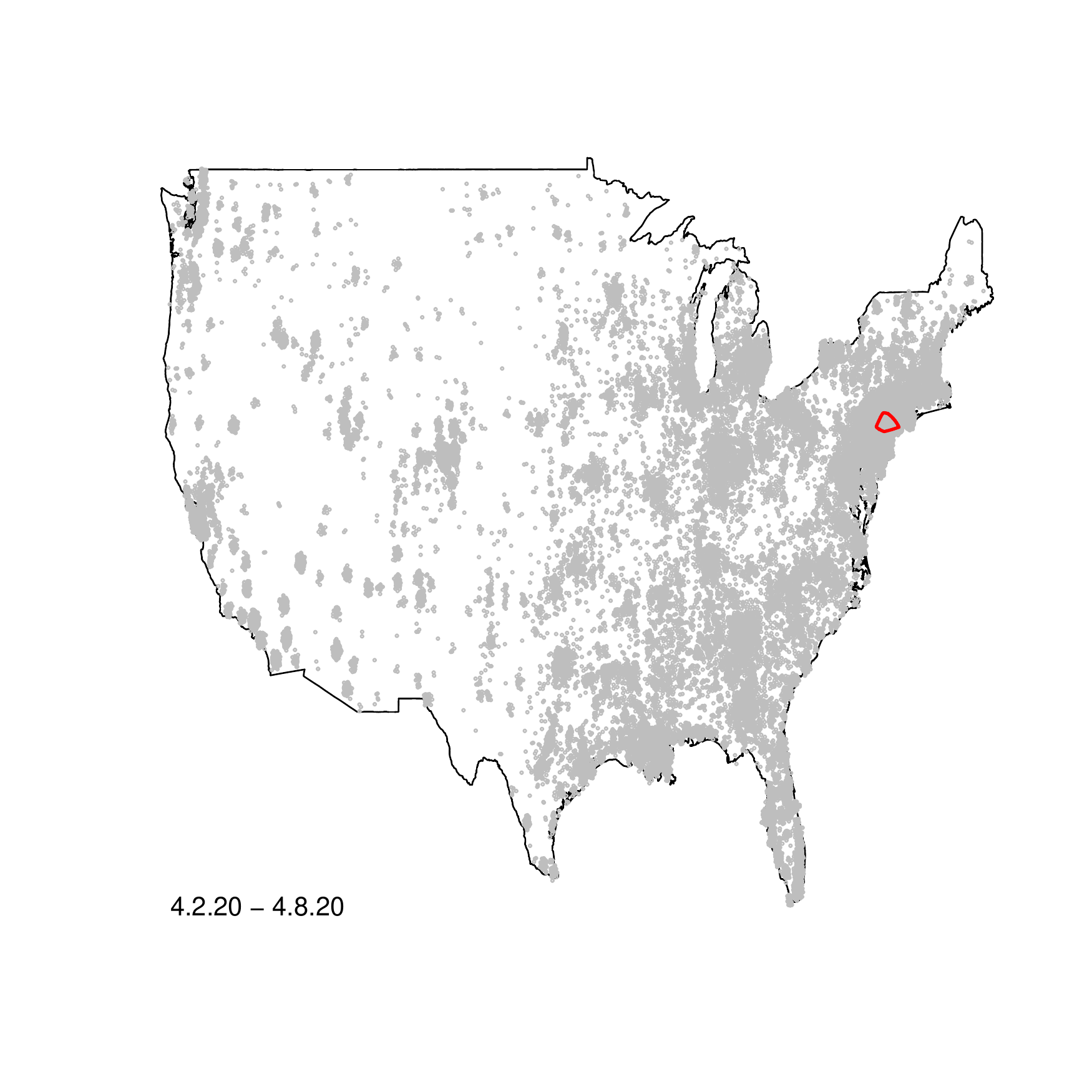}\\
	\includegraphics[height=175pt,width=220pt]{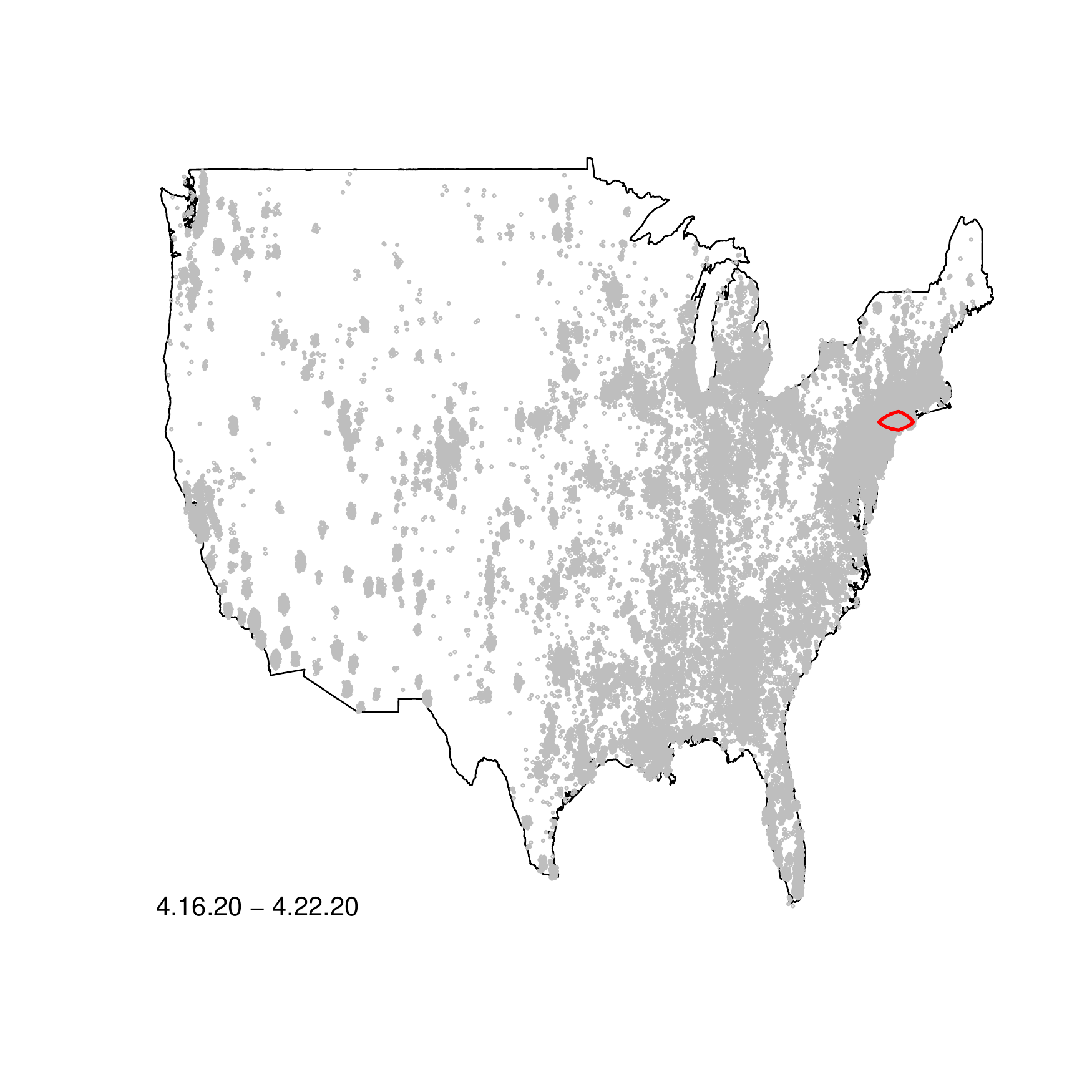}\includegraphics[height=175pt,width=220pt]{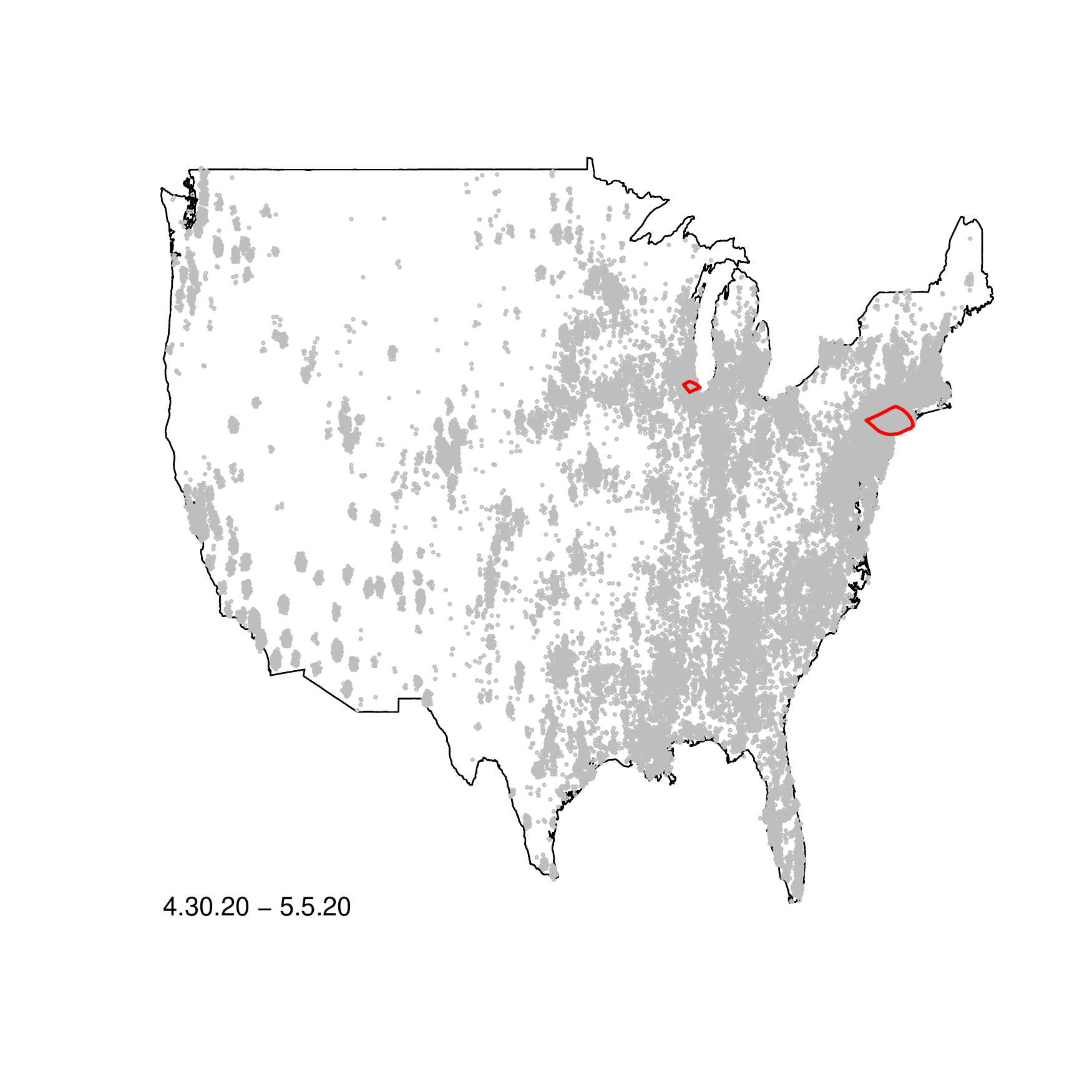}\vspace{-.3cm}\\
	\caption{Time and spatial evolution of HDRs when $\tau=0.8$ for confirmed COVID-19 cases in the United States by week between February 2020 and May 2020.}\label{coveeuu33}
\end{figure}

Figure 2 in Section 2 of SM shows, as before, an interactive representation of weekly HDRs estimators for $\tau=0.8$ from February 27th to May 5th. In this case, the set reconstructions contain at least the 20\% of confirmed cases of COVID-19 however results are practically equal to the obtained when $\tau=0.9$. The only major difference is detected in the last week analyzed when Chicago appears as a new problematic cluster. Again, Figure \ref{coveeuu33} shows the HDRs estimator for the third, sixth, eighth and tenth weeks.

\begin{figure}[h!]  
 		\includegraphics[height=175pt,width=220pt]{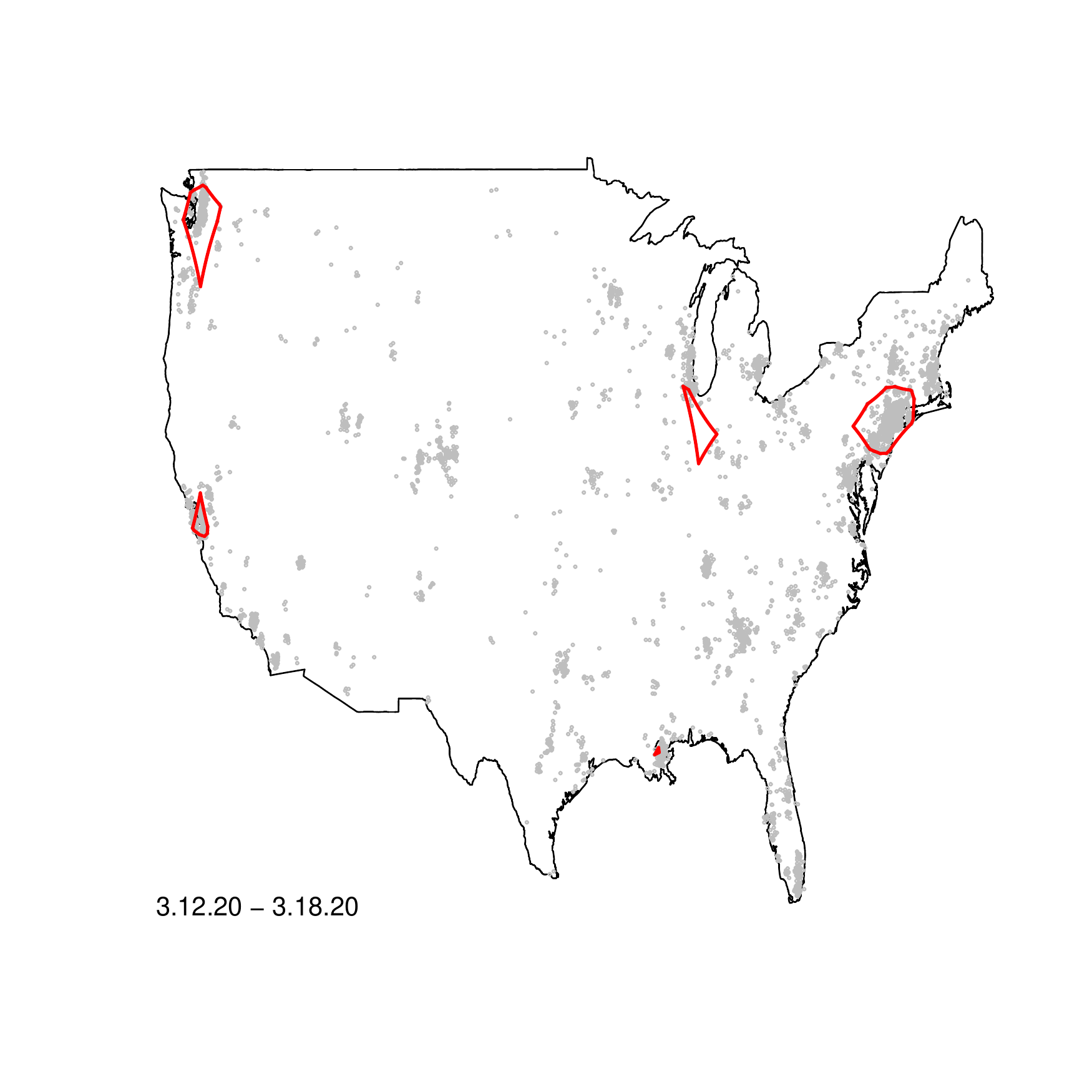}\includegraphics[height=175pt,width=220pt]{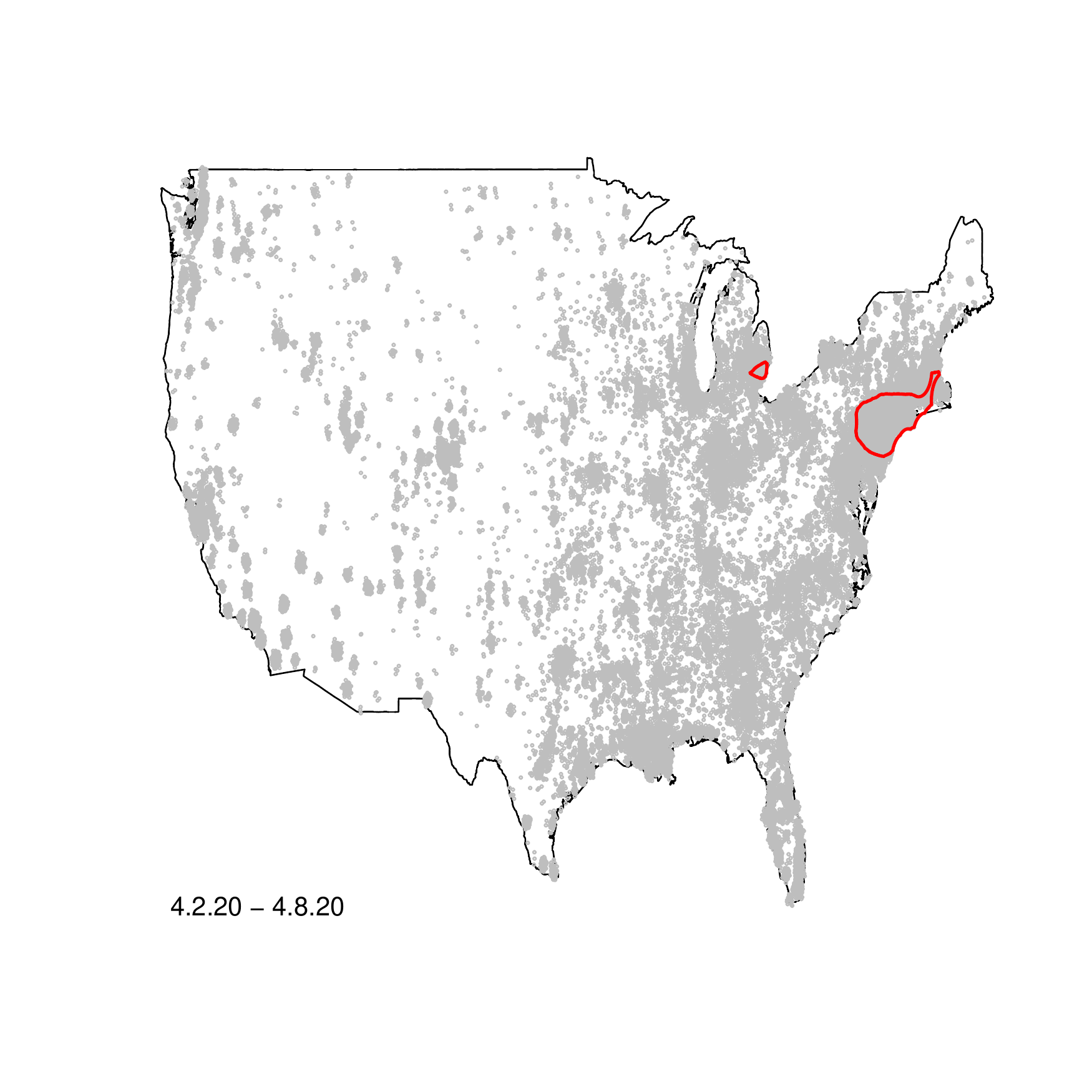}\\
 		\includegraphics[height=175pt,width=220pt]{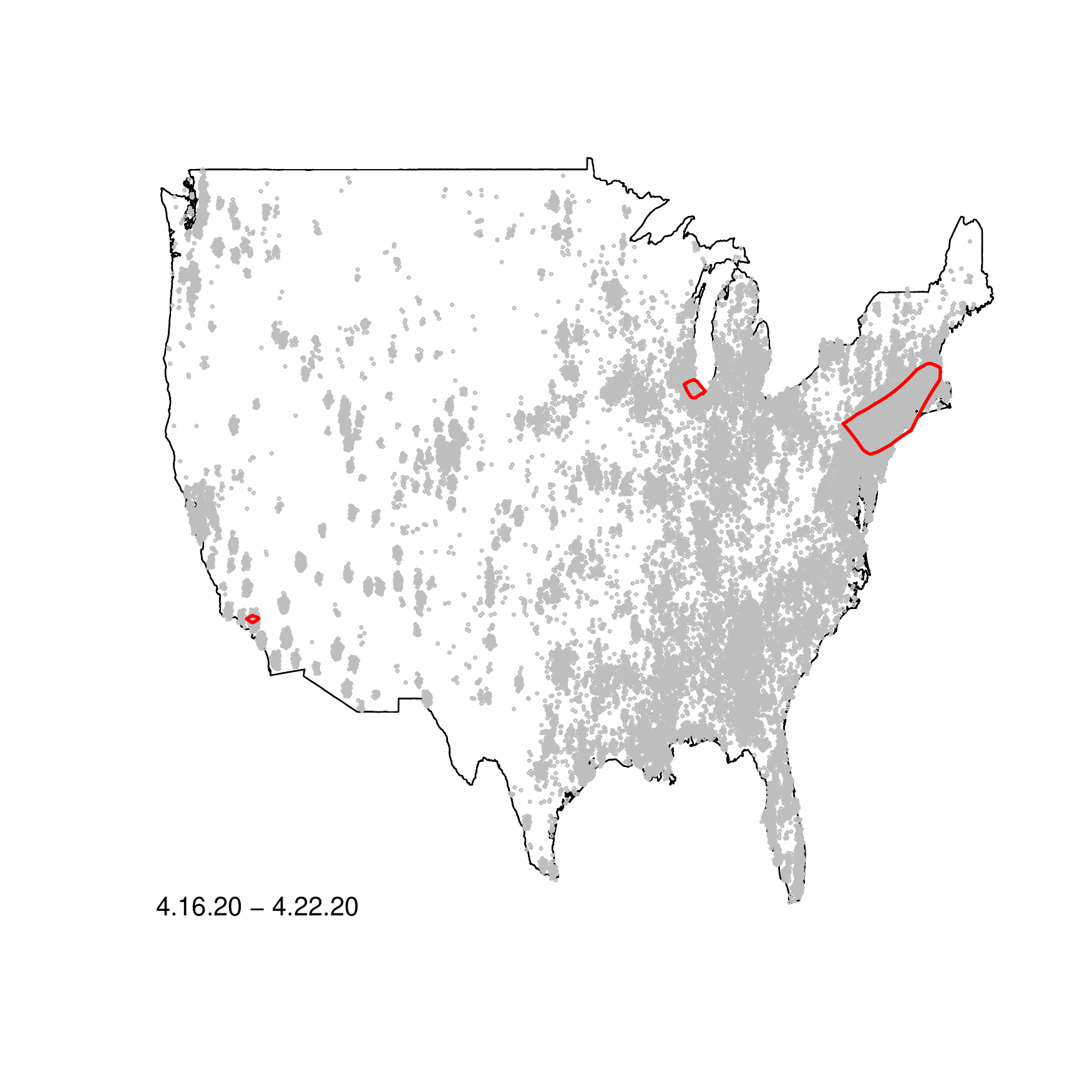}\includegraphics[height=175pt,width=220pt]{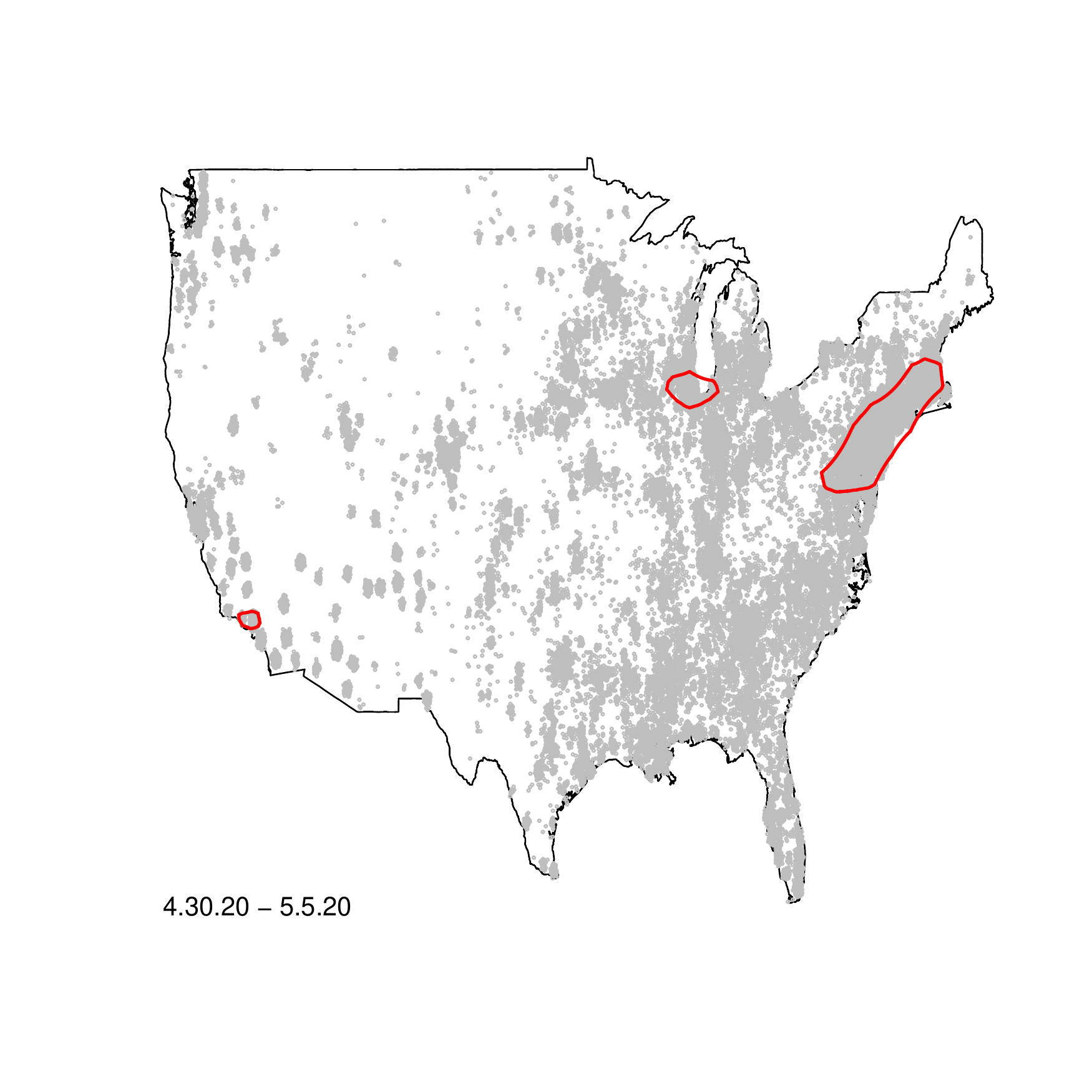}\\
 			\caption{Time and spatial evolution of HDRs when $\tau=0.5$ for confirmed COVID-19 cases in the United States by week between February 2020 and May 2020.}\label{coveeuu22}
  \end{figure}

HDRs estimators obtained for $\tau=0.5$ are presented in the interactive representation in Figure 3 in Section 3 of SM. Therefore, the empirical probability content of estimated clusters is, in this case, $0.5$. As before, the animation covers the period from February 27th to May 5th. In the first of the analyzed weeks, results are not very informative mainly due to the small sample size. Although two clusters are estimated in the second of the weeks, they provide more useful information to manage the pandemic because their locations are clear. One of them is located in Seattle and the another one includes the cities of Boston and New York reaching Detroit. The number of clusters increases in the third week reflecting the expansion of the disease and also its geographical evolution. Specifically, four risky areas are detected on Seattle, San Francisco, Detroit and New York. However, only two areas (Chicago and the east coast of the country) register the $50$\% of cases from fifth to seventh week. Los Angeles is added as a new risky cluster in the last weeks of the performed analysis.  Again, Figure \ref{coveeuu22} shows the HDRs estimator for the third, 
sixth, eighth and tenth weeks.

 Plug-in reconstructions of HDRs have not been represented because, when $\tau=0.9$ and $\tau=0.8$, the estimated number of clusters and their geographical location is exactly the same than the obtained from the Algorithm \ref{al1}. When $\tau=0.5$, an only relevant difference is detected: the cluster sited on New York and Boston in the seventh and eighth week is divided into two connected components by the plug-in method. Therefore, both alternatives provides practically the same HDRs reconstructions when big samples sizes are considered and the biggest modes of the COVID-19 distribution are estimated selecting high values of $\tau$. Although the simplicity of plug-in methods is maybe its main advantage, Algorithm \ref{al1} can be useful if additional information on the HDR shape to be reconstructed is available. Therefore, this algorithm tackles the problem of HDR estimation from a completely different perspective with interesting advantages where geometry plays a fundamental role. In particular, unlike plug-in methods, the simple geometric structure of the estimator boundary makes possible to determine explicitly the distance between different clusters boundaries (arcs of circumference) of the estimated HDR.
 
 Finally, it is important to remark that this type of studies can be really useful to determine the areas where interventions to slow the
 progress of the disease must be carried out urgently or to identify where they have been really effective. The fight should be stronger in which areas where the COVID-19 incidence is clearly higher. Of course, they could be performed considering smaller periods of time (even daily) as long as the amount and quality of data was high enough.

 \section{Conclusions and discussion}\label{conclusions}
 The two main goals of this work are to study the practical performance of the new data-driven algorithm proposed in \cite{rosaa} for estimating a $r-$convex HDR and apply it for reconstructing the HDRs of confirmed cases of COVID-19 in the United States from February to May 2020. The route designed to reach this goal can be summarized
 as follows: (1) Defining formally the concepts of density level set, cluster and HDRs illustrating them through a COVID-19 data set (2) reviewing some of the main geometric aspects of this method, (3) describing in detail and motivating this practical algorithm, (4) checking its behavior through an extensive simulation study comparing it with classical plug-in HDR estimation methods and (5) illustrating its performance in order to reconstruct the clusters of confirmed cases of COVID-19 in the United States where big enough samples sizes are available.
 
 According to the results obtained, the algorithm analyzed in this work presents its best behavior for large sample sizes and the interest is to estimate the biggest modes of the distribution considering big values of $\tau$. The analysis performed in this work, where the Algorithm \ref{al1} and plug-in methods are used to reconstruct the HDRs of COVID-19 from weekly occurrences and detecting temporal changes in the spatial pattern of the virus, is under these assumptions. No remarkable differences were detected between the HDRs estimators provided by the two approaches.  Of course, this type of studies could be performed considering any other time periods or different geographical regions due to the large amount of data available for COVID-19.

  Finally, HDRs estimation theory can be seen as a powerful tool to fight against COVID-19. Although plug-in methods offers quite competitive results, the hybrid algorithm considered in this work provides flexible HDRs reconstructions from a geometrical perspective. Note that the explicit boundary of this estimator is fully determined which is an important characteristic when, for instance, areas or distances must be calculated  between clusters. A nonparametric test for comparing two or more populations in general dimension could be designed using distances between the estimated HDRs. The test statistic could measure, for example, the discrepancy among the HDRs estimators of these populations using the distances between their boundaries, fully known if the hybrid algorithm studied in this work is considered. \vspace{-.8cm}

  $$ $$
   \textbf{Acknowledgments.} P. Saavedra-Nieves acknowledges the financial support of Ministerio de Econom\'ia y Competitividad of the Spanish government under grants MTM2016-76969P and MTM2017-089422-P and ERDF. Author is also grateful to Rosa M. Crujeiras and Alberto Rodríguez-Casal for their help.
 
 \bibliographystyle{Chicago}
 \bibliography{Bibliography-MM-MC}

\end{document}